\documentclass[aps,prb,twocolumn,superscriptaddress]{revtex4-1}
\usepackage{amsmath,amssymb,bm,epsfig,graphicx,longtable}
\begin{document}

\title{Shortcomings of meta-GGA functionals when describing magnetism}
\author{Fabien Tran}
\affiliation{Institute of Materials Chemistry, Vienna University of Technology,
Getreidemarkt 9/165-TC, A-1060 Vienna, Austria}
\author{Guillaume Baudesson}
\affiliation{Institute of Materials Chemistry, Vienna University of Technology,
Getreidemarkt 9/165-TC, A-1060 Vienna, Austria}
\affiliation{Univ Rennes, ENSCR, CNRS, ISCR (Institut des Sciences Chimiques
de Rennes) - UMR 6226, F-35000 Rennes, France}
\author{Jes\'{u}s Carrete}
\affiliation{Institute of Materials Chemistry, Vienna University of Technology,
Getreidemarkt 9/165-TC, A-1060 Vienna, Austria}
\author{Georg K. H. Madsen}
\affiliation{Institute of Materials Chemistry, Vienna University of Technology,
Getreidemarkt 9/165-TC, A-1060 Vienna, Austria}
\author{Peter Blaha}
\affiliation{Institute of Materials Chemistry, Vienna University of Technology,
Getreidemarkt 9/165-TC, A-1060 Vienna, Austria}
\author{Karlheinz Schwarz}
\affiliation{Institute of Materials Chemistry, Vienna University of Technology,
Getreidemarkt 9/165-TC, A-1060 Vienna, Austria}
\author{David J. Singh}
\affiliation{Department of Physics and Astronomy, University of Missouri,
Columbia, Missouri 65211-7010, USA}

\begin{abstract}

Several recent studies have shown that SCAN, a functional belonging to the
meta-generalized gradient approximation (MGGA) family, leads to significantly overestimated
magnetic moments in itinerant ferromagnetic metals. However, this behavior is
not inherent to the MGGA level of approximation since TPSS, for instance,
does not lead to such severe overestimations. In order to provide a broader
view of the accuracy of MGGA functionals for magnetism, we extend the
assessment to more functionals, but also to antiferromagnetic solids.
The results show that to describe magnetism there is overall no real advantage
in using a MGGA functional compared to GGAs. For both types of approximation,
an improvement in ferromagnetic metals is necessarily accompanied by a
deterioration (underestimation) in antiferromagnetic insulators, and vice-versa.
We also provide some analysis in order to understand in more detail the
relation between the mathematical form of the functionals and the results.

\end{abstract}

\maketitle

\section{\label{introduction}Introduction}

The local density approximation\cite{KohnPR65} (LDA) and generalized gradient
approximation\cite{BeckePRA88,PerdewPRB92b} (GGA) of density functional
theory\cite{HohenbergPR64,KohnPR65} (DFT) usually provide a fair description
of the magnetism in itinerant ferromagnetic (FM) $3d$ metals, albeit a slight
overestimation of the magnetic moment can be obtained (see, e.g.,
Refs.~\onlinecite{BarbielliniJPCM90,SinghPRB91b,SharmaJCTC18}).
On the other hand, the LDA and GGA are inaccurate for antiferromagnetic (AFM)
insulators, where the $3d$ electrons are more localized
and the self-interaction error (SIE)\cite{PerdewPRB81} present in
LDA and GGA is more important. As a consequence, the atomic moment around the
transition-metal atom in AFM systems is clearly underestimated.\cite{TerakuraPRB84}

The exchange-correlation (xc) functionals of the meta-GGA (MGGA) level
of approximation\cite{VanVoorhisJCP98,StaroverovJCP03} should
in general be more accurate since they
use an additional ingredient, the kinetic-energy density (KED), which makes possible
to remove a portion of the SIE.\cite{StaroverovPRB04} The strongly constrained
and appropriately normed (SCAN) MGGA functional proposed recently by Sun
\textit{et al}.\cite{SunPRL15} was constructed in such a way that it satisfies all the 17
known mathematical constraints that can be imposed on a MGGA functional,
and was appropriately normed, i.e. made accurate, for particular systems.
The SCAN functional has been shown to be accurate for both molecules and solids,
\cite{SunPRL15,TranJCP16,ZhangNPJCM18,IsaacsPRM18} including systems bound by
noncovalent interactions provided that a dispersion term is added.
\cite{PengPRX16,BrandenburgPRB16,TranPRM19}
On the other hand, it has been realized that SCAN leads to magnetic moments in bulk
FM Fe, Co, and Ni that are by far too large.
\cite{IsaacsPRM18,JanaJCP18a,EkholmPRB18,FuPRL18,FuPRB19,MejiaRodriguezPRB19}
The overestimation of the magnetic moment with SCAN has also been observed in alloys
\cite{RomeroEPJB18,BuchelnikovPRB19} and surface systems.\cite{ShepardJCP19}

Nevertheless, this overestimation of magnetic moments is not inherent to the MGGA,
since other MGGA functionals like TPSS,\cite{TaoPRL03} revTPSS,\cite{PerdewPRL09}
and TM\cite{TaoPRL16} lead to values similar to PBE.\cite{SunPRB11b,JanaJCP18a,FuPRB19}
Interestingly, Mej\'{i}a-Rodr\'{i}guez and Trickey\cite{MejiaRodriguezPRB19}
showed that SCAN-L, a deorbitalized version of SCAN they proposed in
Refs.~\onlinecite{MejiaRodriguezPRA17,MejiaRodriguezPRB18},
leads to a magnetic moment which is similar to PBE,
while the results for the geometry and binding energy of molecules and
solids stay close to the original
SCAN.\cite{MejiaRodriguezPRA17,MejiaRodriguezPRB18,TranJCP18}

Regarding the general performance of MGGA functionals for magnetism in solids,
a few questions remain. For instance, not that many results for the atomic
magnetic moment in AFM systems have been reported. Recent tests on various
oxides have shown that SCAN underestimates the moment in some
cases like MnO or Fe$_{2}$O$_{3}$, but overestimates it in MnO$_{2}$.\cite{SaiGautamPRM18,LongPRM20}
In Refs.~\onlinecite{XiaoPRB14,KylanpaaPRM17},
the FM and AFM phases of VO$_{2}$ were studied with numerous functionals
including TPSS, revTPSS, MGGA\_MS0,\cite{SunJCP12} MGGA\_MS2,\cite{SunJCP13} and
SCAN. It was shown that the latter three functionals lead to moments that are larger
than those predicted by TPSS and revTPSS, especially for the AFM phase. Comparisons with reference
Monte-Carlo results for the AFM phase of VO$_{2}$ indicate that MGGA\_MS0, MGGA\_MS2,
and SCAN should be more accurate.\cite{XiaoPRB14,KylanpaaPRM17}
In Ref.~\onlinecite{LanePRB18}, the high-$T_{c}$ superconductor parent compound
La$_{2}$CuO$_{4}$ were studied with TPSS, revTPSS, and SCAN, the latter giving
a value of the moment of the Cu atom in good agreement with experiment, while
a clear underestimation is obtained with TPSS and revTPSS.
A recent study by Zhang \textit{et al}. has shown that SCAN underestimates the atomic
magnetic moment in MnO, FeO, CoO, and NiO.\cite{Zhang19}
Finally, it has been reported that SCAN leads to a magnetic moment in AFM $\alpha$-Mn
that is much larger than with PBE.\cite{PulkkinenPRB20}

Despite these results for FM and AFM systems, what is missing is a
more systematic study of the relative performance of MGGA functionals for magnetism.
In particular, besides SCAN and (rev)TPSS, not much is known about the
performance of other MGGA functionals. It is also not fully clear
to which extent an increase (e.g., with respect to PBE) of the moment
in FM solids with a given MGGA necessarily translates into an increase for AFM solids.
In the present work, a more systematic comparison of MGGA functionals
for magnetism is presented. FM and AFM systems are considered, as well
as nonmagnetic (NM) ones. The latter may be wrongly described as magnetic with
DFT methods.\cite{TranPRB12,FuPRL18} The search for a possible magnetic ground state
for the supposedly NM systems is restricted to FM.

The paper is organized as follows. A description of the methods is
given in Sec.~\ref{methods}. In Sec.~\ref{results}, the results are presented
and discussed, and Sec.~\ref{summary} gives the summary of this work.

\section{\label{methods}Methods}

Among the plethora of MGGA functionals that exist,\cite{DellaSalaIJQC16}
we selected a few representatives of various types;
empirical vs. non-empirical, old standard vs. modern,
general purpose vs. specialized for a particular property.
These are the following:
BR89,\cite{BeckePRA89} TPSS,\cite{TaoPRL03} revTPSS,\cite{PerdewPRL09}
MGGA\_MS2,\cite{SunJCP13} MVS,\cite{SunPNAS15}
SCAN,\cite{SunPRL15} TM,\cite{TaoPRL16} HLE17,\cite{VermaJPCC17}
SCAN-L,\cite{MejiaRodriguezPRA17,MejiaRodriguezPRB18} and
TASK.\cite{AschebrockPRR19}
Here, we just mention that
HLE17 consists in a simple empirical rescaling of TPSS exchange 
and correlation, which are multiplied by 1.25 and 0.5, respectively,
in order to achieve better results for the band gaps of solids and
excitation energies of molecules.\cite{VermaJPCC17}
The very recent TASK from Aschebrock and K\"{u}mmel,\cite{AschebrockPRR19}
which is an exchange functional that is combined with LDA
correlation,\cite{PerdewPRB92a} also provides accurate band gaps, but in contrast to
HLE17 it was constructed in a nonempirical way without tuning parameters.
All MGGAs except SCAN-L and BR89 are
$t$-MGGAs since they depend on the Kohn-Sham (KS) KED
$t_{\sigma}=\left(1/2\right)\sum_{i=1}^{N_{\sigma}}
\nabla\psi_{i\sigma}^{*}\cdot\nabla\psi_{i\sigma}$
($\sigma$ is the spin index).
SCAN-L is a deorbitalized version of SCAN. A $t$-MGGA is
deorbitalized\cite{PerdewPRB07,MejiaRodriguezPRA17,BienvenuJCTC18} by replacing
$t_{\sigma}$ by an orbital-free (and thus necessarily approximate) expression that
depends on $\rho_{\sigma}$, $\nabla\rho_{\sigma}$, and
$\nabla^{2}\rho_{\sigma}$, and is thereby turned into a $\nabla^{2}\rho$-MGGA,
which is an explicit functional of the electron density.
The BR89 exchange functional of Becke and Roussel, which was proposed as an
accurate approximation to the Hartree-Fock exchange energy,\cite{BeckePRA89}
depends on both $t_{\sigma}$ and $\nabla^{2}\rho_{\sigma}$.
BR89, which is combined in this work with LDA correlation,\cite{PerdewPRB92a}
is tested since it differs radically from the other MGGAs in terms of
construction. Therefore, it may be interesting to see the results obtained
with such a functional.

For comparison, the results obtained with the
$\left(t_{\sigma},\nabla^{2}\rho_{\sigma}\right)$-dependent
modified Becke-Johnson (mBJLDA) potential,\cite{TranPRL09}
LDA,\cite{PerdewPRB92a} and the two GGAs
PBE\cite{PerdewPRL96} and HLE16\cite{VermaJPCL17} are also shown.
HLE16 was constructed specifically for band gaps in a similar way as HLE17, by rescaling
the exchange and correlation parts (with 1.25 and 0.5, respectively) of the
highly parameterized GGA HCTH/407.\cite{BoeseJCP01} The mBJLDA potential was also
proposed specifically for band gap calculations, for which it is currently the most
accurate semilocal method.\cite{TranJPCA17,BorlidoJCTC19,TranJAP19}

The calculations were performed with the all-electron WIEN2k code,\cite{WIEN2k,BlahaJCP20}
which is based on the linearized augmented plane-wave (LAPW)
method.\cite{AndersenPRB75,Singh,KarsaiCPC17}
Among the functionals, HLE17 and TASK were taken
from the library of exchange-correlation functionals Libxc.\cite{MarquesCPC12,LehtolaSX18}
The MGGA functionals are not implemented self-consistently
in WIEN2k, i.e., only the total energy can be calculated.
Nevertheless, it is still possible to calculate the magnetic moment
without the corresponding MGGA potential by using only the total energy.
The fixed spin-moment (FSM) method\cite{SchwarzJPF84}
(used for a part of the calculations presented in
Refs.~\onlinecite{FuPRL18,FuPRB19,MejiaRodriguezPRB19})
can be used to calculate the magnetic moment of FM systems.
The FSM method can not be applied to AFM systems, nevertheless, in the same spirit,
the atomic moment can to some extent be constrained to have a chosen value.
This can be done by adding and subtracting a constant shift $C$
to the spin-up and spin-down xc potentials (of a given GGA functional), respectively,
inside the LAPW sphere surrounding a transition-metal atom:
\begin{equation}
v_{\text{xc+shift},\sigma}^{\text{GGA}}(\bm{r}) =
v_{\text{xc},\sigma}^{\text{GGA}}(\bm{r}) + \sigma C,
\label{vxcshift}
\end{equation}
where $\sigma=1$ ($-1$) for spin-up (spin-down) electrons.
Supposing that for an atom the spin-up electrons are majority, a negative (positive) $C$
should increase (decrease) the magnitude of the spin magnetic moment.
Obviously, in order to keep the AFM state, shifts $C$ of the same magnitude,
but with opposite signs have to be applied to the transition-metal atoms with opposite
sign of the magnetic moment. As with the FSM method for FM solids, the variational
principle is used: for a given MGGA functional, the spin atomic moment
is the one obtained at the value of $C$ which leads to the lowest total energy.

The fact that MGGA functionals are applied non-self-consistently also
means that for both the FSM and $C$-shift methods
the xc potential corresponding to another functional,
typically a GGA, has to be used to generate the orbitals $\psi_{i\sigma}$.
In a recent study\cite{TranJCP19} we showed that the non-self-consistent
calculation of band gaps with MGGAs can be done accurately, provided that
an appropriate GGA potential to generate the orbitals is chosen.
The criterion to choose the potential was based on
the variational principle; among a plethora of GGA potentials, the
one that is chosen is the one yielding orbitals that lead to the
lowest total MGGA energy. As expected, these optimal orbitals also lead to
band gaps that are the closest to the true MGGA band gaps obtained
self-consistently from another code. That just means that in order to reproduce the true
(i.e., self-consistent) MGGA results, one should use
GGA orbitals which, according to the variational principle, are the closest
to the MGGA orbitals. In the present study, the GGA orbitals that are used
are those recommended in Ref.~\onlinecite{TranJCP19}, namely
RPBE\cite{HammerPRB99} (for TPSS, revTPSS, MGGA\_MS2, SCAN, TM, and SCAN-L),
EV93PW91\cite{EngelPRB93,PerdewPRB92b} (for MVS),
and mRPBE\cite{TranJCP19} (for HLE17).
For TASK and BR89, not considered in Ref.~\onlinecite{TranJCP19},
the orbitals generated by the GGAs RPBE and PBE
potentials (among all GGA potentials that
we have tried, those listed in Ref.~\onlinecite{TranJCP19}), respectively,
lead to the lowest total energy.
Therefore, in the present study the TASK and BR89 functionals have
been calculated with the RPBE and PBE orbitals, respectively.

In order to validate our procedure, the magnetic moments
obtained self-consistently with the
VASP\cite{KressePRB96} and GPAW codes,\cite{EnkovaaraJPCM10,FerrighiJCP11}
both are based on the projector augmented wave (PAW) method,\cite{BlochlPRB94b}
will be compared to our results for a few test cases.

Another technical point concerns the definition of the atomic magnetic moment
in AFM solids. Since there is no unique way to define an atom in a molecule or
solid, the region of integration around an atom to calculate the atomic moment
can to some extent be chosen arbitrarily. In solid-state physics, basis sets like
LAPW\cite{AndersenPRB75} or PAW\cite{BlochlPRB94b} use spheres surrounding the
atoms, which are commonly used to define the atoms and to calculate the corresponding
magnetic moment.
As shown in Sec.~\ref{results}, different radii may lead to
quite different values of the atomic moment. A way to define the region
of integration independently of the basis set is to use the quantum theory
of atoms in molecules (QTAIM) of Bader.\cite{Bader90,BaderCR91}
In QTAIM, the volume of an atom (usually called basin) is delimited by a surface
with zero flux in the gradient of the electron density.
The atomic moments of the AFM solids presented in
Sec.~\ref{results} were obtained using the QTAIM as implemented
in the Critic2 code.\cite{OterodelaRozaCPC09,OterodelaRozaCPC14}
There is probably only very few works comparing the value of
the atomic moment obtained from different definitions of the atom,
therefore such a comparison will also be discussed in Sec.~\ref{results}.

\begin{table} 
\caption{\label{geometry}Experimental\cite{BergerhoffJCICS83,BelskyAC02}
lattice constants (in \AA) and angles (in degrees) of the unit cell for
the solids considered in this work. When necessary, the positions of atoms
(in fractional coordinates) are also indicated.
The space group number is indicated in parenthesis.
For the AFM solids, the AFM order leads to a lowering
of the symmetry (second indicated space group).}
\label{table1}
\begin{tabular}{lcccccc} 
\hline
\hline
Solid                     & $a$ & $b$ & $c$ & $\alpha$ & $\beta$ & $\gamma$ \\
\hline
NM \\
Sc (194)                  & 3.309 & 3.309 &  5.273 & 90 & 90   & 120  \\
V (229)                   & 3.028 & 3.028 &  3.028 & 90 & 90   &  90  \\
Y (194)                   & 3.652 & 3.652 &  5.747 & 90 & 90   & 120  \\
Pd (225)                  & 3.881 & 3.881 &  3.881 & 90 & 90   &  90  \\
Pt (225)                  & 3.916 & 3.916 &  3.916 & 90 & 90   &  90  \\
\\
FM \\
Fe (229)                  & 2.867 & 2.867 &  2.867 & 90 & 90   &  90  \\
Co (194)                  & 2.507 & 2.507 &  4.070 & 90 & 90   & 120  \\
Ni (225)                  & 3.523 & 3.523 &  3.523 & 90 & 90   &  90  \\
FeCo (221)                & 2.857 & 2.857 &  2.857 & 90 & 90   &  90  \\
ZrZn$_{2}$ (227)          & 7.396 & 7.396 &  7.396 & 90 & 90   &  90  \\
\multicolumn{7}{l}{\hspace{0.5cm}Zr(1/8,1/8,1/8), Zn(1/2,0,0)} \\
YFe$_{2}$ (227)           & 7.363 & 7.363 &  7.363 & 90 & 90   &  90  \\
\multicolumn{7}{l}{\hspace{0.5cm}Y(1/8,1/8,1/8), Fe(1/2,0,0)} \\
Ni$_{3}$Al (221)          & 3.568 & 3.568 &  3.568 & 90 & 90   &  90  \\
\multicolumn{7}{l}{\hspace{0.5cm}Ni(1/2,1/2,0), Al(0,0,0)} \\
\\
AFM \\
Cr$_{2}$O$_{3}$ (167,146) & 4.953 & 4.953 & 13.588 & 90 & 90   & 120  \\
\multicolumn{7}{l}{\hspace{0.5cm}Cr(0,0,0.3475), O(0.3058,0,1/4)} \\
Fe$_{2}$O$_{3}$ (167,146) & 5.035 & 5.035 & 13.747 & 90 & 90   & 120  \\
\multicolumn{7}{l}{\hspace{0.5cm}Fe(0,0,0.35534), O(0.3056,0,1/4)} \\
MnO (225,166)             & 4.445 & 4.445 &  4.445 & 90 & 90   &  90  \\ 
FeO (225,166)             & 4.334 & 4.334 &  4.334 & 90 & 90   &  90  \\ 
CoO (225,166)             & 4.254 & 4.254 &  4.254 & 90 & 90   &  90  \\ 
NiO (225,166)             & 4.171 & 4.171 &  4.171 & 90 & 90   &  90  \\ 
CuO (15,14)               & 4.684 & 3.423 &  5.129 & 90 & 99.54&  90  \\
\multicolumn{7}{l}{\hspace{0.5cm}Cu(1/4,1/4,0), O(0,0.4184,1/4)} \\
CrSb (194,164)            & 4.122 & 4.122 &  5.464 & 90 & 90   & 120  \\
CrSb$_{2}$ (58,14)        & 6.028 & 6.874 &  3.272 & 90 & 90   &  90  \\
\multicolumn{7}{l}{\hspace{0.5cm}Cr(0,0,0), Sb(0.1835,0.3165,0.32)} \\
\hline
\hline
\end{tabular} 
\end{table} 

The solids that are considered for the present study are listed in
Table~\ref{geometry} along with their experimental
geometry\cite{BergerhoffJCICS83,BelskyAC02} used for the calculations.
The set is divided into five NM, seven FM, and nine AFM solids.
We mention that for Fe,
Fu and Singh\cite{FuPRL18,FuPRB19} considered the effect of the lattice
constant on the magnetic moment. It can be non-negligible if a functional
leads to an inaccurate lattice constant that is far from the experimental one.
For instance, compared to the value obtained at the experimental lattice constant,
the LDA magnetic moment is smaller by $\sim0.2~\mu_{\text{B}}$ when it is
calculated at the corresponding LDA lattice constant.
For the present work, no optimization of the geometry was done, i.e., the calculations
were performed at the same (experimental) geometry with all functionals.
The reasons are the following. First, the effect of geometry and functional
on the magnetic moment would be entangled, which would lead to a more
complicated analysis and discussion of the results. Second, some of the tested
functionals lead to extremely poor lattice constants
(see Sec.~\ref{Discussion}), so that it would not make sense to calculate a
property at such inaccurate geometry.

\section{\label{results}Results}

\subsection{\label{orbitals_volume}Choice of orbitals and atomic region}

\begin{table*}
\caption{\label{AFM1}Spin atomic magnetic moment $\mu_{\text{S}}$
(in $\mu_{\text{B}}$) of AFM MnO, FeO, CoO, and NiO.
The results in the first three columns were obtained with WIEN2k
using different atomic sphere sizes (their radii, in bohr, are indicated)
for calculating $\mu_{\text{S}}$. The results in the last three columns were
obtained with three different codes and using the Bader volume for
calculating $\mu_{\text{S}}$. The WIEN2k results for the MGGAs were
obtained with the $C$-shift method [Eq.~(\ref{vxcshift})] and using either
the RPBE or PBE (results in parenthesis) orbitals. All VASP and GPAW
results were obtained self-consistently. The calculations were done at the
geometry specified in Table~\ref{geometry}.}
\begin{ruledtabular}
\begin{tabular}{lcccccc}
\multicolumn{1}{l}{} &
\multicolumn{3}{c}{WIEN2k} &
\multicolumn{3}{c}{Bader volume} \\
\cline{2-4}\cline{5-7}
\multicolumn{1}{l}{} &
\multicolumn{1}{c}{Small sphere} &
\multicolumn{1}{c}{Medium sphere} &
\multicolumn{1}{c}{Large sphere} &
\multicolumn{1}{c}{WIEN2k} &
\multicolumn{1}{c}{VASP} &
\multicolumn{1}{c}{GPAW} \\
\hline
MnO \\
Sphere radius & 2.05 & 2.25 & 2.45                             \\
PBE           & 4.19 & 4.31 & 4.38 & 4.39        & 4.38 & 4.37 \\
TPSS          & 4.21 & 4.33 & 4.40 & 4.41 (4.42) & 4.40 & 4.40 \\
SCAN          & 4.32 & 4.44 & 4.51 & 4.53 (4.53) & 4.50 & 4.49 \\
\\
FeO \\
Sphere radius & 2.00 & 2.20 & 2.40                             \\
PBE           & 3.39 & 3.45 & 3.48 & 3.48        & 3.46 & 3.48 \\
TPSS          & 3.43 & 3.49 & 3.53 & 3.52 (3.52) & 3.49 & 3.51 \\
SCAN          & 3.53 & 3.59 & 3.63 & 3.62 (3.62) & 3.59 & 3.60 \\
\\
CoO \\
Sphere radius & 1.95 & 2.15 & 2.35                             \\
PBE           & 2.42 & 2.45 & 2.45 & 2.45        & 2.43 & 2.45 \\
TPSS          & 2.45 & 2.50 & 2.51 & 2.50 (2.39) & 2.48 & 2.51 \\
SCAN          & 2.55 & 2.59 & 2.61 & 2.60 (2.42) & 2.59 & 2.61 \\
\\
NiO \\
Sphere radius & 1.90 & 2.10 & 2.30                             \\
PBE           & 1.38 & 1.38 & 1.37 & 1.37        & 1.32 & 1.36 \\
TPSS          & 1.47 & 1.47 & 1.46 & 1.46 (1.46) & 1.42 & 1.45 \\
SCAN          & 1.62 & 1.62 & 1.61 & 1.60 (1.60) & 1.59 & 1.60 \\
\end{tabular}
\end{ruledtabular}
\end{table*}

Before discussing the relative performance of the functionals, we show in
Table~\ref{AFM1} some results for MnO, FeO, CoO, and NiO in order to
illustrate the influence of self-consistency and choice of integration region
(i.e., definition of the atom) on the spin atomic magnetic moment $\mu_{S}$.
As mentioned in Sec.~\ref{methods} and discussed in more detail in
Ref.~\onlinecite{TranJCP19}, the GGA RPBE potential is the optimal one
for the MGGAs TPSS and SCAN. The importance of using the orbitals
generated by the RPBE potential is visible in the case of CoO;
compared to using the PBE orbitals (the usual default choice)
$\mu_{S}$ is larger by about $0.1$ and $0.2~\mu_{\text{B}}$ for TPSS
and SCAN, respectively. Such differences are not negligible, and in fact
we can also see that using the RPBE orbitals brings the WIEN2k results into
agreement with those obtained self-consistently with VASP and GPAW codes.
We just note that for NiO there is a discernible discrepancy between VASP and
the two other codes. After looking into this issue, we came to the conclusion that
the problem may be due to VASP projectors that are unadapted for the
particular case of NiO.
For MnO, FeO, and NiO, using either PBE or RPBE orbitals does not matter at all. 
Indeed, we found that
in the case of CoO the optimal choice of GGA orbitals
(as listed in Sec.~\ref{methods}) is critical to avoid values of
$\mu_{S}$ that are to small by 0.1$-$$0.2~\mu_{\text{B}}$
as would be obtained with the PBE orbitals.
The other cases where using the optimal orbitals
(instead of the standard PBE) is also important concern a few of the
AFM and FM solids when the MGGA HLE17 is used,
for which the optimal potential is mRPBE.

From the results in Table~\ref{AFM1}, the other main observation
is that the atomic volume inside which the atomic moment $\mu_{S}$
is calculated may have some influence as well.
WIEN2k calculations were done with three different radii
for the atomic sphere, which were chosen to lie within a reasonable range from a
physical point of view, in particular not too small in order to avoid core leakage.
The value of $\mu_{S}$ calculated from within the sphere varies the most for
MnO; from the smallest sphere (2.05~bohr) to the largest (2.45~bohr)
$\mu_{S}$ increases by about $0.2~\mu_{\text{B}}$, which is rather significant.
On the other hand, there is no change in $\mu_{S}$ for NiO
(since Ni has the largest nuclear charge $Z$
and therefore the most localized $3d$ electrons), but
of course reducing the sphere size further would at some point lead to a
decrease of the magnetic moment. For FeO and CoO, the variation of $\mu_{S}$
is intermediate between MnO and NiO. The other important point to note is that
in all cases the magnetic moment obtained with the largest sphere agrees
with the one obtained using the Bader volume.

Since the Bader volume is uniquely defined and the corresponding
$\mu_{S}$ agrees with the value obtained with the largest LAPW atomic sphere, using it
as the atomic region to calculate $\mu_{S}$ can be considered as a pretty sound choice.
Therefore, the comparison of the functionals for the atomic magnetic moment in
AFM solids will be based on the values obtained with the Bader volume.

\subsection{\label{functionals}Comparison of functionals}

\subsubsection{\label{FM_solids}Ferromagnetic solids}

\begin{table*}
\caption{\label{FM}Spin magnetic moment $\mu_{S}$ (in $\mu_{\text{B}}$ per formula unit)
of FM solids. The experimental values for Fe, Co, and Ni are also spin magnetic
moments. The results in parenthesis for FeCo are the atomic moments
(defined according to the Bader volume)
on Fe and Co. The results for the MGGA functionals were obtained with the FSM method.
The calculations were done at the geometry specified in Table~\ref{geometry}.
The largest discrepancies with respect to experiment are underlined.}
\begin{ruledtabular}
\begin{tabular}{lccccccc} 
Method  & Fe   & Co   & Ni   & FeCo & YFe$_{2}$ & ZrZn$_{2}$ & Ni$_{3}$Al \\
\hline
LDA  & 2.20 & 1.59 & 0.62 & 4.51(2.76,1.75) & 3.20 & \underline{0.67} & \underline{0.71} \\
PBE         & 2.22 & 1.62 & 0.64 & 4.55(2.81,1.75) & 3.38      & \underline{0.90}       & \underline{0.77}       \\
HLE16          & \underline{2.74} & 1.73 & 0.64 & 4.77(3.06,1.71) & 2.91      & \underline{1.72}       & \underline{0.57}       \\
mBJLDA          & 2.51 & 1.69 & 0.73 & 4.60(2.87,1.73) & 3.60      & \underline{0.95}       & \underline{0.85}       \\
TPSS       & 2.23 & 1.65 & 0.66 & 4.63(2.86,1.77) & 3.66      & \underline{0.82}       & \underline{0.80}       \\
revTPSS    & 2.29 & 1.67 & 0.68 & 4.67(2.89,1.79) & 3.71      & \underline{0.83}       & \underline{0.83}       \\
MGGA\_MS2    & 2.30 & 1.74 & 0.73 & 4.80(2.96,1.84) & \underline{3.81}      & \underline{0.98}       & \underline{0.95}       \\
MVS  & \underline{2.71} & \underline{1.80} & \underline{0.76} & \underline{4.88}(3.01,1.87) & 3.69      & \underline{1.04}       & \underline{0.84}       \\
SCAN       & 2.63 & \underline{1.79} & \underline{0.76} & \underline{4.86}(2.99,1.87) & \underline{3.88}      & \underline{1.08}       & \underline{0.95}       \\
TM      & 2.25 & 1.68 & 0.69 & 4.69(2.89,1.79) & 3.67      & \underline{0.88}       & \underline{0.85}       \\
HLE17       & 2.67 & 1.72 & 0.65 & 4.74(3.02,1.71) & 3.60      & \underline{1.27}       & \underline{0.66}       \\
TASK            & \underline{2.75} & \underline{1.83} & \underline{0.76} & \underline{4.91}(3.02,1.89) & 3.45      & \underline{1.37} & \underline{0.89}       \\
SCAN-L  & 2.13 & 1.65 & 0.68 & 4.62(2.85,1.77) & 3.26      & \underline{1.03}       & \underline{0.84}       \\
BR89      & 2.45 & 1.67 & 0.66 & 4.64(2.86,1.78) & 3.74      &  \underline{0.83}       &  \underline{0.76}       \\
Expt.    &
1.98,\footnotemark[1]2.05,\footnotemark[2]2.08\footnotemark[3] &
1.52,\footnotemark[3]1.58,\footnotemark[2]\footnotemark[4]1.55-1.62\footnotemark[1] &
0.52,\footnotemark[3]0.55\footnotemark[2]\footnotemark[5] &
4.54\footnotemark[6] &
2.90\footnotemark[7] &
0.17\footnotemark[8] &
0.23\footnotemark[9] \\
\end{tabular}
\end{ruledtabular}
\footnotetext[1]{Ref.~\onlinecite{ChenPRL95}.}
\footnotetext[2]{Ref.~\onlinecite{Scherz03}.}
\footnotetext[3]{Ref.~\onlinecite{ReckPR69}.}
\footnotetext[4]{Ref.~\onlinecite{MoonPR64}.}
\footnotetext[5]{Ref.~\onlinecite{MookJAP66}.}
\footnotetext[6]{Ref.~\onlinecite{DiFabrizioPRB89}.}
\footnotetext[7]{Ref.~\onlinecite{BuschowJAP70}.}
\footnotetext[8]{Refs.~\onlinecite{UhlarzPRL04,YellandPRB05}.}
\footnotetext[9]{Ref.~\onlinecite{DeBoerJAP69}.}
\end{table*}

We start the discussion on the comparison of the functionals with the FM solids.
The results for the spin magnetic moment $\mu_{S}$ (per formula unit)
are shown in Table~\ref{FM}. It is known that the GGAs (and sometimes also the LDA)
slightly overestimate the magnitude of $\mu_{S}$ in itinerant metals like Fe, Co, or
Ni.\cite{BarbielliniJPCM90,SinghPRB91b}
For these systems, the overestimation with LDA and the standard GGA PBE is in the range
$0.05$$-$$0.2~\mu_{\text{B}}$. The other GGA considered in this work,
HLE16, leads to unpredictable results, since it yields a
moment that is much larger (by $0.5~\mu_{\text{B}}$) than the one predicted by PBE for Fe,
but to identical values for Ni, while the increase is $0.1~\mu_{\text{B}}$
for Co. However, such behavior with HLE16 is not that surprising since,
as shown in Ref.~\onlinecite{TranJPCA17} and discussed in Sec.~\ref{Discussion},
it has a strong enhancement factor
that leads to an xc potential with very large oscillations and therefore
possibly unexpected results. The results obtained for the compounds show that
LDA and PBE are accurate for FeCo, but overestimate $\mu_{S}$ for
YFe$_{2}$ and significantly for ZrZn$_{2}$ and Ni$_{3}$Al.
However, for the latter two systems
spin fluctuations, which require a treatment beyond standard DFT,
are supposed to significantly reduce the measured moment (see discussion in
Ref.~\onlinecite{MazinPRB04} for ZrZn$_{2}$ and in
Refs.~\onlinecite{AguayoPRL04,OrtenziPRB12} for Ni$_{3}$Al).
The results with HLE16 are again disparate; compared to PBE, $\mu_{S}$
is increased for FeCo and ZrZn$_{2}$, but reduced
for YFe$_{2}$ and Ni$_{3}$Al. HLE16 leads to the best agreement with
experiment for YFe$_{2}$, but to the worst for ZrZn$_{2}$.

Turning to the results obtained with the MGGA methods, we mention again that
several studies have already reported that SCAN, which is highly successful
in solid-state physics for total-energy calculations,
\cite{PengPRX16,ZhangNPJCM18,IsaacsPRM18,KovacsJCP19,TranPRM19,YangPRB19}
clearly overestimates the magnetic moment in itinerant metals.
\cite{IsaacsPRM18,JanaJCP18a,EkholmPRB18,FuPRL18,FuPRB19,MejiaRodriguezPRB19}
Those studies considered mostly Fe, Co, and Ni which were considered.
For the intermetallic ferromagnets considered here, the overestimation of $\mu_{S}$ with SCAN
is also substantial. In fact, among all xc methods SCAN leads to one of the largest
overestimations except for Fe and ZrZn$_{2}$.
This makes SCAN an inaccurate functional for
itinerant metals in general. The deorbitalized SCAN-L leads to (much) smaller value of
$\mu_{S}$ compared to its parent functional, confirming the results from
Ref.~\onlinecite{MejiaRodriguezPRB19}. Interestingly, among all methods SCAN-L
leads to one of the smallest magnetic moments for Fe.
Thus one may suppose that the orbital dependence in SCAN,
which is thought to be crucial for reducing the SIE, is at the same time
problematic for itinerant systems.

Among the other MGGAs, TPSS, revTPSS, TM, and BR89 lead on average to the
smallest overestimations of the magnetic moment and to values that
are only slightly larger (usually by less than $0.1~\mu_{\text{B}}$) than PBE.
Note that, for ZrZn$_{2}$, TPSS and revTPSS give $\mu_{S}=0.82-0.83~\mu_{\text{B}}$,
which is smaller than $0.90~\mu_{\text{B}}$ obtained with PBE.
Concerning the other MGGA functionals, very large overestimations of the
magnetic moment are obtained with MGGA\_MS2 (YFe$_{2}$ and Ni$_{3}$Al),
MVS (Fe, Co, Ni, and FeCo), HLE17 (ZrZn$_{2}$), and TASK (for all
systems except YFe$_{2}$).
On the other hand, HLE17 leads to a moment of $0.66~\mu_{\text{B}}$ for
Ni$_{3}$Al, which is smaller than the PBE value of $0.77~\mu_{\text{B}}$.
Thus, HLE17 behaves in an erratic way just like HLE16 does. 
The mBJLDA potential clearly overestimates $\mu_{S}$ in all cases,
but never leads to one of the most extreme values.

In summary, all functionals lead to overestimations of the magnetic moment,
at least if the effects due to spin fluctuations are ignored.
LDA, the GGA PBE, and the MGGAs TPSS, revTPSS, TM, SCAN-L, and BR89 lead to the smallest deviations
with respect to experiment, while TASK, SCAN, and MVS give the largest overestimations.
The results obtained with the GGA HLE16 and MGGA HLE17 are erratic.
In passing, we note that hybrid functionals also lead to
magnetic moments in metals that are greatly overestimated, as shown
in Refs.~\onlinecite{PaierJCP06,JangJM11,JangJPSJ12,JanthonJCTC14,GaoSSC16,TranPRM18,JanaJCP18a,JanaJCP20}
for the screened HSE06.\cite{HeydJCP03,KrukauJCP06}
The same conclusion applies to the GLLB-SC potential\cite{KuismaPRB10}
(see Ref.~\onlinecite{TranPRM18} for results). Furthermore,
we would like to provide two additional informations related to SCAN-L:
(a) Among the other orbital-free KED
used in Refs.~\onlinecite{MejiaRodriguezPRA17,TranJCP18} for deorbitalizing SCAN,
we also tested GEA2L; it gives magnetic moments that are quasi-identical to SCAN-L.
(b) A deorbitalization of TASK leads to a reduction of the
magnetic moment that is roughly similar to SCAN-L. As discussed later in
Sec.~\ref{Discussion}, this is due to a particular feature in the analytical forms
of the SCAN and TASK functionals and differences between the iso-orbital indicator
$\alpha$ and its deorbitalized version $\alpha_{L}$.

\begin{table} 
\caption{\label{FM_E}Magnetic energy $-\Delta E_{\text{tot}}$
(in meV per formula unit) of FM solids.
The results for the MGGA functionals were obtained with the FSM method.
The calculations were done at the geometry
specified in Table~\ref{geometry}.}
\begin{ruledtabular}
\begin{tabular}{lccccccc} 
Method          & Fe   & Co   & Ni   & FeCo & YFe$_{2}$ & ZrZn$_{2}$ & Ni$_{3}$Al \\
\hline
LDA             &  446 &  200 &   51 &  987 & 469       &   8        &  9   \\
PBE             &  565 &  256 &   62 & 1260 & 658       &  47        & 22   \\
HLE16           & 2050 &  909 &  132 & 3534 & 4186      & 247        & 24   \\
TPSS           &  640 &  290 &   71 & 1447 & 762       &  25        & 34   \\
revTPSS        &  678 &  315 &   77 & 1542 & 829       &  25        & 38   \\
MGGA\_MS2       &  868 &  414 &  109 & 2074 & 1123       &  68        & 87   \\
MVS     & 1417 &  685 &  129 & 2734 & 2036      &  97        & 50   \\
SCAN      & 1061 &  557 &  132 & 2503 & 1434      & 137        & 82   \\
TM        &  711 &  333 &   84 & 1632 & 889       &  46        & 48   \\
HLE17    & 1491 &  647 &  107 & 2773 & 2205      & 149        & 31   \\
TASK            & 1630 &  789 &  148 & 3314 & 2620      & 225        & 67   \\
SCAN-L    &  623 &  273 &   74 & 1473 & 694       & 109        & 44   \\
BR89      &  771 &  372 &   79 & 1663 & 928       & 39         & 30   \\
\end{tabular} 
\end{ruledtabular}
\end{table}

Table~\ref{FM_E} shows the results obtained for the magnetic energy,
defined as the difference
\begin{equation}
\Delta E_{\text{tot}}=E_{\text{tot}}^{\text{FM}}-E_{\text{tot}}^{\text{NM}}
\end{equation}
between the total energies of the FM and NM (i.e., spin-unpolarized)
states of the system. A negative value indicates that the FM
state is more stable than the NM state,
which is the case here for all solids and functionals.
Note that no results are shown for mBJLDA, since it is only a potential with
no corresponding xc energy functional.\cite{KarolewskiJCTC09,GaidukJCP09}
Considering all functionals except HLE16 and HLE17, there is a clear correlation
between the magnetic moment and magnetic energy; the functionals leading to
the largest values of $\mu_{S}$ (TASK, SCAN, and MVS)
also lead to the largest values of $\Delta E_{\text{tot}}$.
This trend was observed in Refs.~\onlinecite{FuPRL18,FuPRB19,MejiaRodriguezPRB19}
and is connected with the magnetic susceptibility.
However, HLE16 and HLE17 do not really follow this trend.
For instance, for Ni and YFe$_{2}$
HLE16 gives the smallest magnetic moment, but the largest value for
$\Delta E_{\text{tot}}$. Similar observations can be made with HLE17.

\begin{figure*}
\includegraphics[scale=0.68]{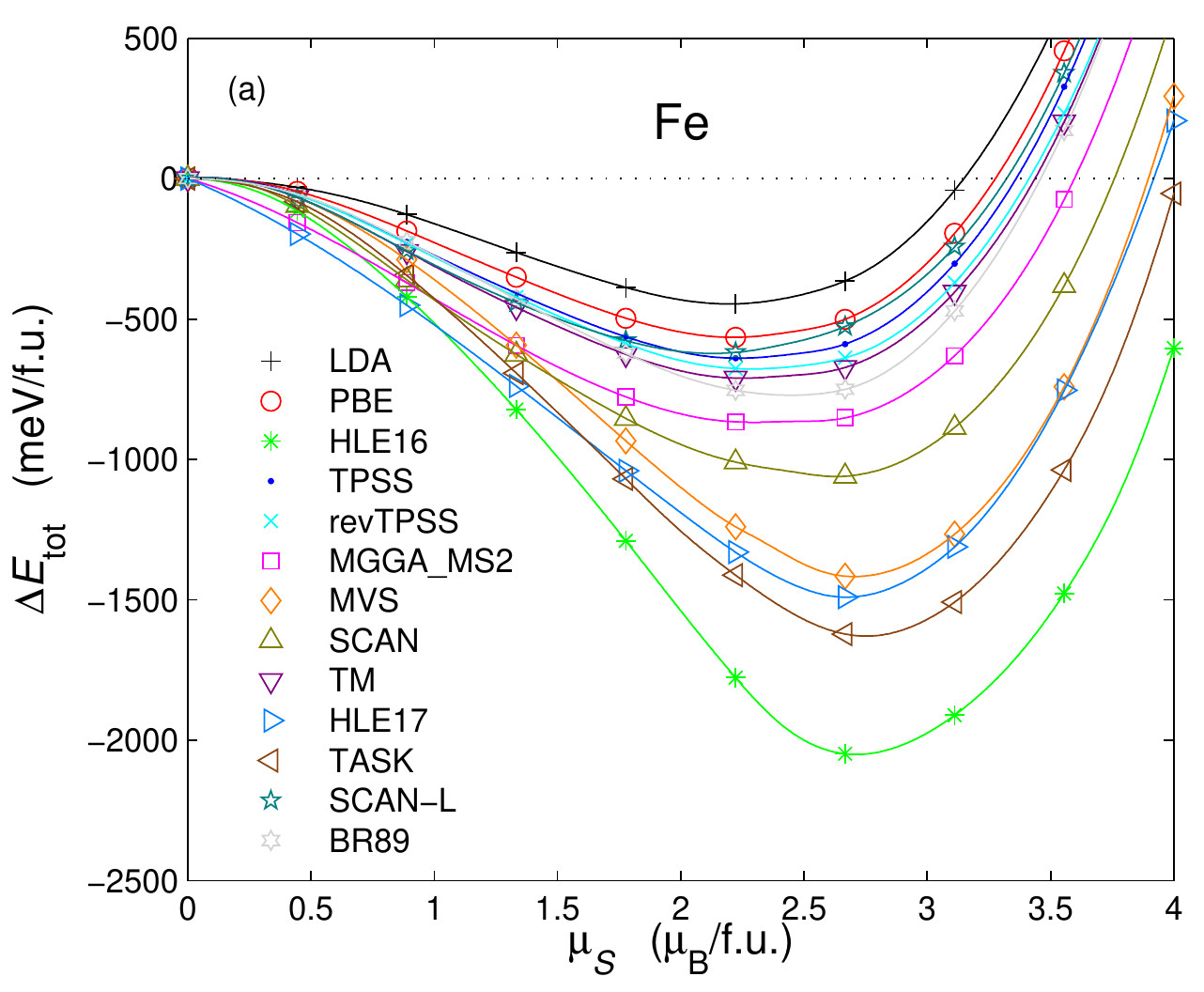}
\includegraphics[scale=0.68]{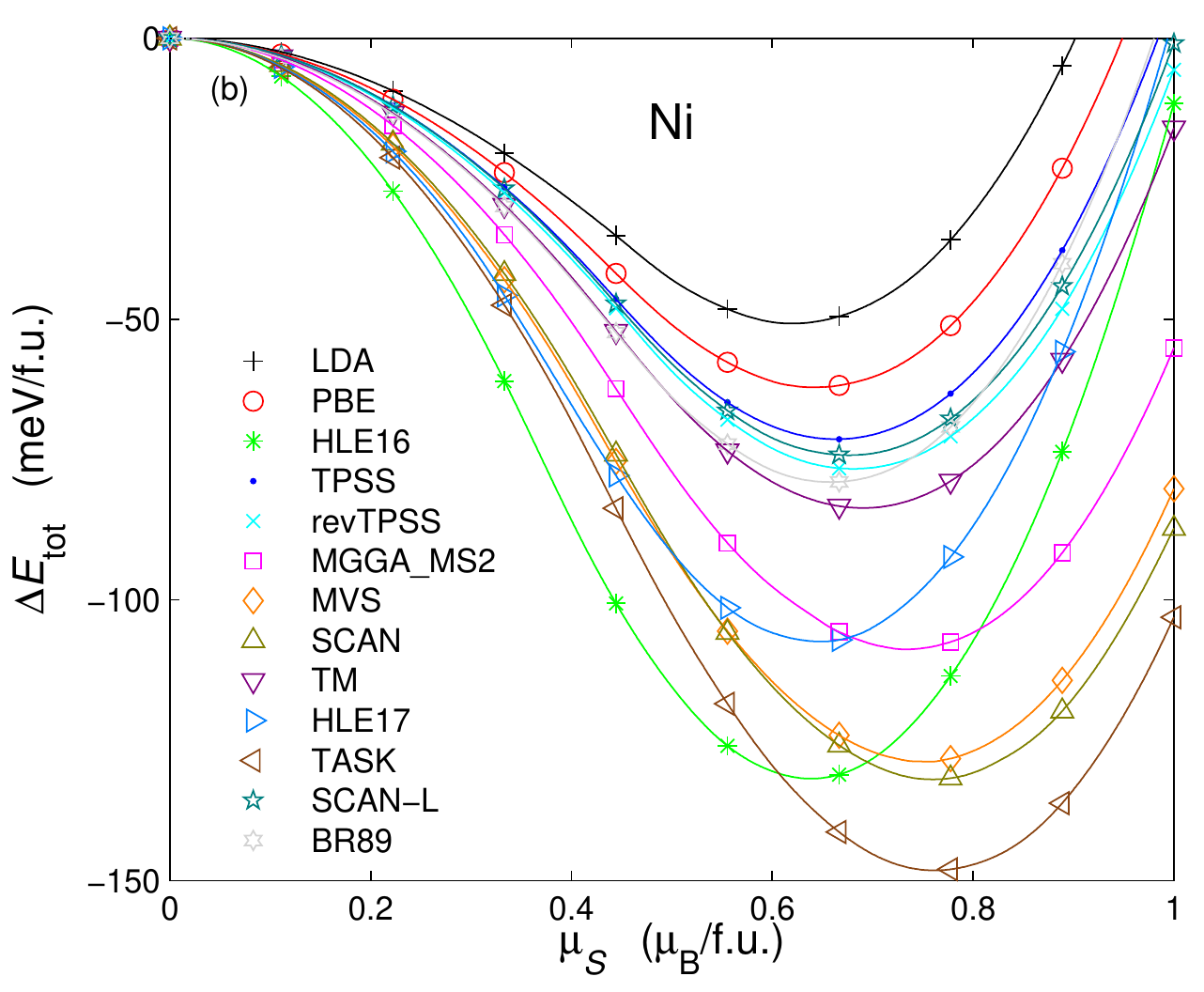}
\includegraphics[scale=0.68]{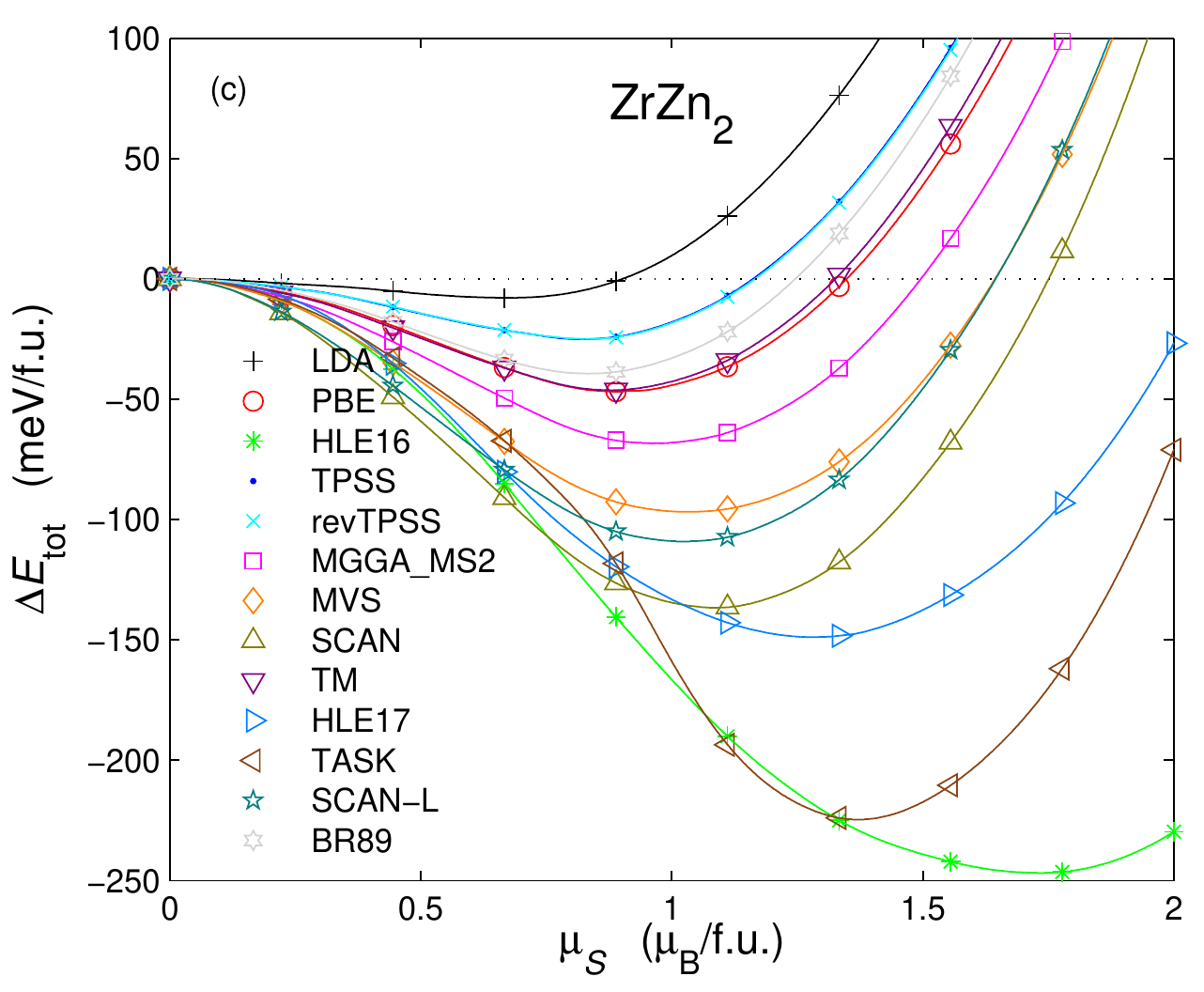}
\includegraphics[scale=0.68]{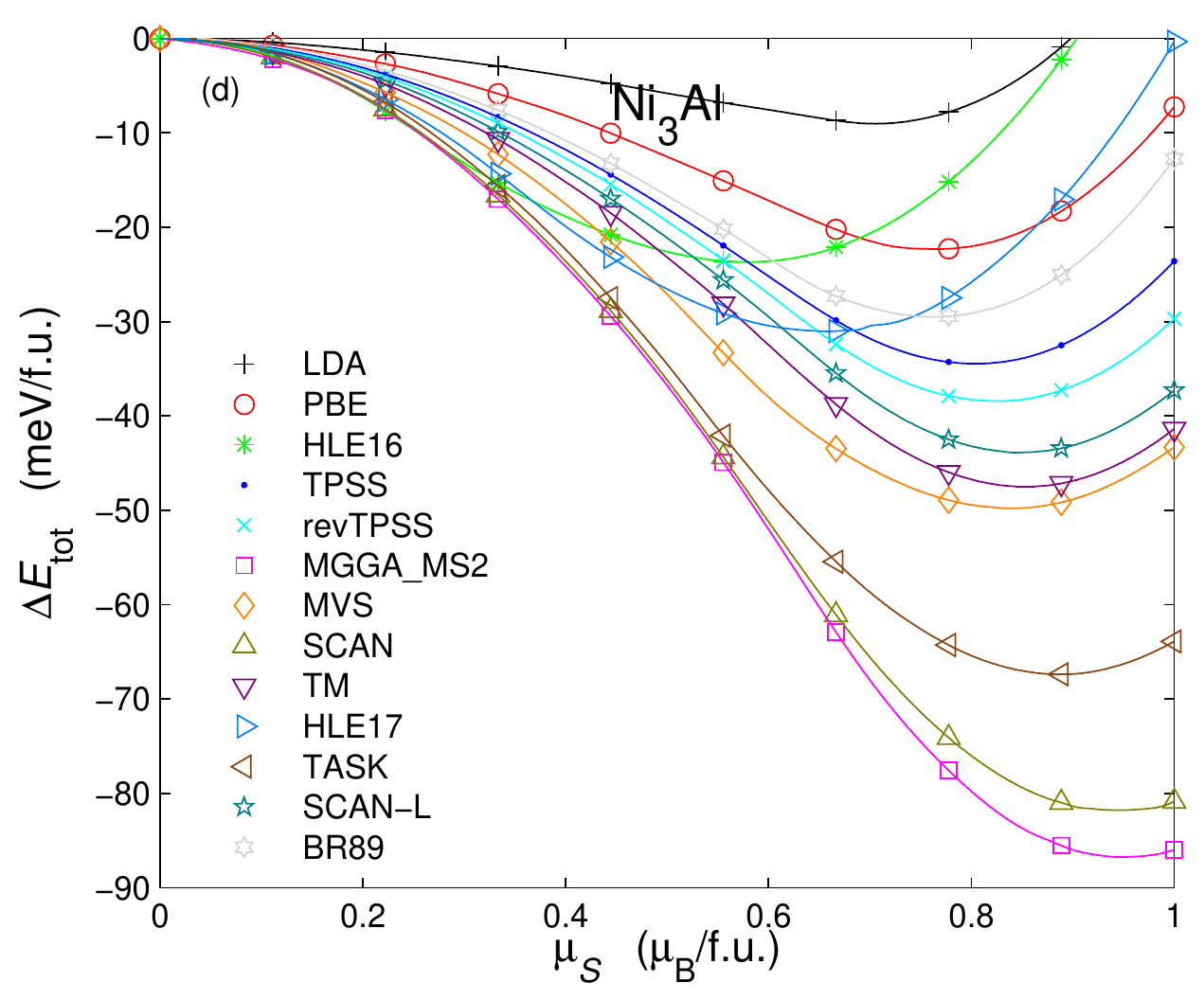}
\caption{\label{fig_M_E_FM}Magnetic energy $\Delta E_{\text{tot}}$ as a function
of the magnetic moment $\mu_{S}$ in Fe (a), Ni (b), ZrZn$_{2}$ (c), and Ni$_{3}$Al (d).}
\end{figure*}

Figure~\ref{fig_M_E_FM} shows $\Delta E_{\text{tot}}$ as a function
of $\mu_{S}$ in the cases of Fe, Ni, ZrZn$_{2}$, and Ni$_{3}$Al.
As discussed above, a larger magnetic moment usually corresponds to a deeper
minimum. However, we can see that this is not really the case with the HLE16
and HLE17 functionals for Ni and Ni$_{3}$Al.

\subsubsection{\label{NM_solids}Nonmagnetic solids}

\begin{table}
\caption{\label{NM}Spin magnetic moment $\mu_{S}$ (in $\mu_{\text{B}}$ per formula unit)
of (supposedly) NM solids. A non-zero $\mu_{S}$ means a FM ground state.
The results for the MGGA functionals were obtained
with the FSM method. The calculations were done at the geometry specified in
Table~\ref{geometry}.}
\begin{ruledtabular}
\begin{tabular}{lccccc} 
Method    & Sc   & V    & Y    &  Pd  & Pt   \\
\hline
LDA       & 0.00 & 0.00 & 0.00 & 0.08 & 0.01 \\
PBE       & 0.41 & 0.00 & 0.00 & 0.24 & 0.00 \\
HLE16     & 2.86 & 0.78 & 2.86 & 0.36 & 0.51 \\
mBJLDA    & 0.53 & 0.00 & 0.00 & 0.39 & 0.47 \\
TPSS      & 0.39 & 0.00 & 0.00 & 0.29 & 0.01 \\
revTPSS   & 0.37 & 0.01 & 0.01 & 0.30 & 0.00 \\
MGGA\_MS2 & 0.58 & 0.02 & 0.56 & 0.44 & 0.00 \\
MVS       & 0.69 & 0.00 & 0.71 & 0.41 & 0.55 \\
SCAN      & 0.62 & 0.55 & 0.60 & 0.44 & 0.08 \\
TM        & 0.37 & 0.00 & 0.00 & 0.34 & 0.00 \\
HLE17     & 0.86 & 0.61 & 0.76 & 0.38 & 0.42 \\
TASK      & 0.80 & 0.64 & 0.79 & 0.43 & 0.37 \\
SCAN-L    & 0.49 & 0.01 & 0.46 & 0.26 & 0.00 \\
BR89      & 0.50 & 0.00 & 0.00 & 0.36 & 0.04 \\
\end{tabular}
\end{ruledtabular}
\end{table}

We now turn to NM solids, but consider only elemental transition metals.
In Table~\ref{NM} we present those cases for which DFT can predict a FM ground
state instead of the experimental NM ground state.
Note that Refs.~\onlinecite{FuPRL18,MejiaRodriguezPRB19} reported that SCAN
leads to a FM ground state for V and Pd.
As we can see, in many cases disagreement with experiment is obtained.
The worst cases are Sc and Pd for which all methods except LDA lead to a non-zero magnetic
moment. LDA gives $\mu_{S}=0$ and $\mu_{S}<0.1$~$\mu_{\text{B}}$
for Sc and Pd, respectively. For the other systems,
the functionals which usually lead to the correct NM state are LDA, PBE, TPSS, revTPSS,
TM, and BR89. They were giving the least overestimations of
the magnetic moment in FM systems. Note that the GGA HLE16 leads to extreme
values, 2.86~$\mu_{\text{B}}$, for Sc and Y, whereas the other functionals give
values below 0.9~$\mu_{\text{B}}$ for these two systems. Besides HLE16,
TASK and HLE17 lead to the largest magnetic moments on average.

\begin{figure}
\includegraphics[width=\columnwidth]{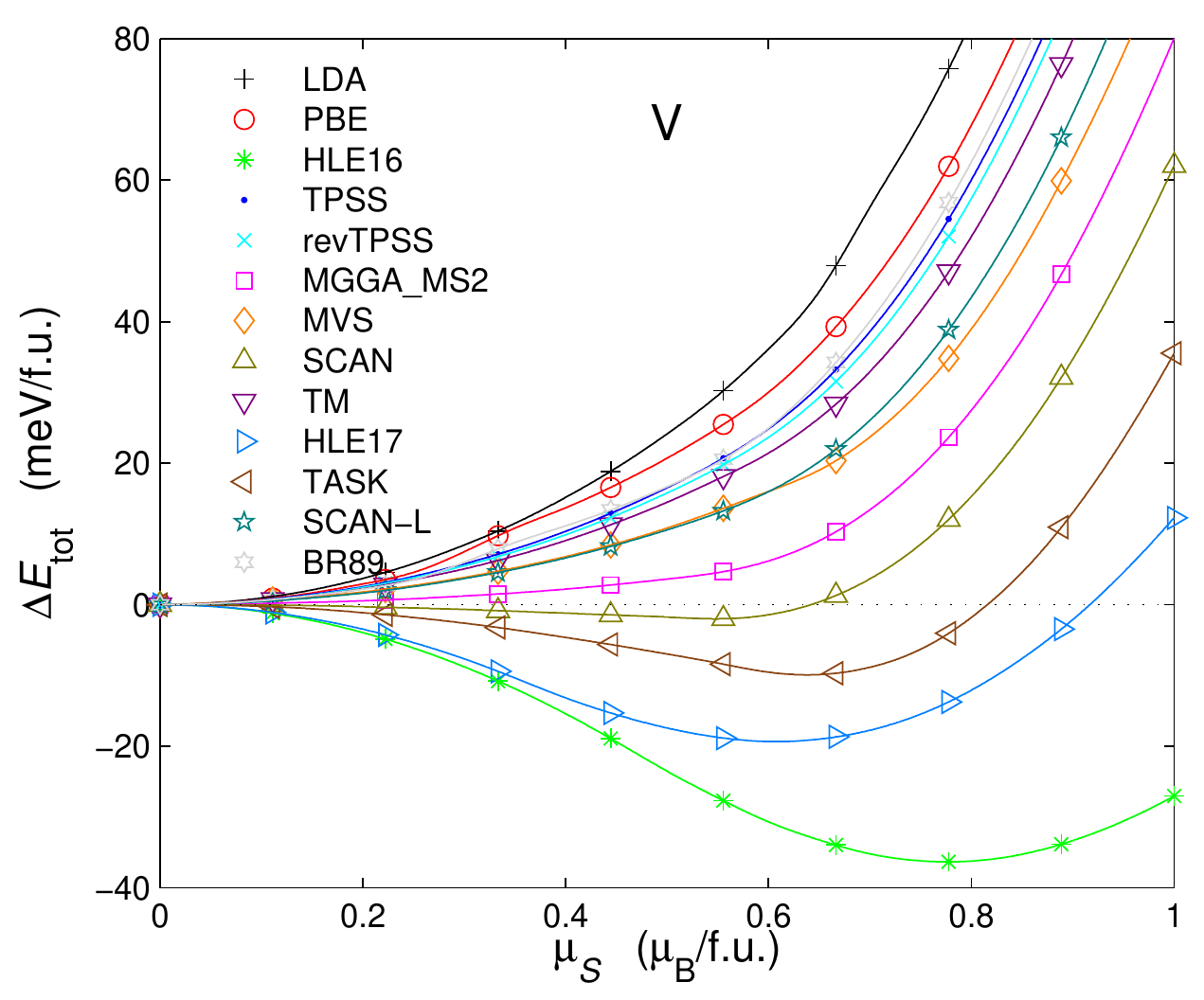}
\caption{\label{fig_M_E_NM}Magnetic energy $\Delta E_{\text{tot}}$ as a function
of the magnetic moment $\mu_{S}$ in V.}
\end{figure}

Figure~\ref{fig_M_E_NM} shows the magnetic energy for V as a function of $\mu_{S}$.
HLE16 leads to the deepest minimum, which was also the case for several of
the FM solids, as seen above. SCAN leads to a very shallow minimum but a quite large moment
of 0.55~$\mu_{\text{B}}$, while SCAN-L and the other common MGGAs retain
a NM state.

We mention that we also considered the possibility of an AFM ground state
in Mo instead of the NM one. Mo belongs to the same group as Cr which
is (incommensurate) AFM. By using a simple cubic two-atom CsCl cell,
one functional, HLE16, leads to an AFM ground state
with an atomic moment of 0.54~$\mu_{\text{B}}$ in the Bader volume.
The other functionals lead to the correct NM phase.

\subsubsection{\label{AFM_solids}Antiferromagnetic solids}

\begin{table*}
\caption{\label{AFM2}Calculated spin atomic magnetic moment $\mu_{S}$
of the transition-metal atom
(in $\mu_{\text{B}}$ and defined according to the Bader volume) of AFM solids
compared to experimental values of the total atomic magnetic moment $\mu_{S}+\mu_{L}$.
The orbital moment $\mu_{L}$ is estimated to be in the range
0.6-1~$\mu_{\text{B}}$ for FeO,
\cite{SvanePRL90,TranPRB06,RadwanskiPB08,SchronJPCM13}
1-1.6~$\mu_{\text{B}}$ for CoO,
\cite{SvanePRL90,SolovyevPRL98,ShishidouJPSJ98,NeubeckJPCS01,JauchPRB02,GhiringhelliPRB02,RadwanskiPB04,TranPRB06,RadwanskiPB08,BoussendelPRB10,SchronJPCM13}
0.3-0.45~$\mu_{\text{B}}$ for NiO,\cite{SvanePRL90,FernandezPRB98,NeubeckJPCS01,RadwanskiPB04,RadwanskiPB08}
and much smaller in other oxides. No values of $\mu_{L}$ for CrSb and CrSb$_{2}$ could
be found in the literature.
The results for the MGGA functionals were obtained with the $C$-shift method
[Eq.~(\ref{vxcshift})].
The calculations were done at the geometry specified
in Table~\ref{geometry}. The values which are in clear disagreement with
experiment are underlined.}
\begin{ruledtabular}
\begin{tabular}{lccccccccc} 
Method          & MnO   & FeO  & CoO  & NiO  & CuO  & Cr$_{2}$O$_{3}$ & Fe$_{2}$O$_{3}$ & CrSb & CrSb$_{2}$ \\
\hline
LDA             & \underline{4.33}  & 3.42 & \underline{2.38} & \underline{1.20} & \underline{0.12} & 2.53            & \underline{3.42}            & 2.74 & \underline{2.64}        \\
PBE             & \underline{4.39}  & 3.48 & \underline{2.45} & \underline{1.37} & \underline{0.37} & 2.62            & \underline{3.61}            & 2.90 & \underline{2.75}       \\
HLE16           & \underline{4.69}  & 3.67 & 2.59 & 1.45 & \underline{0.37} & \underline{3.13}            & 4.08            & \underline{4.10} & \underline{4.04}       \\
mBJLDA          & 4.57  & 3.64 & 2.72 & 1.74 & 0.72 & 2.74            & 4.14            & 2.94 & \underline{2.73}       \\
TPSS           & \underline{4.41}  & 3.52 & 2.50 & 1.46 & 0.45 & 2.63            & 3.74            & 2.97 & \underline{2.81}       \\
revTPSS        & \underline{4.42}  & 3.53 & 2.51 & 1.46 & 0.45 & 2.64            & 3.78            & 3.00 & \underline{2.84}       \\
MGGA\_MS2        & 4.48  & 3.59 & 2.56 & 1.58 & 0.59 & 2.71            & 3.95            & 3.10 & \underline{2.95}       \\
MVS   & 4.55  & 3.66 & 2.64 & 1.60 & 0.47 & 2.76            & 4.07            & 3.44 & \underline{3.28}       \\
SCAN            & 4.53  & 3.62 & 2.60 & 1.60 & 0.57 & 2.73            & 4.01            & 3.32 & \underline{3.18}       \\
TM             & \underline{4.42}  & 3.53 & 2.52 & 1.49 & 0.48 & 2.64            & 3.78            & 2.96 & \underline{2.79}       \\
HLE17           & 4.62  & 3.65 & 2.63 & 1.56 & 0.49 & \underline{2.92}            & 4.05            & \underline{3.83} & \underline{3.74}       \\
TASK            & 4.63  & 3.70 & 2.67 & 1.60 & 0.50 & \underline{2.90}            & 4.18            & \underline{3.71} & \underline{3.61} \\
SCAN-L           & \underline{4.43}  & 3.50 & 2.49 & 1.50 & 0.48 & 2.65            & \underline{3.71}            & 2.96 & \underline{2.79}       \\
BR89             & 4.47              & 3.53 & \underline{2.44} & \underline{1.42} & 0.43 & 2.69            & 3.81          &  3.12 &   \underline{2.98}     \\
Expt.    &
4.58\footnotemark[1] &
3.32,\footnotemark[2]4.2,\footnotemark[3]4.6\footnotemark[4] &
3.35,\footnotemark[5]3.8,\footnotemark[2]\footnotemark[6]3.98\footnotemark[7] &
1.9,\footnotemark[1]\footnotemark[2]2.2\footnotemark[8]\footnotemark[9] &
0.65\footnotemark[10] &
2.44,\footnotemark[11]2.48,\footnotemark[12]2.76\footnotemark[13] &
4.17,\footnotemark[14]4.22\footnotemark[15] &
3.0\footnotemark[16]&
1.94\footnotemark[17]\\
\end{tabular}
\end{ruledtabular}
\footnotetext[1]{Ref.~\onlinecite{CheethamPRB83}.} 
\footnotetext[2]{Ref.~\onlinecite{RothPR58}.} 
\footnotetext[3]{Ref.~\onlinecite{BattleJPC79}.} 
\footnotetext[4]{Ref.~\onlinecite{FjellvagJSSC96}.} 
\footnotetext[5]{Ref.~\onlinecite{KhanPRB70}.} 
\footnotetext[6]{Ref.~\onlinecite{HerrmannRonzaudJPC78}.} 
\footnotetext[7]{Ref.~\onlinecite{JauchPRB01}.} 
\footnotetext[8]{Ref.~\onlinecite{FernandezPRB98}.} 
\footnotetext[9]{Ref.~\onlinecite{NeubeckJAP99}.} 
\footnotetext[10]{Ref.~\onlinecite{ForsythJPC88}.} 
\footnotetext[11]{Ref.~\onlinecite{GolosovaJAC17}.} 
\footnotetext[12]{Ref.~\onlinecite{BrownJPCM02}.} 
\footnotetext[13]{Ref.~\onlinecite{CorlissJAP65}.} 
\footnotetext[14]{Ref.~\onlinecite{BaronSSS05}.} 
\footnotetext[15]{Ref.~\onlinecite{HillCM08}.} 
\footnotetext[16]{Ref.~\onlinecite{TakeiPR63}.} 
\footnotetext[17]{Ref.~\onlinecite{HolsethACS70}.} 
\end{table*}

The results for the spin atomic moment $\mu_{S}$ in AFM solids are shown in
Table~\ref{AFM2}. We mention again that the Bader volume is used for the region
defining the atomic moment. We also mention that for a comparison with experiment
there is a possible non-negligible orbital contribution $\mu_{L}$ to the experimental
value, and estimates are given in the caption of Table~\ref{AFM2}.

It is well known that LDA and PBE have the tendency to underestimate the moment in AFM
oxides, as we observe here for most oxides. An exception is
Cr$_{2}$O$_{3}$, since the LDA/PBE results lie in the range
of the experimental values. In the cases of FeO and CoO, it is not possible to
make a quantitative comparison with experiment since the range of
experimental values and estimations for $\mu_{L}$ are large.
For the intermetallic compounds CrSb and CrSb$_{2}$, where the magnetic moment on the Cr
atom is considered (note that in CrSb$_{2}$ the moment on the Sb atom is
non-zero, but tiny), PBE seems to lead to good agreement for CrSb, but to a
very large overestimation of $\sim0.8$~$\mu_{\text{B}}$ for CrSb$_{2}$.
Such an overestimation by PBE for CrSb$_{2}$ has already been noted by Kuhn
\textit{et al}.,\cite{KuhnPRB13} who showed that by adding an on-site Hubbard
correction to the Cr atom, using the around mean field (AMF) version of PBE+$U$
\cite{CzyzykPRB94} with $U=2.7$~eV and $J=0.3$~eV, leads to a
reduction of the moment from 2.57 to 2.03~$\mu_{\text{B}}$
(inside the Cr atomic sphere of radius 2.32~bohr).
We could reproduce this trend with PBE+$U$(AMF)
(we get $\mu_{S}=2.34~\mu_{\text{B}}$ inside the Bader volume).
However, when using
the fully localized limit (FLL)\cite{CzyzykPRB94} variant of PBE+$U$
(with same $U$ and $J$) the moment increases by $\sim0.5$~$\mu_{\text{B}}$
with respect to PBE, and therefore worsens the agreement with experiment.
These results with PBE+$U$ are not surprising since DFT+$U$(AMF) is known to
be better adapted than DFT+$U$(FLL) for (near)-metallic systems which are not as
correlated as TM oxides.\cite{PetukhovPRB03,MohnPRL01}

Compared to PBE, the GGA HLE16 significantly increases the magnetic moment
for all systems except CuO, for which the PBE and HLE16 moments are
curiously identical. The increase in $\mu_{S}$ is the largest
for CrSb and CrSb$_{2}$ where it is clearly above 1~$\mu_{\text{B}}$.
Actually, among all methods HLE16 leads to (nearly) the largest value of $\mu_{S}$
for all systems except CoO, NiO, and CuO.

As observed in Sec.~\ref{FM_solids} for the FM solids, the MGGAs TPSS, revTPSS, TM,
SCAN-L, and BR89 lead to results that are relatively similar to PBE in most cases.
These functionals lead to moments that are moderately larger than PBE, and the
largest increase ($\sim0.2~\mu_{\text{B}}$) occurs for Fe$_{2}$O$_{3}$,
CrSb, and CrSb$_{2}$ with BR89. For the AFM solids considered here there is
basically no case where a MGGA leads to a moment
smaller than the PBE value, whereas there were many cases
for the FM systems. All other MGGAs lead to magnetic moments that are increased further,
and the largest values of $\mu_{S}$ (disregarding the HLE16 results)
are obtained in most cases by either TASK (MnO, FeO, CoO, and Fe$_{2}$O$_{3}$),
HLE17 (Cr$_{2}$O$_{3}$, CrSb, and CrSb$_{2}$), or mBJLDA (NiO and CuO).

Due to the large uncertainties in the experimental values, a quantitative
ranking of the theoretical methods is hardly possible.
Overall, we can say that MGGAs perform better than standard PBE. However, in
some cases TPSS, revTPSS, TM, SCAN-L, and BR89 seem to be too weak, with magnetic
moments that are still too small compared to experiment. On the other hand,
HLE17 and TASK, as well as the GGA HLE16, lead to moments that are by far too large for
Cr$_{2}$O$_{3}$ and the weakly correlated systems CrSb and CrSb$_{2}$.
For the latter the overestimation of $\mu_{S}$ is in the range 1$-$2~$\mu_{S}$
with all functionals except PBE+$U$(AMF).
Finally, as already seen for the FM solids, HLE16 behaves erratically,
since it leads to the smallest moment for CuO, but to the largest moment for some
of the other AFM systems.

\section{\label{Discussion}Discussion}

A quite general observation that can be made from the results presented in
Sec.~\ref{functionals} is that if a MGGA functional increases (let us say
with respect to PBE) magnetism in a system, then it will most likely do it in
other magnetic systems, too. However, clear exceptions
were noted with HLE16 and HLE17, which lead to rather
erratic results. The other general conclusion is that all tested MGGAs lead
in most cases to magnetic moments which are larger than the PBE values.
In order to provide insight for some of the results,
for instance by establishing a relation between the mathematical form
of the xc functional and the magnetic moment $\mu_{S}$, we consider the
xc magnetic energy density
\begin{equation}
\Delta\epsilon_{\text{xc}}(\mathbf{r})=
\epsilon_{\text{xc}}^{\text{(A)FM}}(\mathbf{r}) -
\epsilon_{\text{xc}}^{\text{NM}}(\mathbf{r}),
\label{dexc}
\end{equation}
where
$\epsilon_{\text{xc}}(\rho_{\uparrow},\rho_{\downarrow},
\nabla\rho_{\uparrow},\nabla\rho_{\downarrow},
\nabla^{2}\rho_{\uparrow},\nabla^{2}\rho_{\downarrow},
t_{\uparrow},t_{\downarrow})$ is the xc-energy density defined as follows:
\begin{equation}
E_{\text{xc}} = \int\epsilon_{\text{xc}}(\mathbf{r})d^{3}r.
\end{equation}
In Eq.~(\ref{dexc}), $\epsilon_{\text{xc}}^{\text{(A)FM}}$ and
$\epsilon_{\text{xc}}^{\text{NM}}$ were calculated in the (A)FM and NM phases, respectively.
$\Delta\epsilon_{\text{xc}}$ is expected to be mainly negative in magnetic systems.

\begin{figure}
\begin{picture}(8.6,4.3)(0,0)
\put(0,0){\epsfxsize=4.2cm \epsfbox{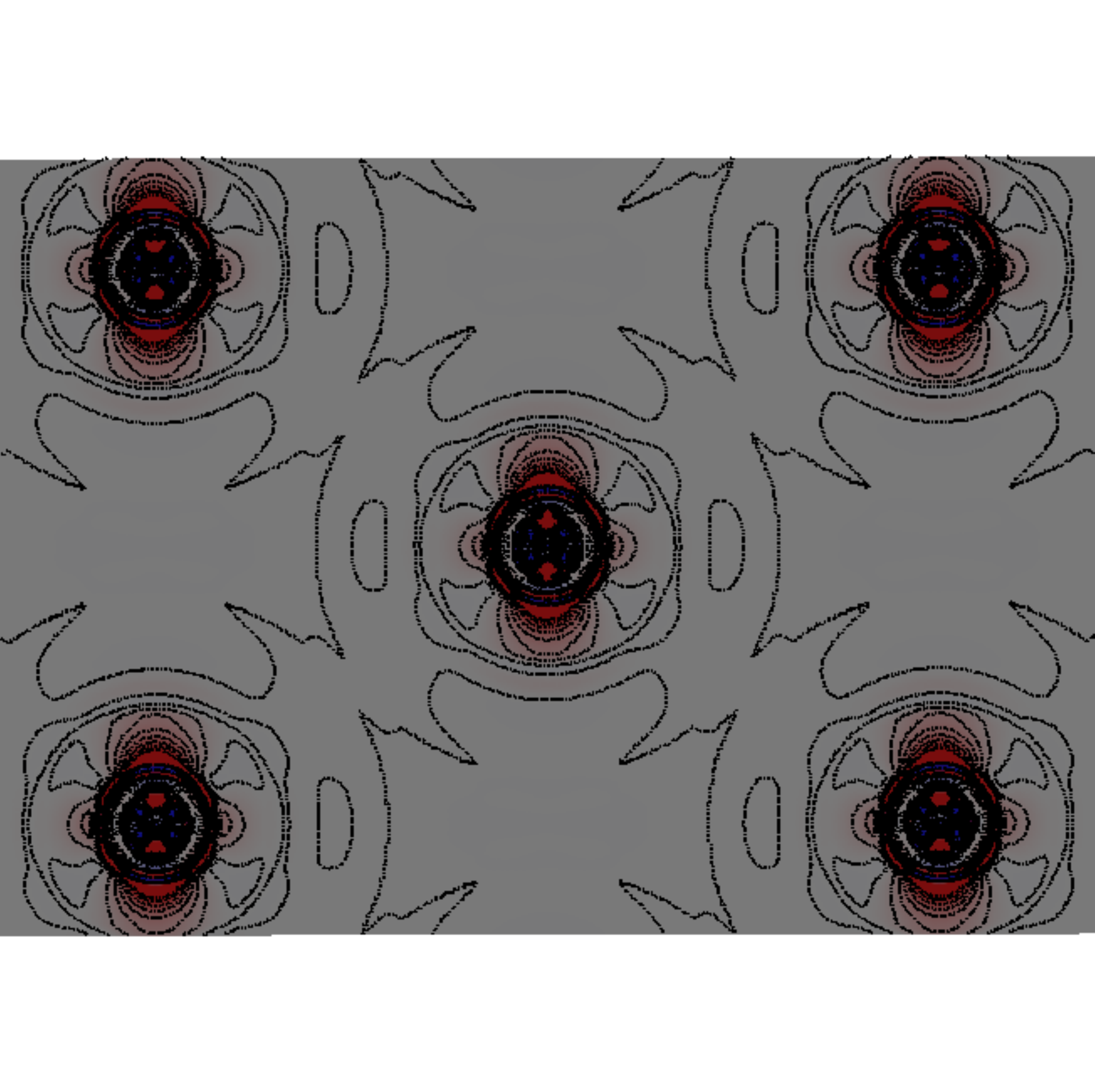}}
\put(4.3,0){\epsfxsize=4.2cm \epsfbox{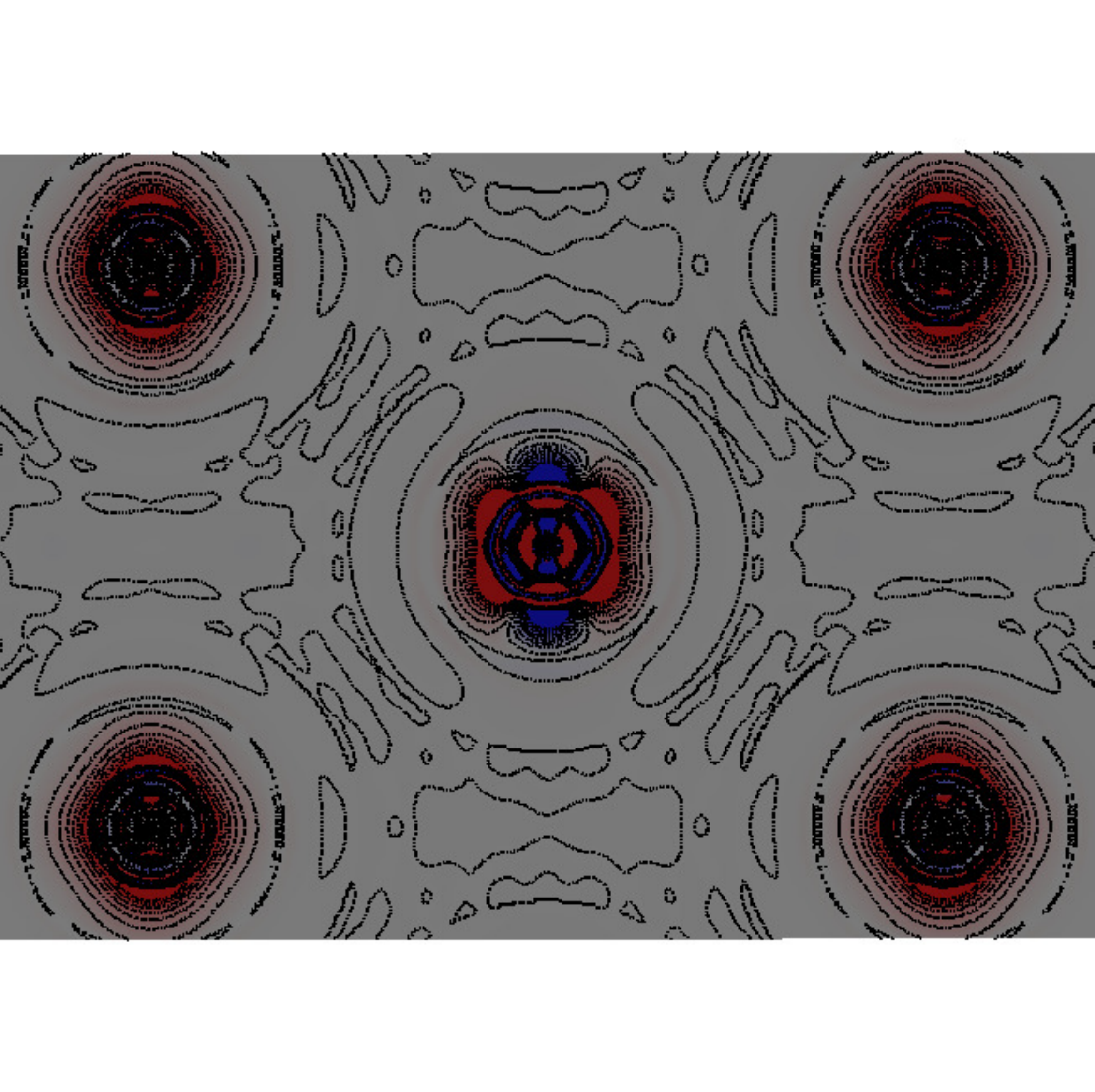}}
\put(0.1,3.8){Fe}
\put(4.4,3.8){FeCo}
\end{picture}
\caption{\label{fig_exc_Fe_FeCo}Difference
$\Delta\epsilon_{\text{xc}}^{\text{SCAN-L}}-\Delta\epsilon_{\text{xc}}^{\text{SCAN}}$
between the xc magnetic energy density obtained
with SCAN and SCAN-L within a (110) plane in Fe (left panel) and FeCo (right panel, the
middle atom is Co). The FM states correspond to $\mu_{S}=2.0$ and 4.5~$\mu_{\text{B}}$
for Fe and FeCo, respectively.
Blue and red regions correspond to negative and positive values, respectively.
The regions with the most intense blue/red colors correspond to absolute
values above 0.02~Ry/bohr$^{3}$.}
\end{figure}

\begin{figure}
\includegraphics[width=\columnwidth]{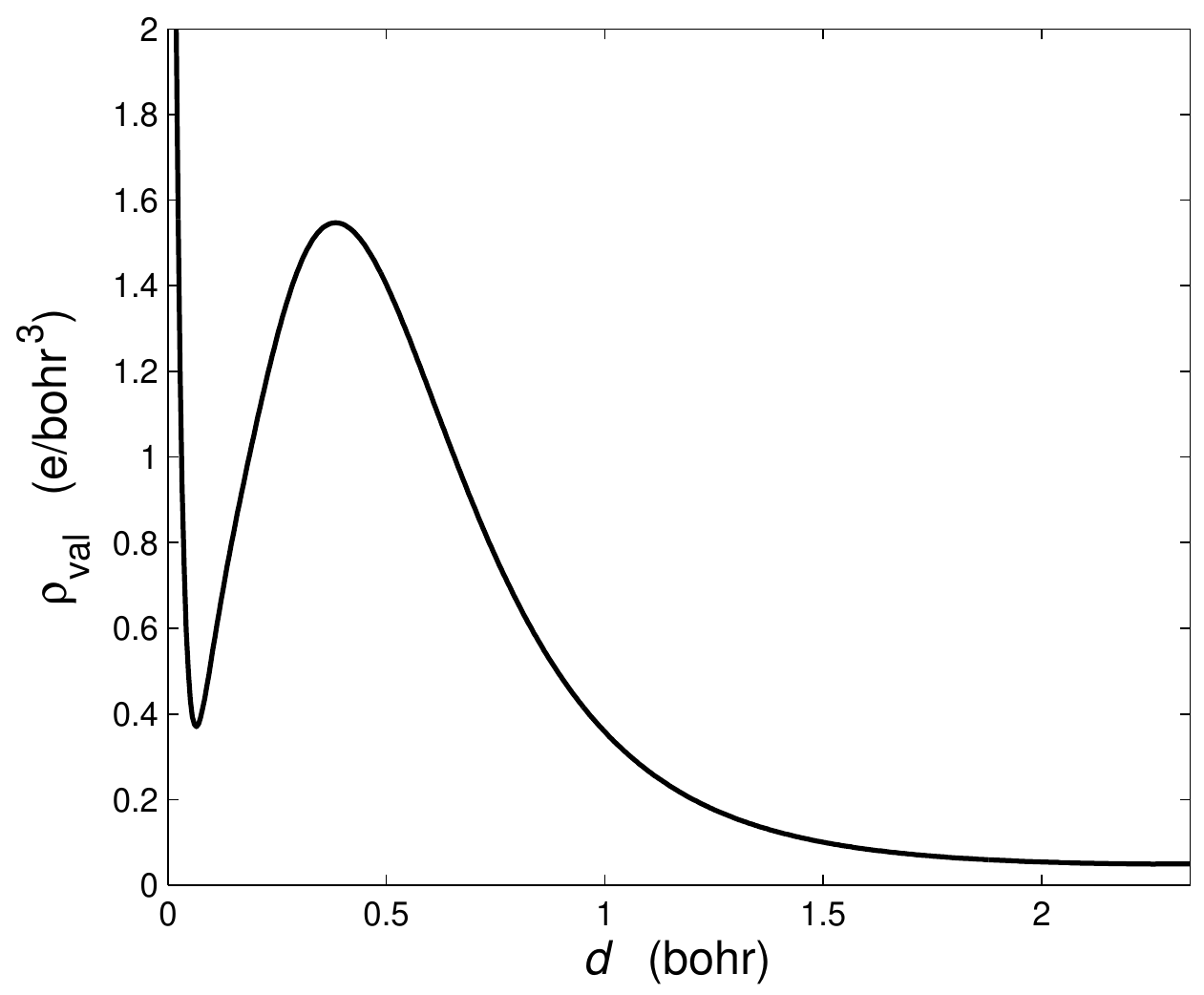}
\caption{\label{fig_rho3d}Valence electron density
$\rho_{\text{val}}=\rho_{\text{val}\uparrow}+\rho_{\text{val}\downarrow}$
in FM Fe plotted from the atom at (0,0,0) until the mid-distance to the
atom at (1/2,1/2,1/2). The maximum near $d=0.5$~bohr is due to the
$3d$ electrons and the spike at the nucleus due to the $4s$ electrons.}
\end{figure}

Figure~\ref{fig_exc_Fe_FeCo} shows the difference in $\Delta\epsilon_{\text{xc}}$
between SCAN and SCAN-L
($\Delta\epsilon_{\text{xc}}^{\text{SCAN-L}}-\Delta\epsilon_{\text{xc}}^{\text{SCAN}}$)
in the cases of the FM systems Fe and FeCo. Note that for a meaningful
comparison, $\epsilon_{\text{xc}}^{\text{FM}}$ is calculated at the same value
of $\mu_{S}$ for both functionals (2.0 and 4.5~$\mu_{\text{B}}$ for Fe
and FeCo, respectively). As discussed in Sec.~\ref{functionals}, SCAN-L
reduces $\mu_{S}$ with respect to its parent SCAN. According to
Fig.~\ref{fig_exc_Fe_FeCo}, this is mainly due to the large values of
$\Delta\epsilon_{\text{xc}}^{\text{SCAN-L}}-\Delta\epsilon_{\text{xc}}^{\text{SCAN}}$
close to the atoms where
$\Delta\epsilon_{\text{xc}}^{\text{SCAN-L}}$ is overall less negative
than $\Delta\epsilon_{\text{xc}}^{\text{SCAN}}$ (this was checked by integrating
only over the atomic region).
The contribution to the integral of
$\Delta\epsilon_{\text{xc}}^{\text{SCAN-L}}-\Delta\epsilon_{\text{xc}}^{\text{SCAN}}$
follows the $3d$ electron density, which in Fe has its maximum already at 0.5~bohr
and quickly decays beyond 1.5~bohr, as shown in Fig.~\ref{fig_rho3d}.
The contribution from the interstitial region is one order of magnitude smaller
and has opposite sign (negative), which is due to reverse polarization of the $4s$
electrons.\cite{MoonPR64,MookJAP66} Note that in general the difference
$\Delta\epsilon_{\text{xc}}^{F1}-\Delta\epsilon_{\text{xc}}^{F2}$
around an atom between two
functionals $F1$ and $F2$ is not uniformly positive or negative;
there are lobes (which differentiate orbitals) with opposite signs.
This is visible for the Co atom in FeCo, for instance.
Of course, which lobes are the most visible also depends on the plane that is chosen
for the plot. In Ref.~\onlinecite{MejiaRodriguezPRB19} the case of Fe was explained
by looking in detail at the differences between the
iso-orbital indicator $\alpha_{\sigma}$ and its deorbitalized version
$\alpha_{L,\sigma}$ (defined later) in the region around the nucleus corresponding
to a sphere with a radius of 1.5~bohr, i.e., where
$\Delta\epsilon_{\text{xc}}^{\text{SCAN-L}}-\Delta\epsilon_{\text{xc}}^{\text{SCAN}}$
is the largest. Later in the text we also provide a more detail discussion on
the difference between $\alpha_{\sigma}$ and $\alpha_{L,\sigma}$ in Fe.

\begin{figure}
\begin{picture}(8.6,11.2)(0,0)
\put(0,7.1){\epsfxsize=4.2cm \epsfbox{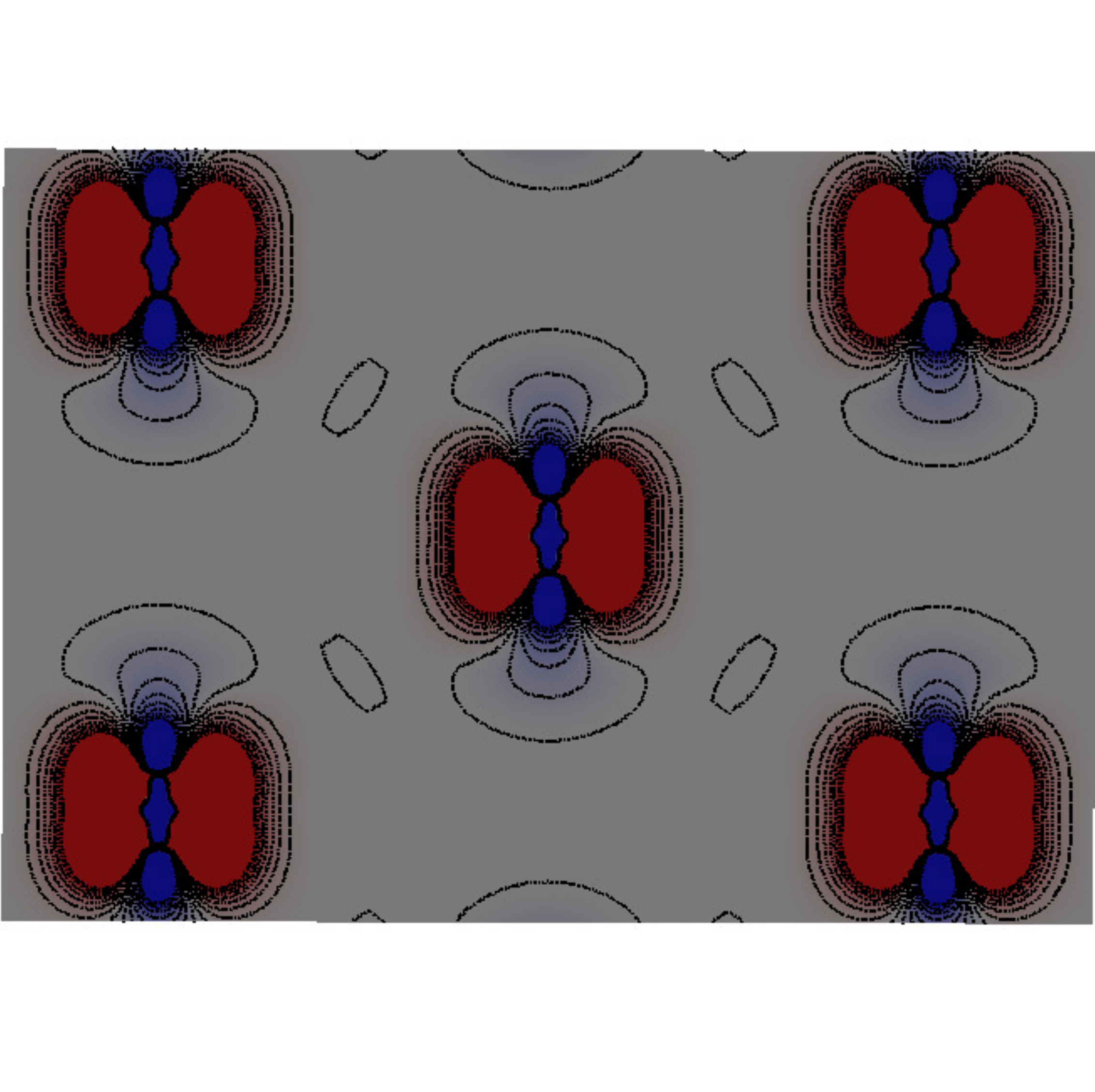}}
\put(4.3,7.1){\epsfxsize=4.2cm \epsfbox{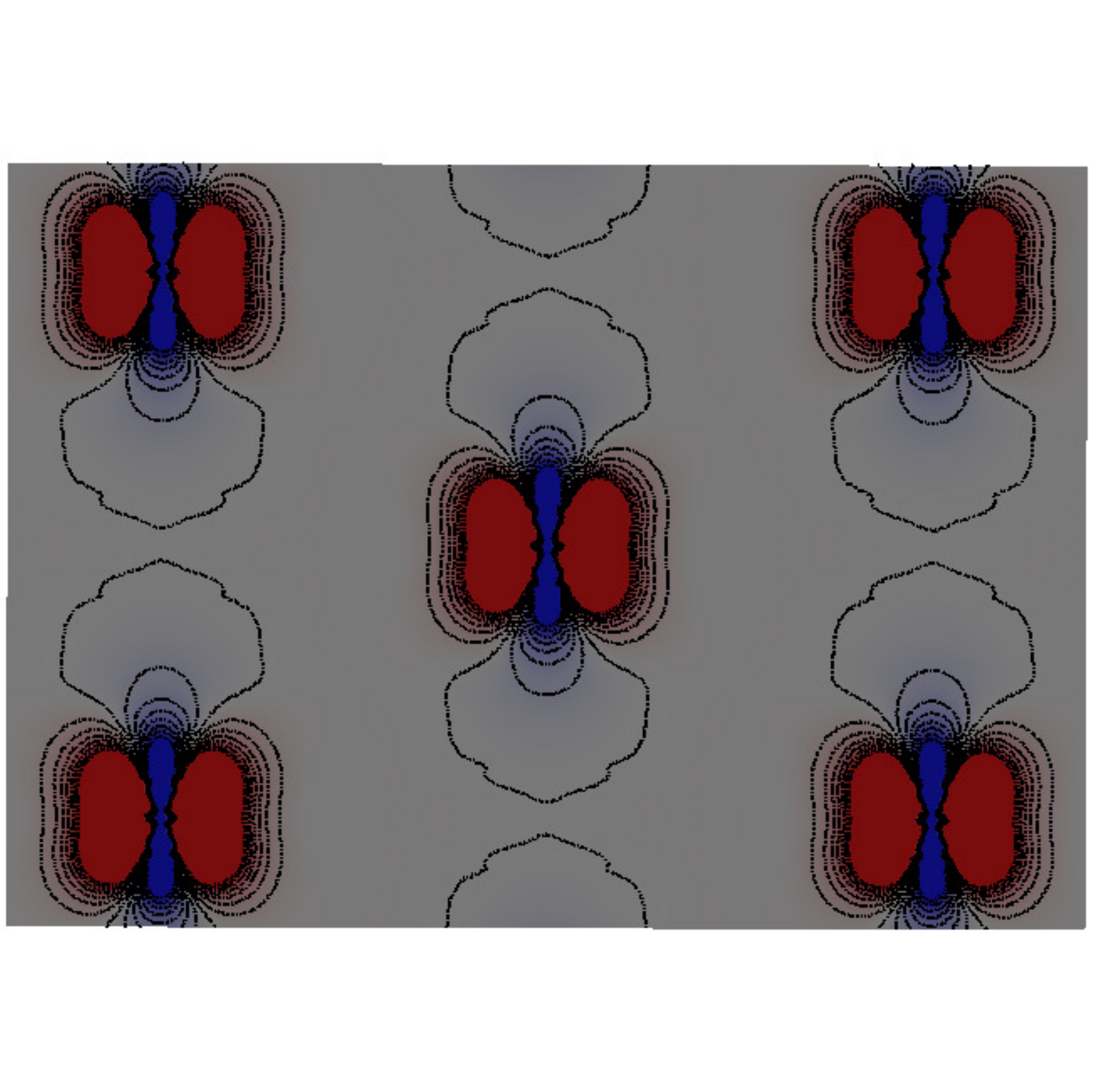}}
\put(0,3.6){\epsfxsize=4.2cm \epsfbox{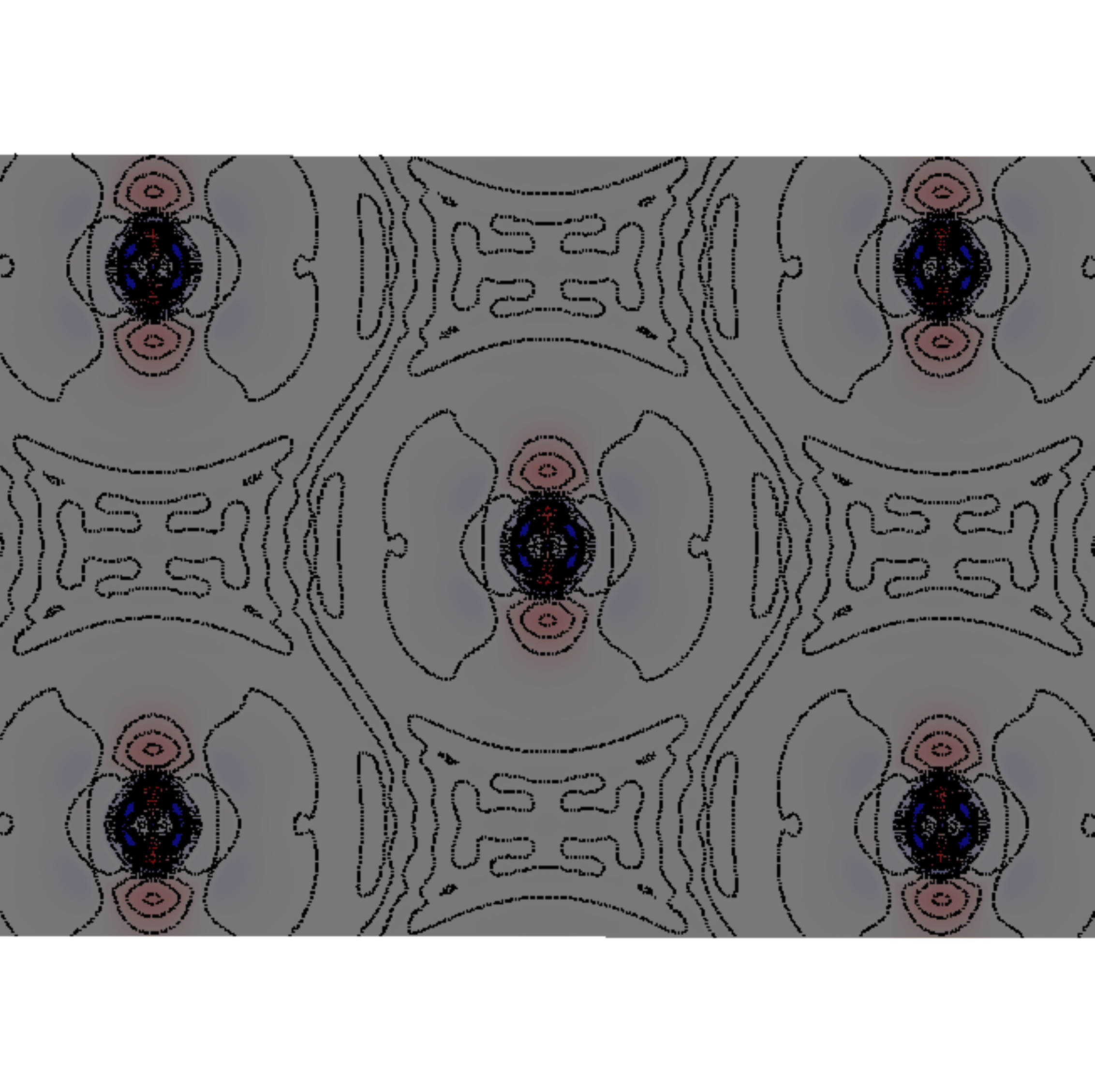}}
\put(4.3,3.6){\epsfxsize=4.2cm \epsfbox{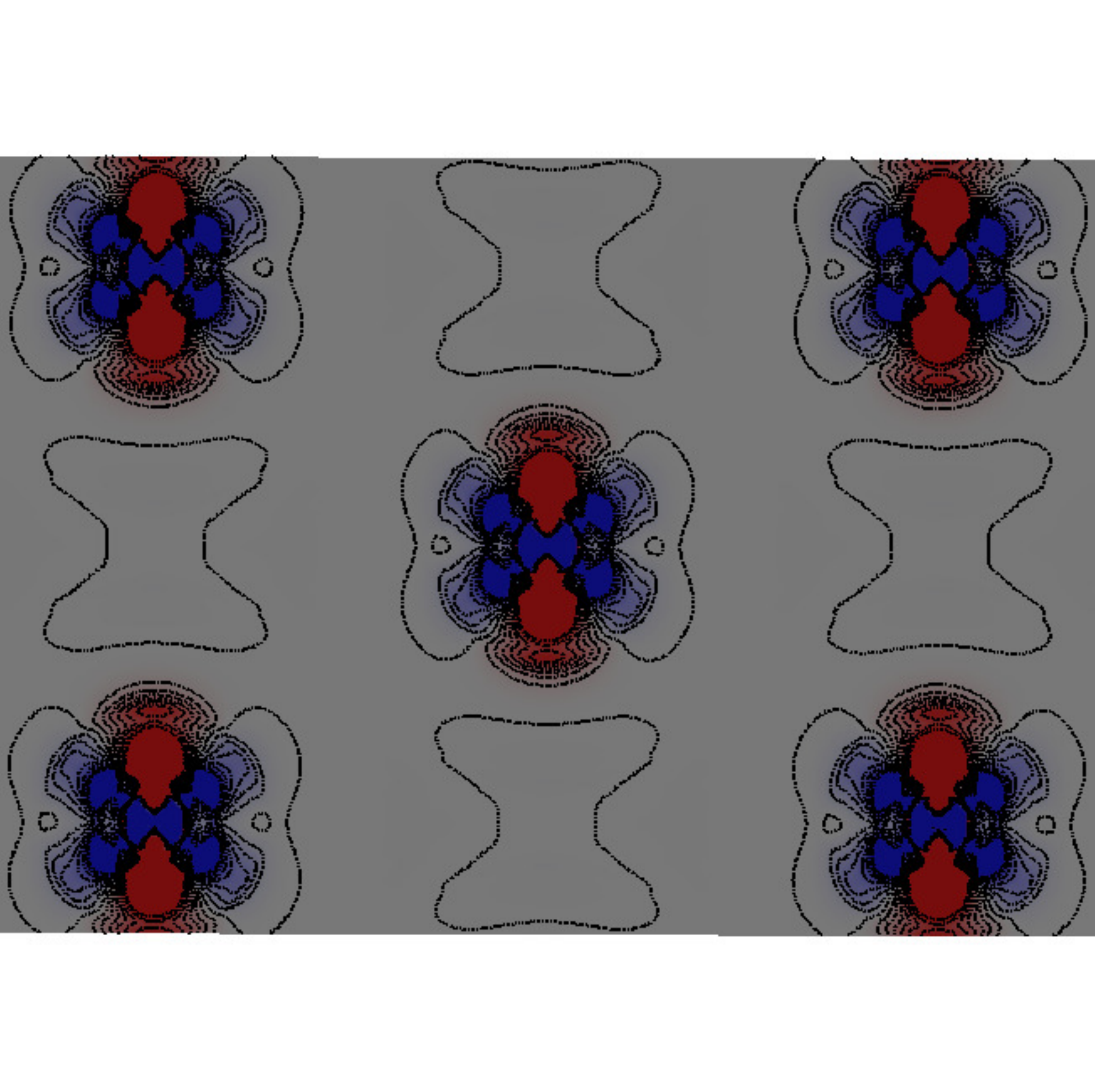}}
\put(0,0){\epsfxsize=4.2cm \epsfbox{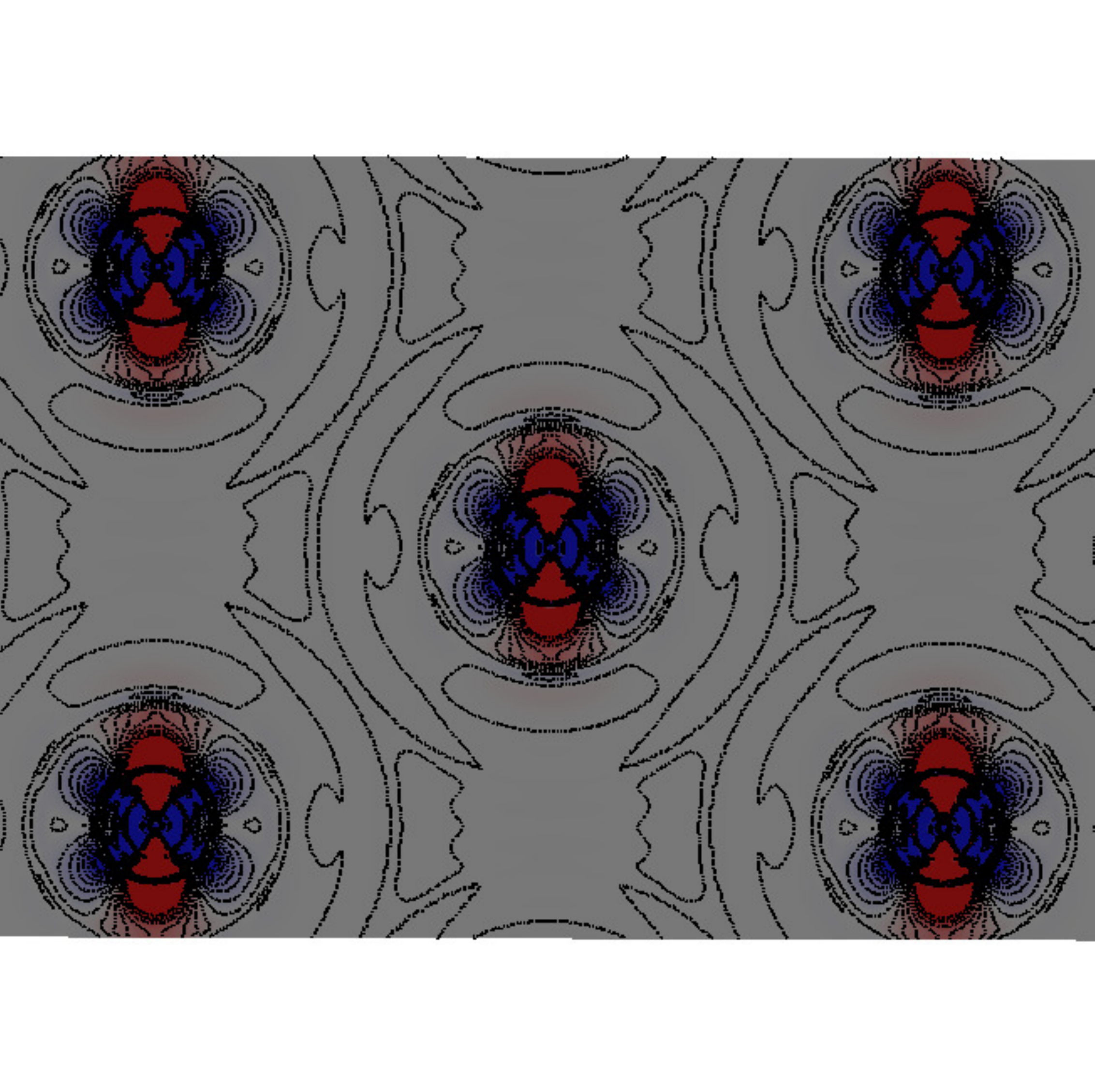}}
\put(4.3,0){\epsfxsize=4.2cm \epsfbox{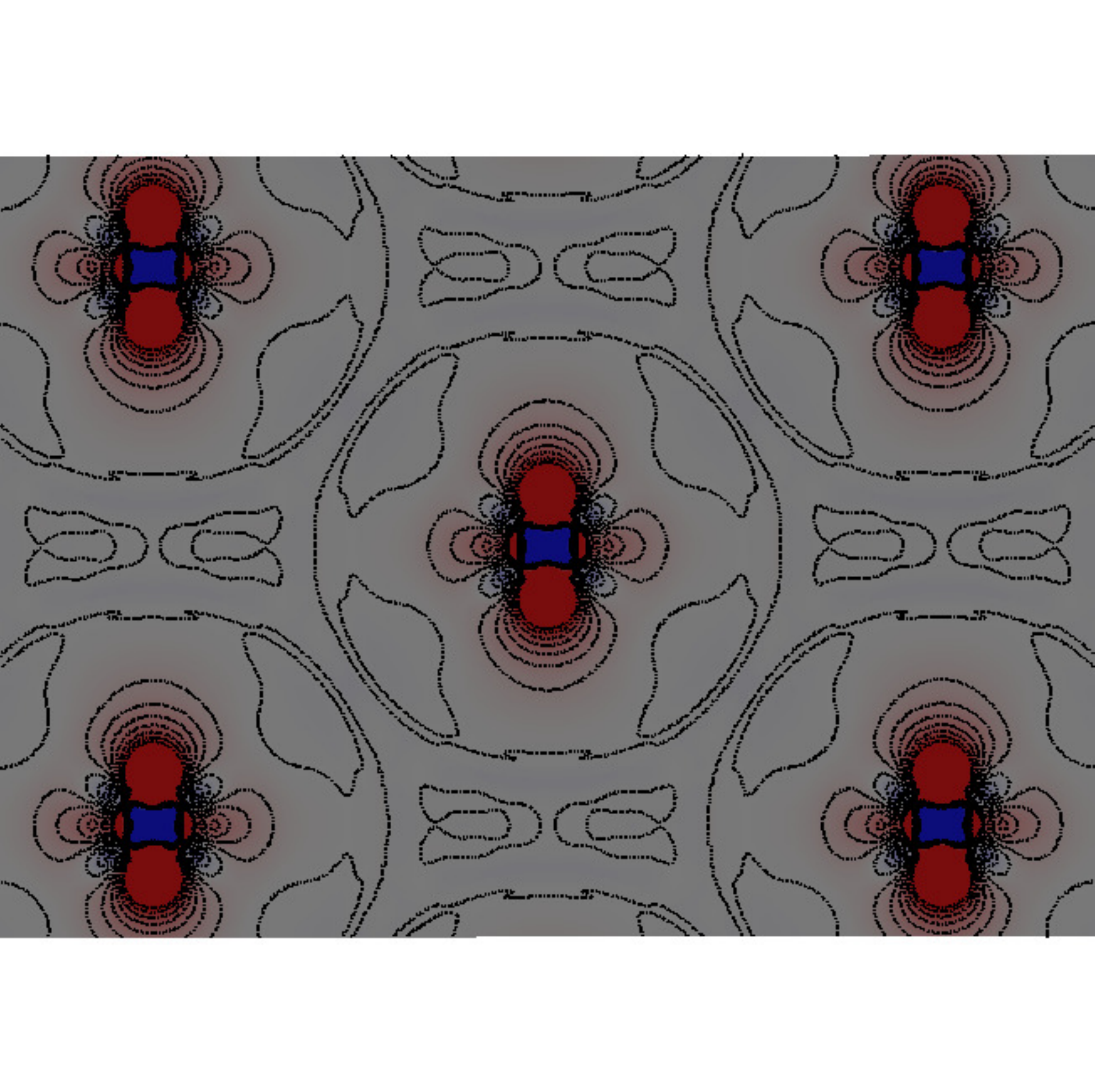}}
\put(0.1,10.8){PBE$-$HLE16}
\put(4.4,10.8){PBE$-$HLE17}
\put(0.1,7.3){PBE$-$TPSS}
\put(4.4,7.3){PBE$-$MVS}
\put(0.1,3.7){PBE$-$SCAN}
\put(4.4,3.7){PBE$-$TASK}
\end{picture}
\caption{\label{fig_exc_Fe_1}Difference
$\Delta\epsilon_{\text{xc}}^{\text{PBE}}-\Delta\epsilon_{\text{xc}}^{F}$
between the xc magnetic energy density
within a (110) plane in Fe obtained with PBE and another functional $F$.
The FM state corresponds to $\mu_{S}=2.0~\mu_{\text{B}}$.
Blue and red regions correspond to negative and positive values, respectively.
The regions with the most intense blue/red colors correspond to absolute
values above 0.02~Ry/bohr$^{3}$.}
\end{figure}
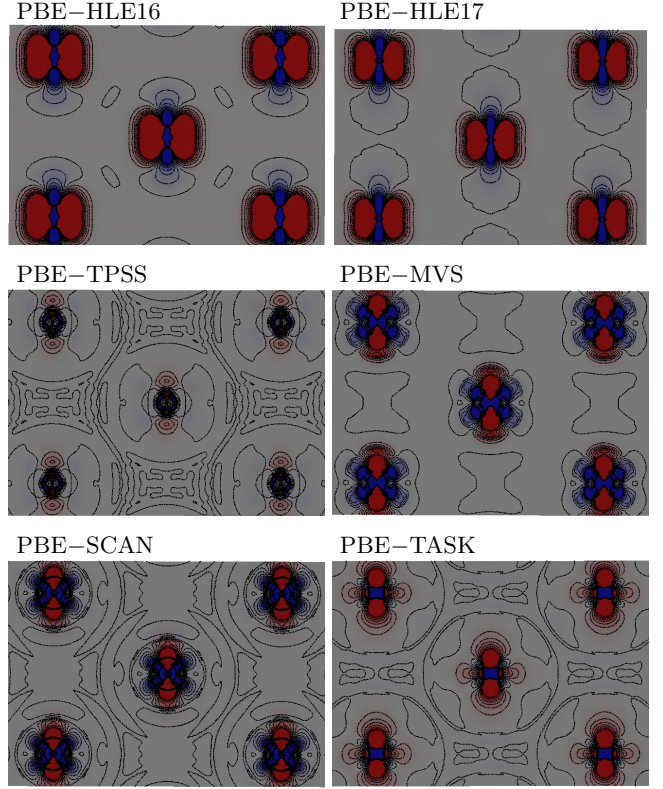

\begin{figure}
\includegraphics[width=0.49\columnwidth]{Fe_PBE-HLE17_dexc2_002_h.pdf}
\includegraphics[width=0.49\columnwidth]{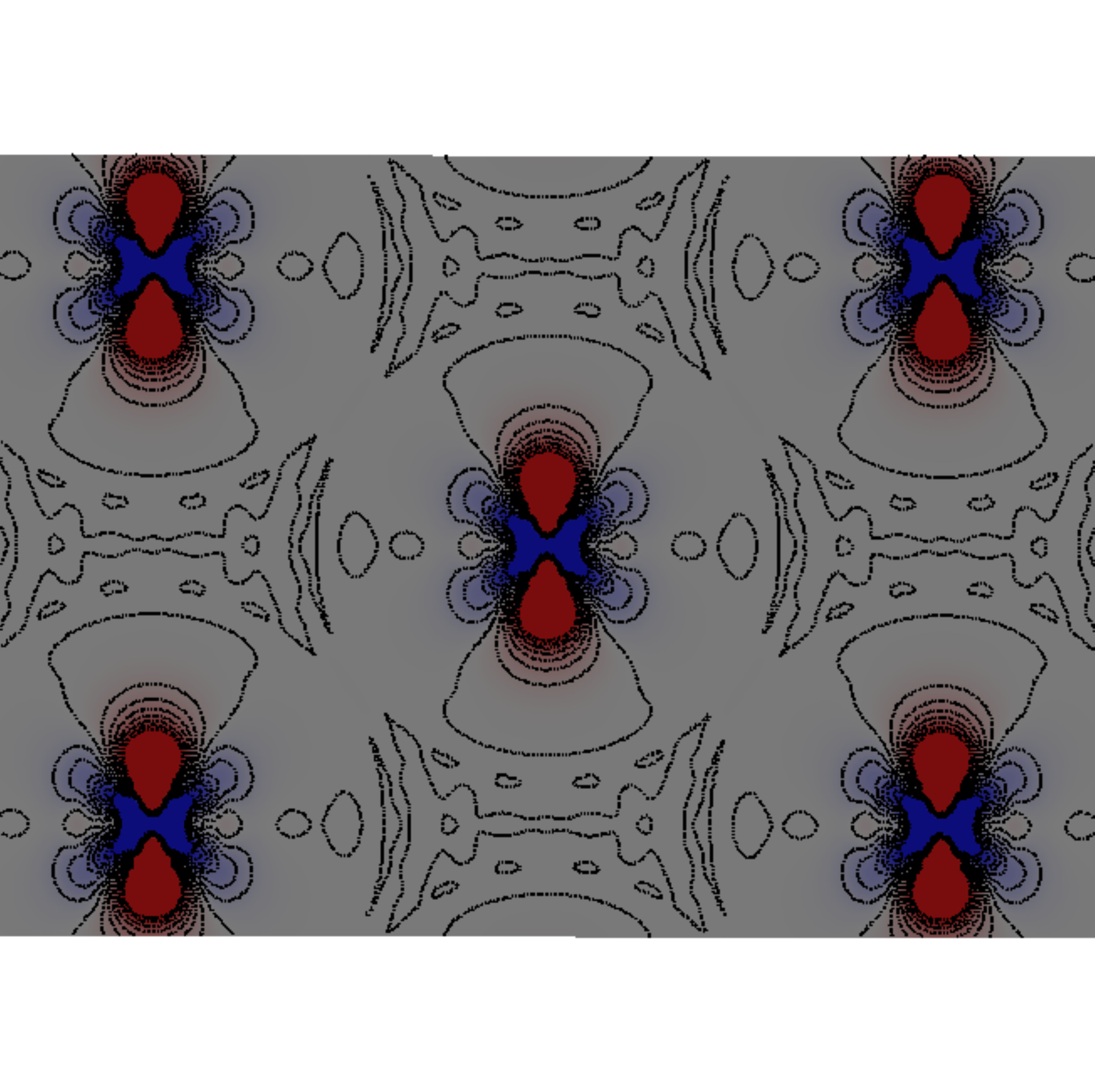}
\caption{\label{fig_exc_Fe_2}Difference
$\Delta\epsilon_{\text{xc}}^{\text{PBE}}-\Delta\epsilon_{\text{xc}}^{\text{HLE17}}$
between the xc magnetic energy density within a (110) plane in Fe obtained with
PBE and HLE17. $\Delta\epsilon_{\text{xc}}^{\text{HLE17}}$ is evaluated
with the density/KED generated from either the
mRPBE (left panel) or the PBE potential (right panel).
The FM state corresponds to $\mu_{S}=2.0~\mu_{\text{B}}$.
Blue and red regions correspond to negative and positive values, respectively.
The regions with the most intense blue/red colors correspond to absolute
values above 0.02~Ry/bohr$^{3}$.}
\end{figure}

The MGGAs MVS, SCAN, HLE17, and TASK, and the GGA HLE16 lead to
magnetic moments and magnetic energies
that are usually clearly larger than the PBE value. For these
functionals Fig.~\ref{fig_exc_Fe_1} shows for Fe the corresponding xc magnetic
energy density in comparison to the one obtained with PBE. We can see that the
empirical HLE16 and HLE17 functionals behave in a similar way and lead to
a xc magnetic energy density that is more negative
than PBE in larger regions of space (the red regions) and with different
orientations of the lobes compared to the other functionals.
However, it is important to mention that  these differences between HLE16/HLE17
and the other functionals are actually mostly due to their corresponding xc
potential (i.e., to self-consistency effects), which lead to different
shape/occupation of the orbitals.
This is demonstrated in Fig.~\ref{fig_exc_Fe_2} which compares
$\Delta\epsilon_{\text{xc}}^{\text{PBE}}-\Delta\epsilon_{\text{xc}}^{\text{HLE17}}$
when $\Delta\epsilon_{\text{xc}}^{\text{HLE17}}$ is
calculated with the density and KED obtained from either the mRPBE or the PBE potential.
Very different patterns are obtained, and in the latter case
$\Delta\epsilon_{\text{xc}}^{\text{PBE}}-\Delta\epsilon_{\text{xc}}^{\text{HLE17}}$
is very similar to
$\Delta\epsilon_{\text{xc}}^{\text{PBE}}-\Delta\epsilon_{\text{xc}}^{\text{MVS}}$
or
$\Delta\epsilon_{\text{xc}}^{\text{PBE}}-\Delta\epsilon_{\text{xc}}^{\text{SCAN}}$,
for instance. Nevertheless, despite the seemingly large influence of the
density/KED on $\Delta\epsilon_{\text{xc}}^{\text{HLE17}}$, the results for $\mu_{S}$ and
$\Delta E_{\text{tot}}$ change little. Indeed, using the density/KED generated from
the PBE potential leads to $\mu_{S}=2.71~\mu_{\text{B}}$ and
$\Delta E_{\text{tot}}=-1441$~meV/f.u., which is quite similar to the
results from Tables~\ref{FM} and \ref{FM_E}
obtained with the mRPBE density/KED
($\mu_{S}=2.67~\mu_{\text{B}}$ and $\Delta E_{\text{tot}}=-1491$~meV/f.u.).

Besides HLE16/HLE17, TASK and TPSS (or TM which is similar) lead to
negative regions that dominate the most and the least, respectively.
This corroborates with the magnetic moment that is among the largest
(smallest) with TASK (TPSS/TM).

\begin{figure}
\includegraphics[width=\columnwidth]{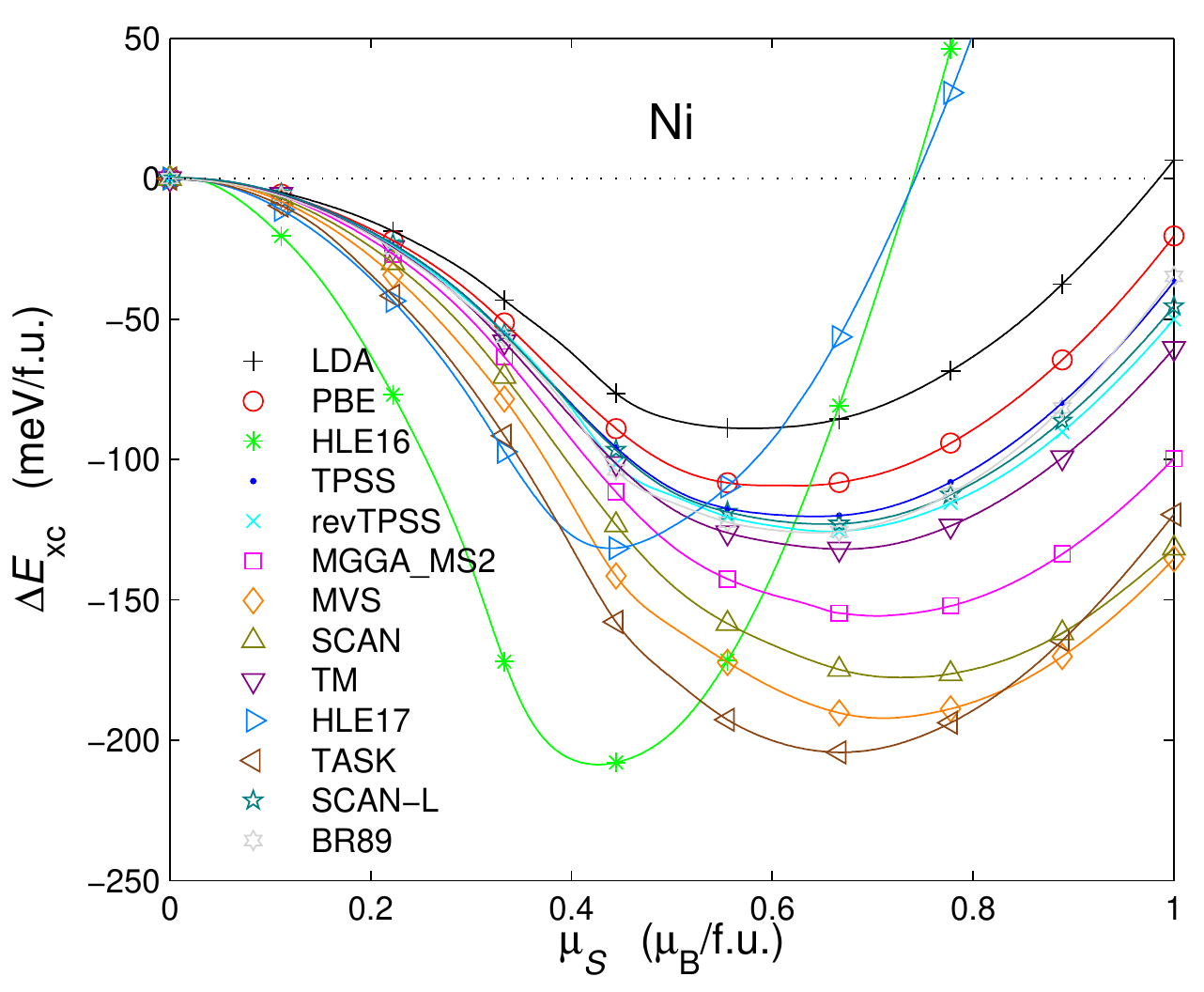}
\caption{\label{fig_Exc_Ni}xc magnetic energy $\Delta E_{\text{xc}}$ as a function
of the magnetic moment $\mu_{S}$ in Ni.}
\end{figure}

Taking Ni as an example, Fig.~\ref{fig_Exc_Ni} shows $\Delta E_{\text{xc}}$ as a
function of $\mu_{S}$. The shape and order of magnitude of the curves look
rather similar to those of the total magnetic energy $\Delta E_{\text{tot}}$
[see Fig.~\ref{fig_M_E_FM}(b)], which indicates that the other terms
(kinetic energy and Coulomb) play a less important role. However, note that
the magnitude of $\Delta E_{\text{xc}}$ is about 50~meV/f.u. larger than
$\Delta E_{\text{tot}}$. As already seen with $\Delta E_{\text{tot}}$,
the HLE16 and HLE17 functionals behave very
differently from the other functionals. They lead to a minimum of the
$\Delta E_{\text{xc}}$ curve which is at a much smaller value of the magnetic moment,
however this effect is much less pronounced for the total magnetic energy 
$\Delta E_{\text{tot}}$ [Fig.~\ref{fig_M_E_FM}(b)].

\begin{figure}
\begin{picture}(8.6,4.6)(0,0)
\put(0,0){\epsfxsize=4.2cm \epsfbox{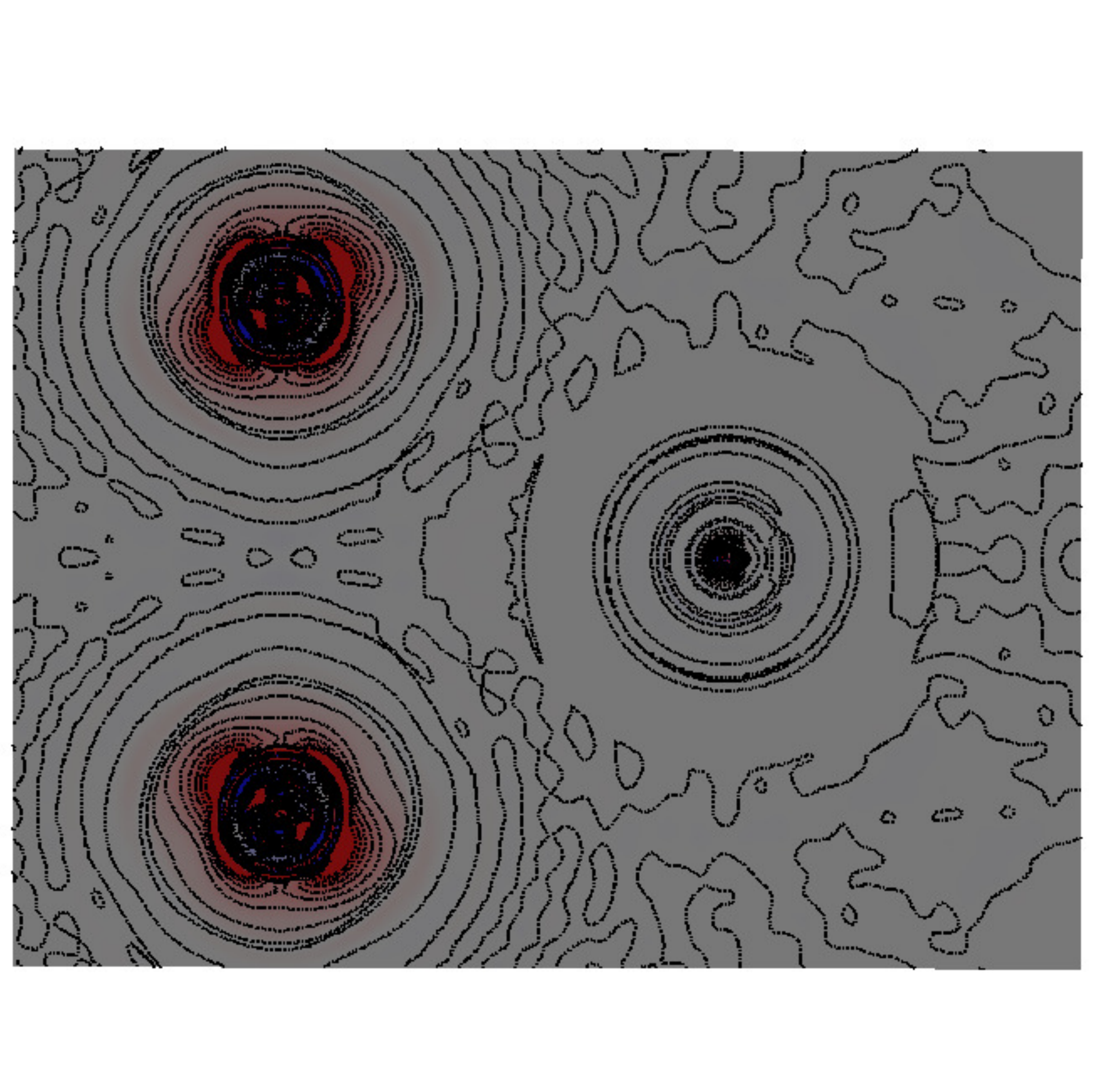}}
\put(4.3,0){\epsfxsize=4.2cm \epsfbox{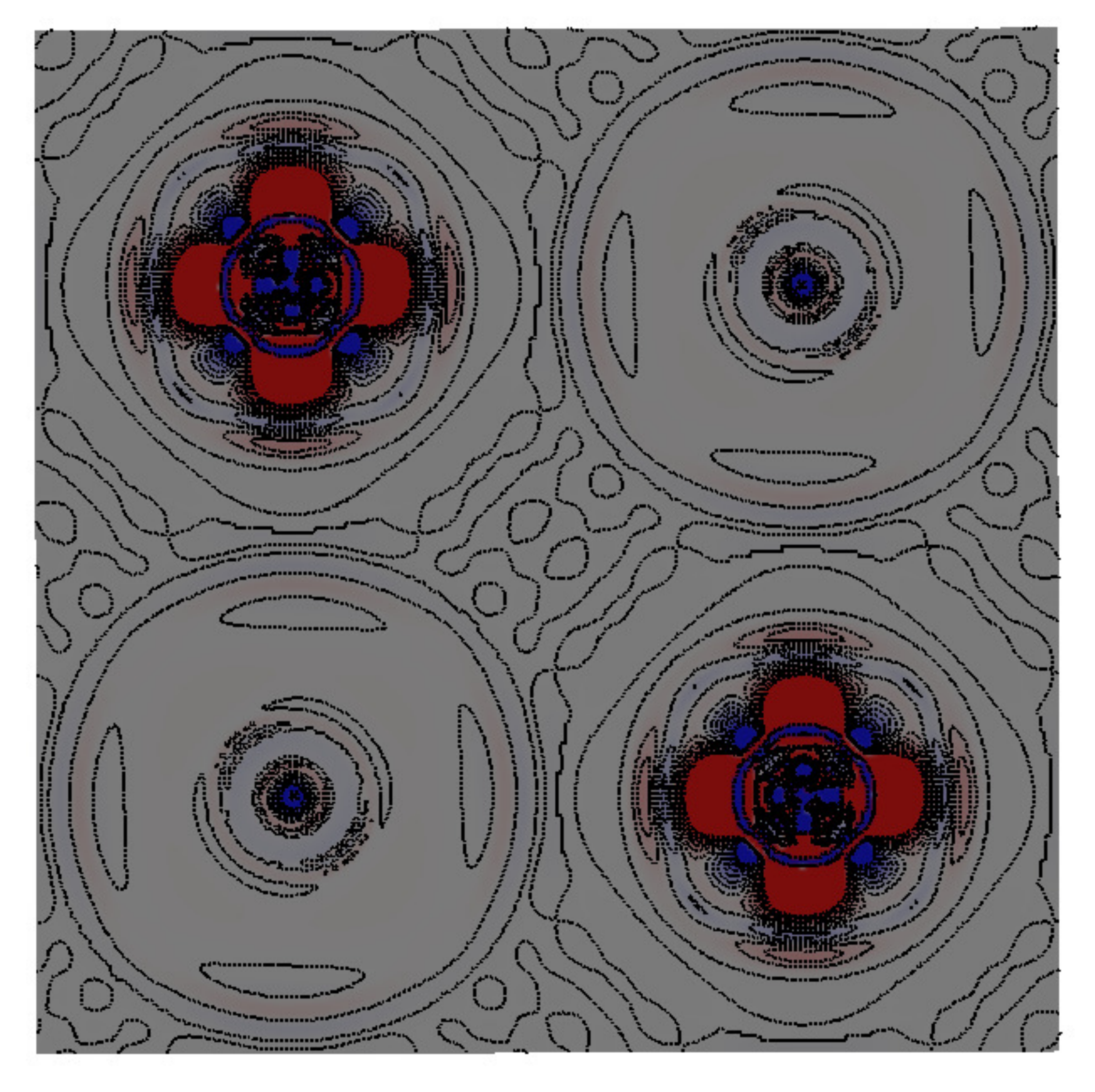}}
\put(0.1,3.8){CrSb}
\put(4.4,4.2){NiO}
\end{picture}
\caption{\label{fig_exc_CrSb_NiO_1}Difference
$\Delta\epsilon_{\text{xc}}^{\text{SCAN-L}}-\Delta\epsilon_{\text{xc}}^{\text{SCAN}}$
between the xc magnetic energy density obtained
with SCAN and SCAN-L within a (110) plane in CrSb
(left panel, the left atoms are Cr) and within a (100) plane in NiO
(right panel, the upper left atom is Ni).
The AFM states correspond to an atomic
moment (defined according to the Bader volume) 
of 3.0~$\mu_{\text{B}}$ (Cr) and 1.5~$\mu_{\text{B}}$ (Ni).
Blue and red regions correspond to negative and positive values, respectively.
The regions with the most intense blue/red colors correspond to absolute
values above 0.01~Ry/bohr$^{3}$.}
\end{figure}

\begin{figure}
\begin{picture}(8.6,12)(0,0)
\put(0,7.6){\epsfxsize=4.2cm \epsfbox{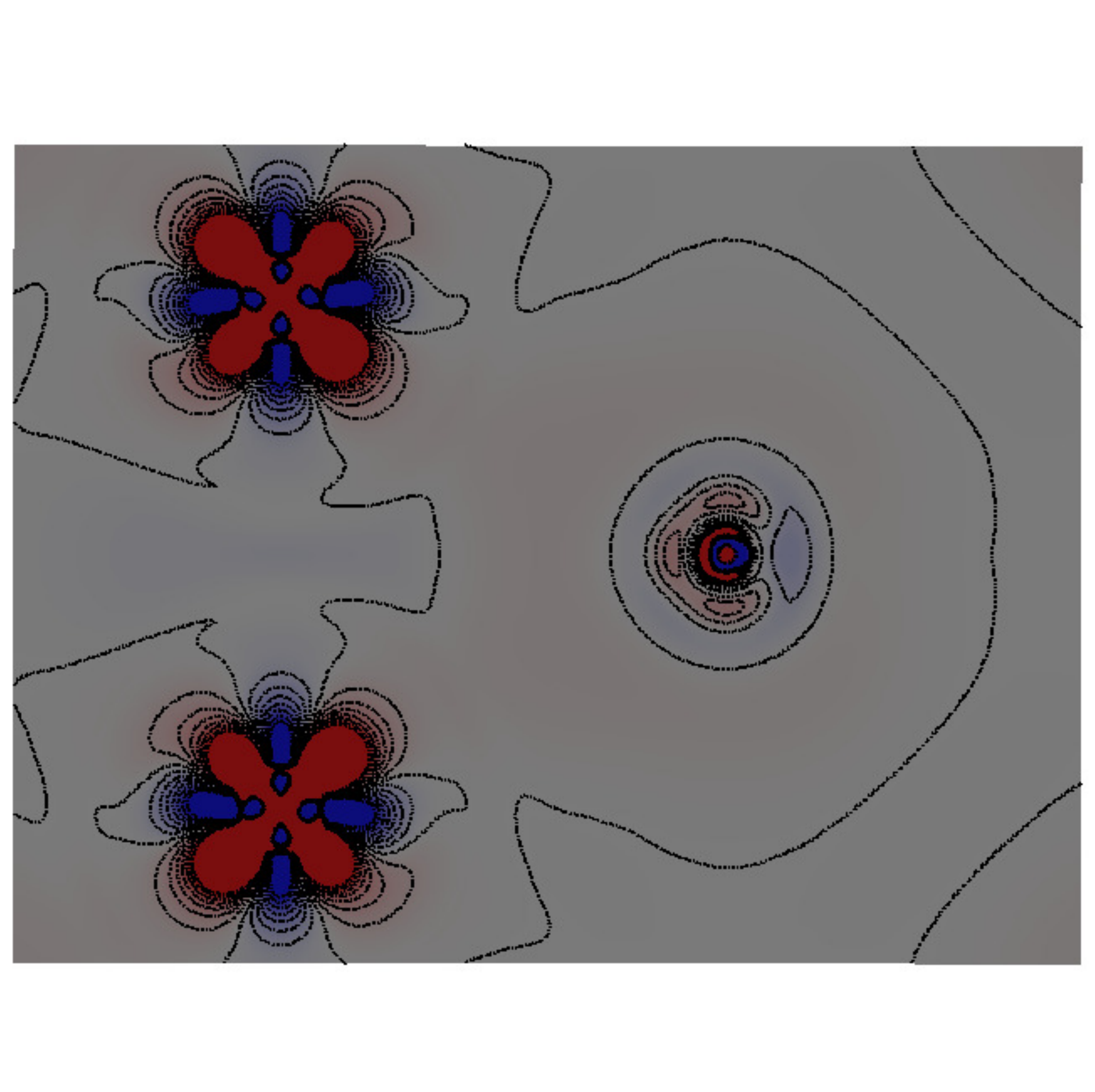}}
\put(4.3,7.55){\epsfxsize=4.2cm \epsfbox{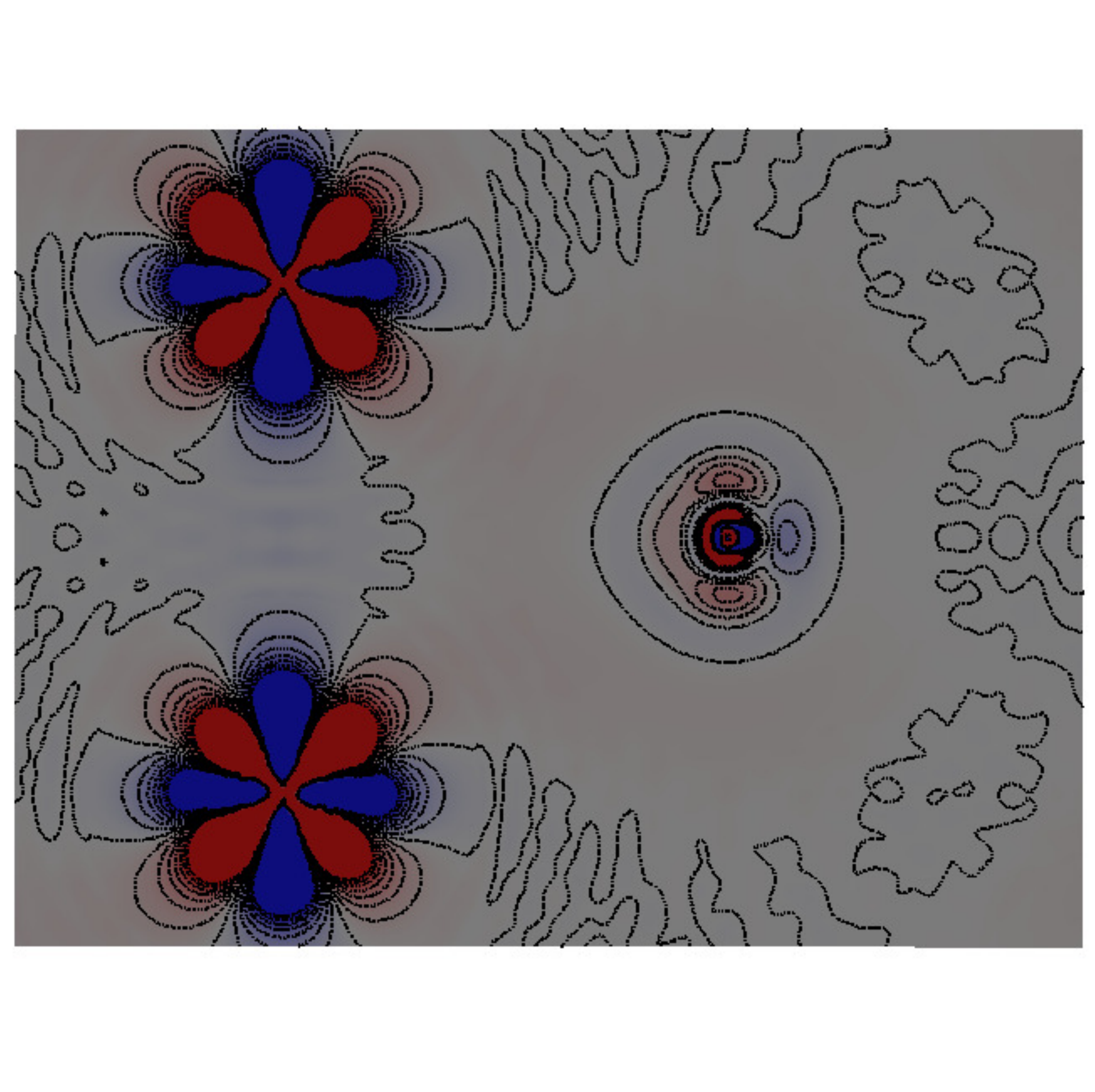}}
\put(0,3.8){\epsfxsize=4.2cm \epsfbox{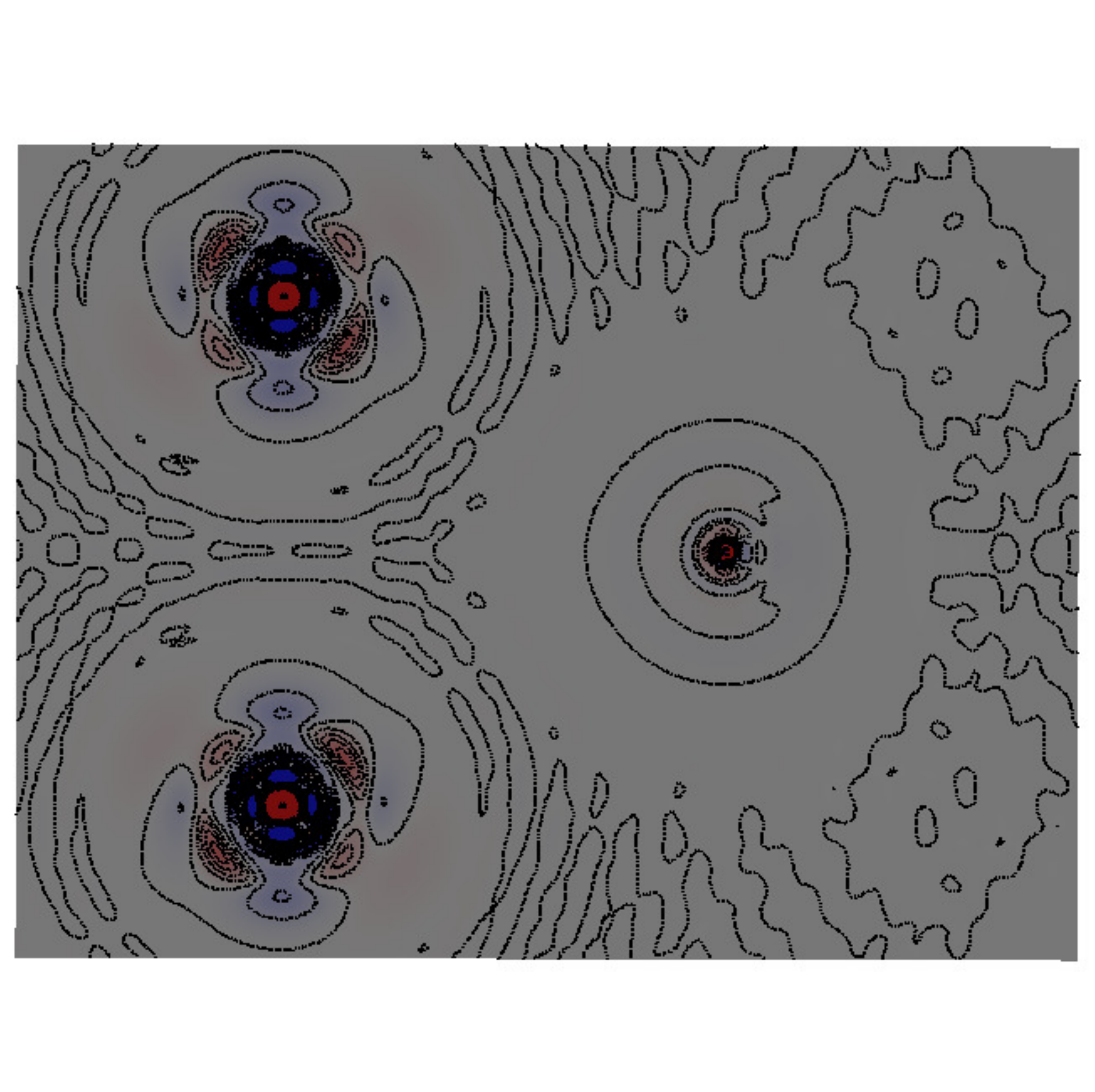}}
\put(4.3,3.8){\epsfxsize=4.2cm \epsfbox{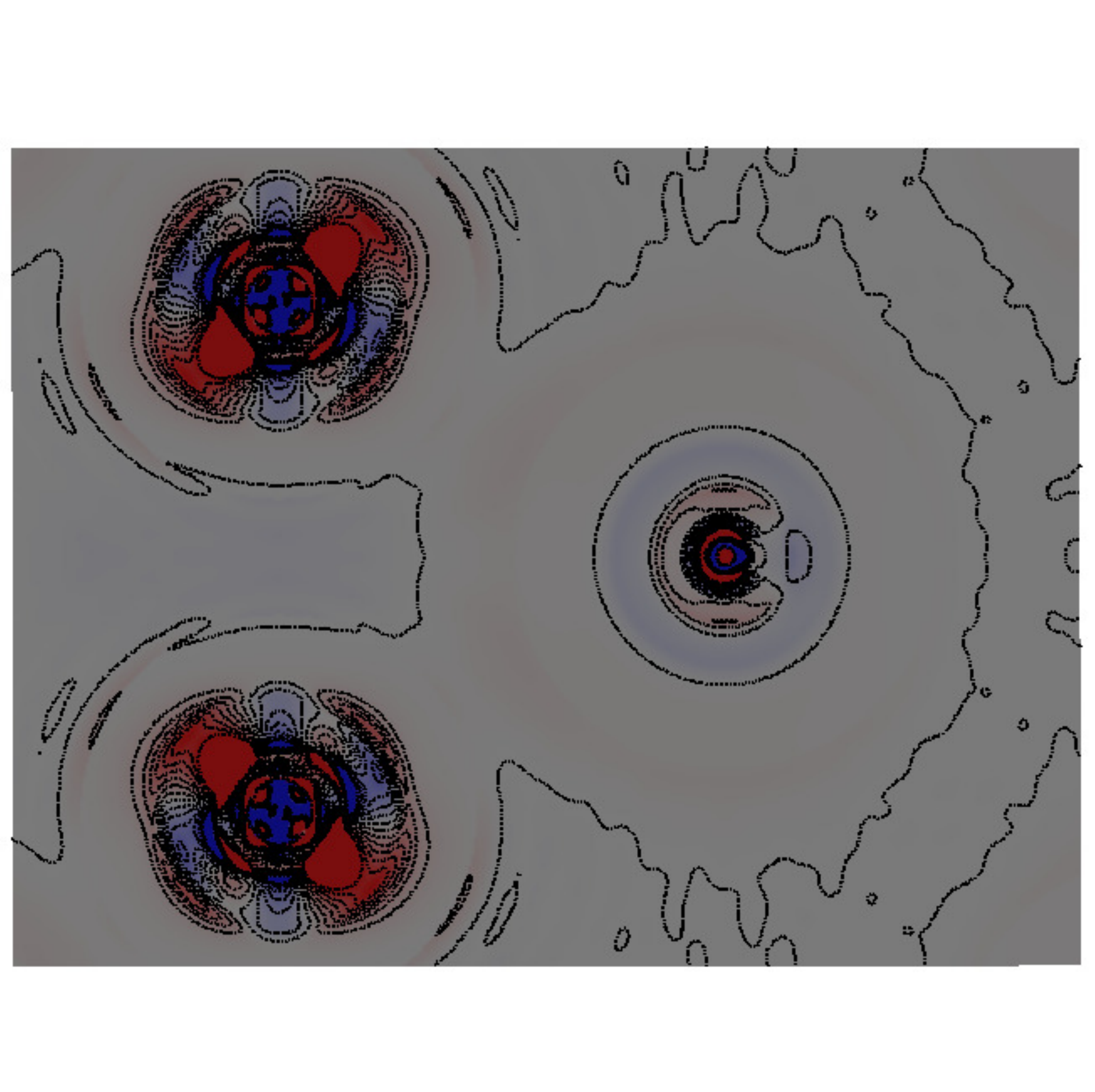}}
\put(0,0.0){\epsfxsize=4.2cm \epsfbox{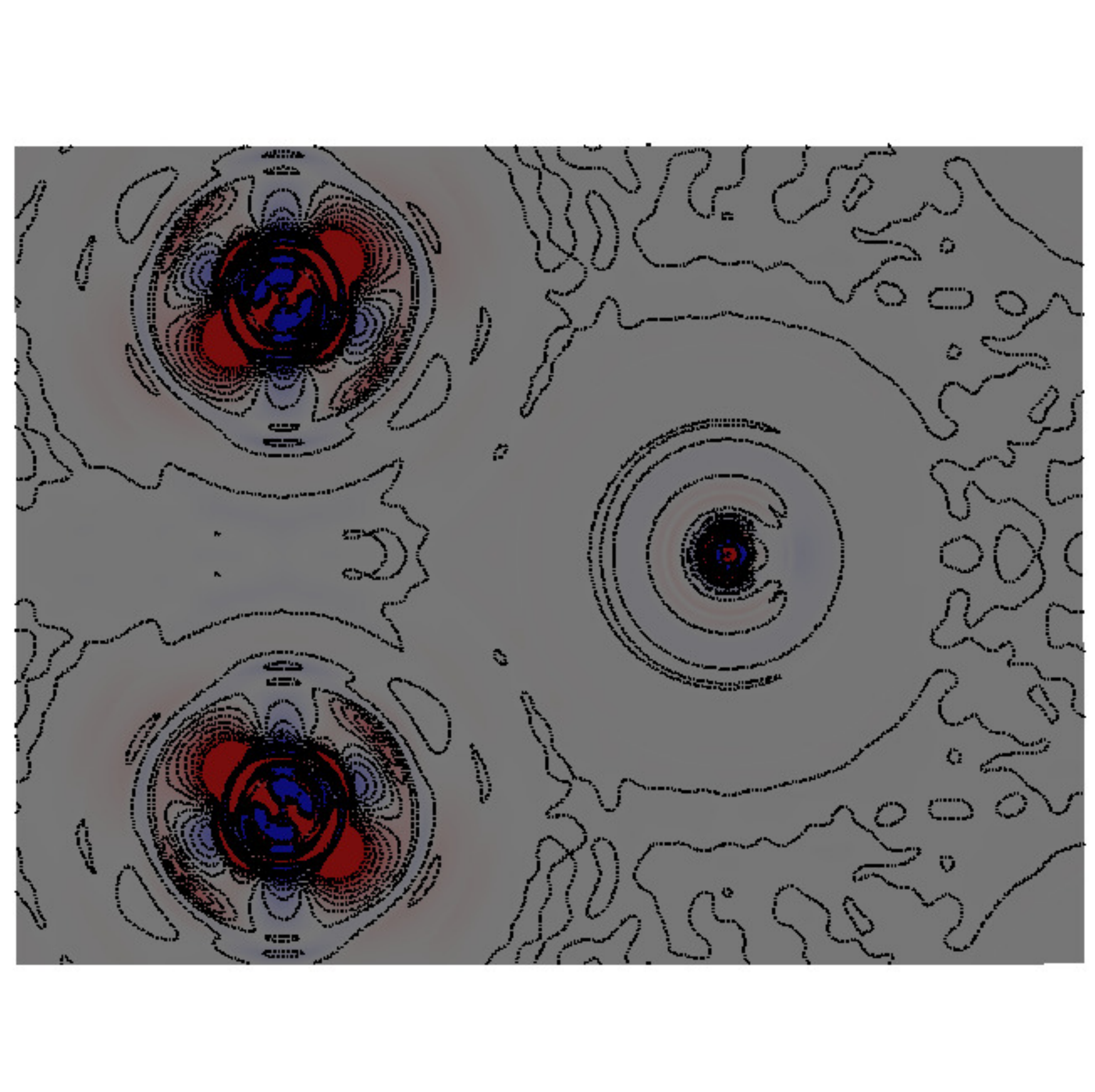}}
\put(4.3,0.0){\epsfxsize=4.2cm \epsfbox{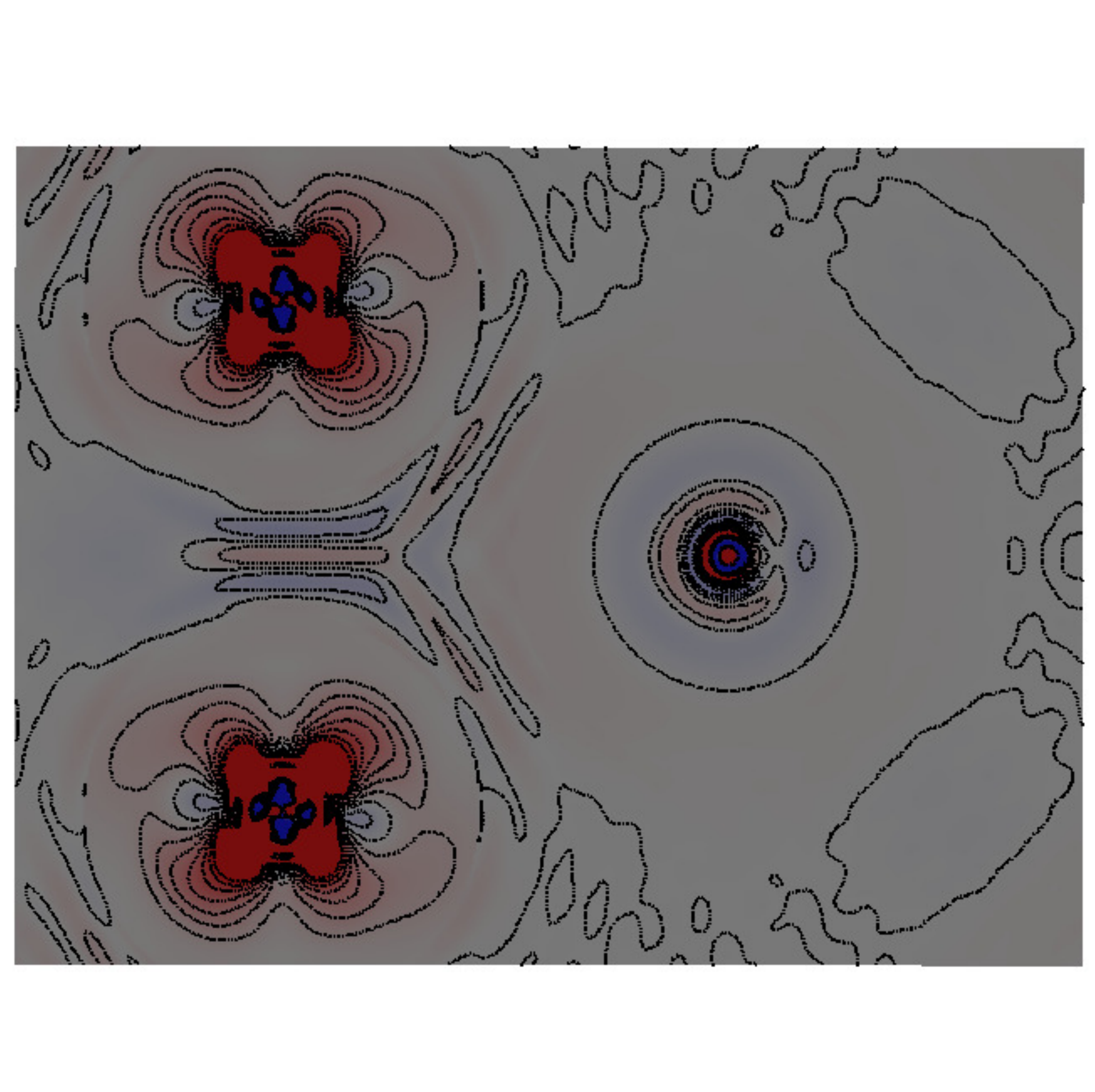}}
\put(0.1,11.3){PBE$-$HLE16}
\put(4.4,11.3){PBE$-$HLE17}
\put(0.1,7.55){PBE$-$TPSS}
\put(4.4,7.55){PBE$-$MVS}
\put(0.1,3.75){PBE$-$SCAN}
\put(4.4,3.75){PBE$-$TASK}
\end{picture}
\caption{\label{fig_exc_CrSb_2}Difference
$\Delta\epsilon_{\text{xc}}^{\text{PBE}}-\Delta\epsilon_{\text{xc}}^{F}$
between the xc magnetic energy density within a (110) plane in CrSb
obtained with PBE and another functional $F$. The AFM state corresponds to a
Cr (left atoms) atomic moment of $\mu_{S}=3.0$~$\mu_{\text{B}}$ (defined
according to the Bader volume). Blue and red regions correspond to negative
and positive values, respectively.
The regions with the most intense blue/red colors correspond to absolute
values above 0.01~Ry/bohr$^{3}$.}
\end{figure}
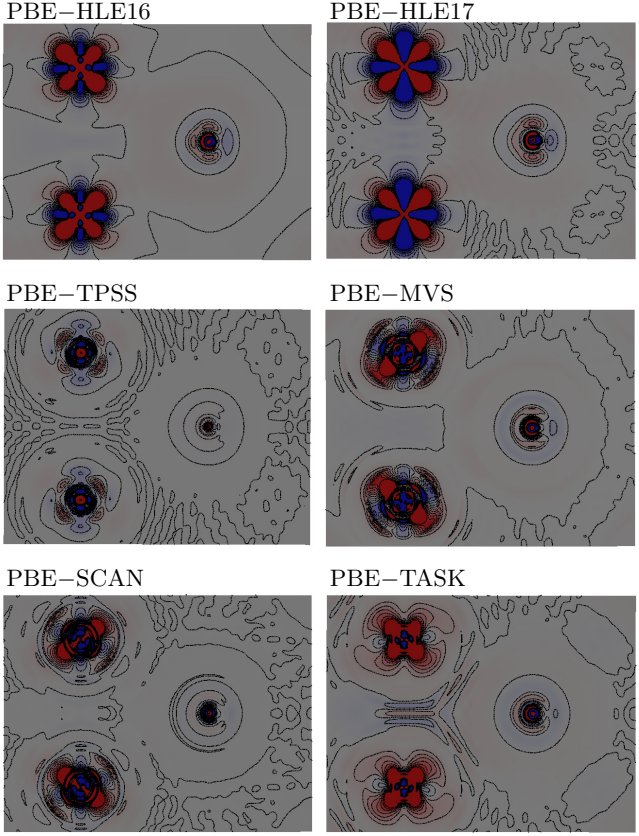

Figures~\ref{fig_exc_CrSb_NiO_1} and \ref{fig_exc_CrSb_2} show $\Delta\epsilon_{\text{xc}}$
in the AFM systems CrSb and NiO. As for the FM systems, the difference
$\Delta\epsilon_{\text{xc}}^{\text{SCAN-L}}-\Delta\epsilon_{\text{xc}}^{\text{SCAN}}$
in CrSb and NiO evidences the fact that SCAN leads to larger atomic moments
than SCAN-L, since $\Delta\epsilon_{\text{xc}}^{\text{SCAN}}$ is overall more negative than
$\Delta\epsilon_{\text{xc}}^{\text{SCAN-L}}$ on the transition-metal atoms.
For CrSb on Fig.~\ref{fig_exc_CrSb_2}, we can see that
$\Delta\epsilon_{\text{xc}}^{\text{PBE}}-\Delta\epsilon_{\text{xc}}^{F}$ is
mostly positive on the Cr atom. In the case of HLE17, the regions
corresponding to positive and negative values are in this plane (visually) roughly equal;
neverthless, around the Cr atoms the positive values of
$\Delta\epsilon_{\text{xc}}^{\text{SCAN-L}}-\Delta\epsilon_{\text{xc}}^{\text{SCAN}}$
clearly dominate and represent about 70\% of the integrated value
(which is positive) of
$\Delta\epsilon_{\text{xc}}^{\text{SCAN-L}}-\Delta\epsilon_{\text{xc}}^{\text{SCAN}}$
in the unit cell.
In the case of TASK, the positive region largely dominates in this plane.
We note again that HLE16, HLE17, and TASK lead to atomic moments of
4.10, 3.83, and 3.71~$\mu_{\text{B}}$, respectively, which are
much larger than 2.90~$\mu_{\text{B}}$ from PBE.
As discussed above for FM Fe, the magnitude of 
$\Delta\epsilon_{\text{xc}}^{\text{PBE}}-\Delta\epsilon_{\text{xc}}^{F}$
is the smallest for TPSS, which leads to a magnetic moment,
2.97~$\mu_{\text{B}}$, very close to PBE.

It may also be interesting to compare the mathematical form of the
xc-enhancement factor $F_{\text{xc}}$ of the various functionals,
which is defined as
\begin{equation}
F_{\text{xc}}(\mathbf{r}) =
\frac{\epsilon_{\text{xc}}(\mathbf{r})}{\epsilon_{\text{x}}^{\text{LDA}}(\mathbf{r})},
\label{Fxc}
\end{equation}
where (in spin-unpolarized formulation)
$\epsilon_{\text{x}}^{\text{LDA}}=-\left(3/4\right)\left(3/\pi\right)^{1/3}\rho^{4/3}$
is the exchange energy density from LDA.\cite{KohnPR65}
$F_{\text{xc}}$ is usually expressed as a function of the Wigner-Seitz radius
$r_{s}=\left(3/\left(4\pi\rho\right)\right)^{1/3}$,
reduced density gradient
$s=\left\vert\nabla\rho\right\vert/\left(2\left(3\pi^{2}\right)^{1/3}\rho^{4/3}\right)$,
and iso-orbital indicator $\alpha=\left(t-t^{\text{W}}\right)/t^{\text{TF}}$,
where $t^{\text{TF}}=\left(3/10\right)\left(3\pi^{2}\right)^{2/3}\rho^{5/3}$ and
$t^{\text{W}}=\left\vert\nabla\rho\right\vert^{2}/\left(8\rho\right)$ are
the Thomas-Fermi\cite{ThomasPCPS27,FermiRANL27} and
von Weizs\"{a}cker\cite{vonWeizsackerZP35} KED, respectively.

\begin{figure}
\includegraphics[width=\columnwidth]{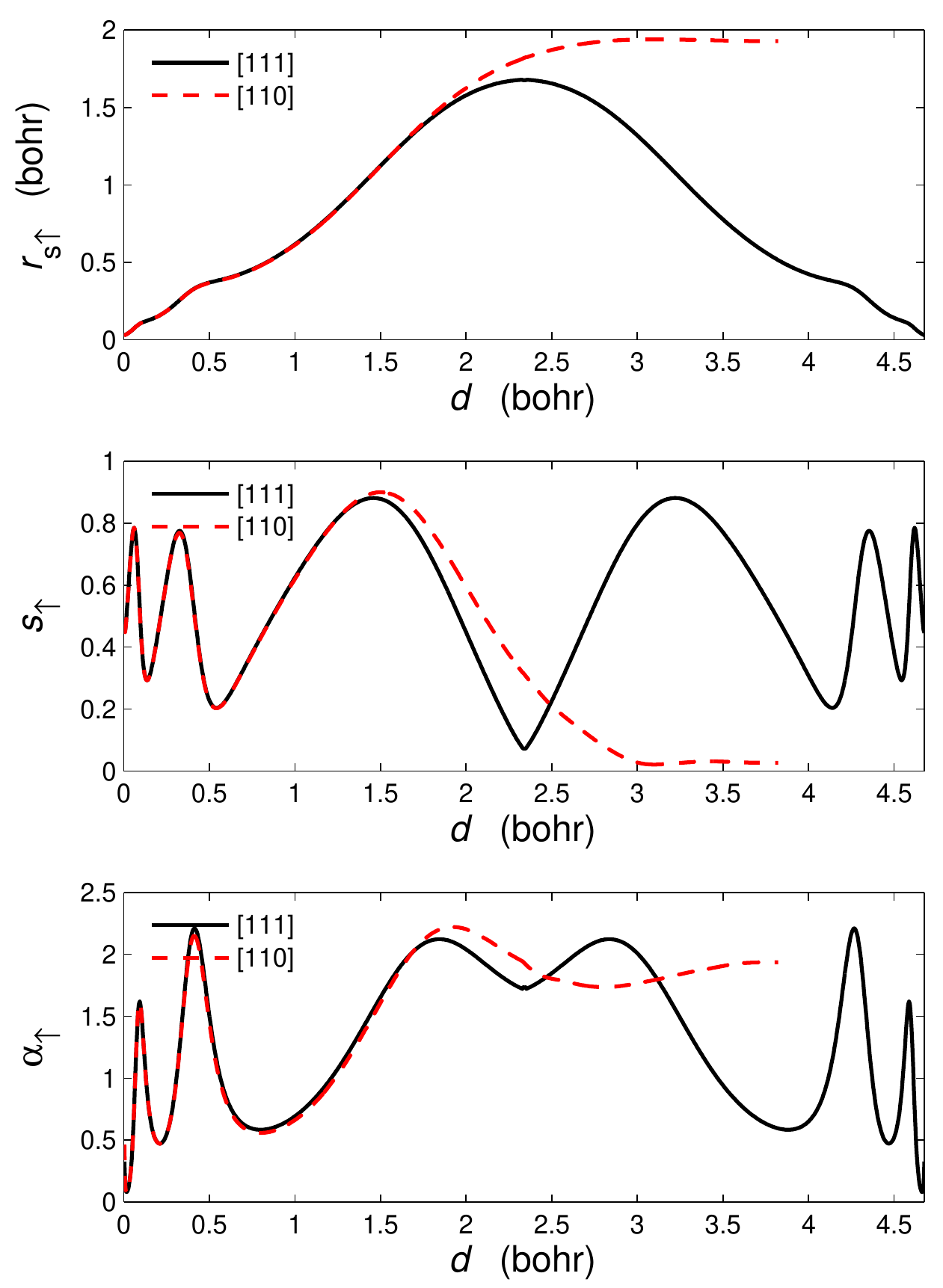}
\caption{\label{fig_rs_s_alpha_bccfe}$r_{s\sigma}$, $s_{\sigma}$, and
$\alpha_{\sigma}$ for $\sigma=\uparrow$ (majority spin) as a function of
the distance $d$ for FM Fe plotted along the direction from
(0,0,0) to (1/2,1/2,1/2) or (1/2,1/2,0).}
\end{figure}

In order to give an idea of the typical values of $r_{s}$, $s$, and $\alpha$
encountered in dense solids, and to make a relationship with
the enhancement factors, Fig.~\ref{fig_rs_s_alpha_bccfe} shows plots of
these quantities in FM Fe. Here, $r_{s}$ ranges from 0 to 2, $s$ from
0 to 1, and $\alpha$ from 0.5 to 2.5.

\begin{figure*}
\includegraphics[scale=0.42]{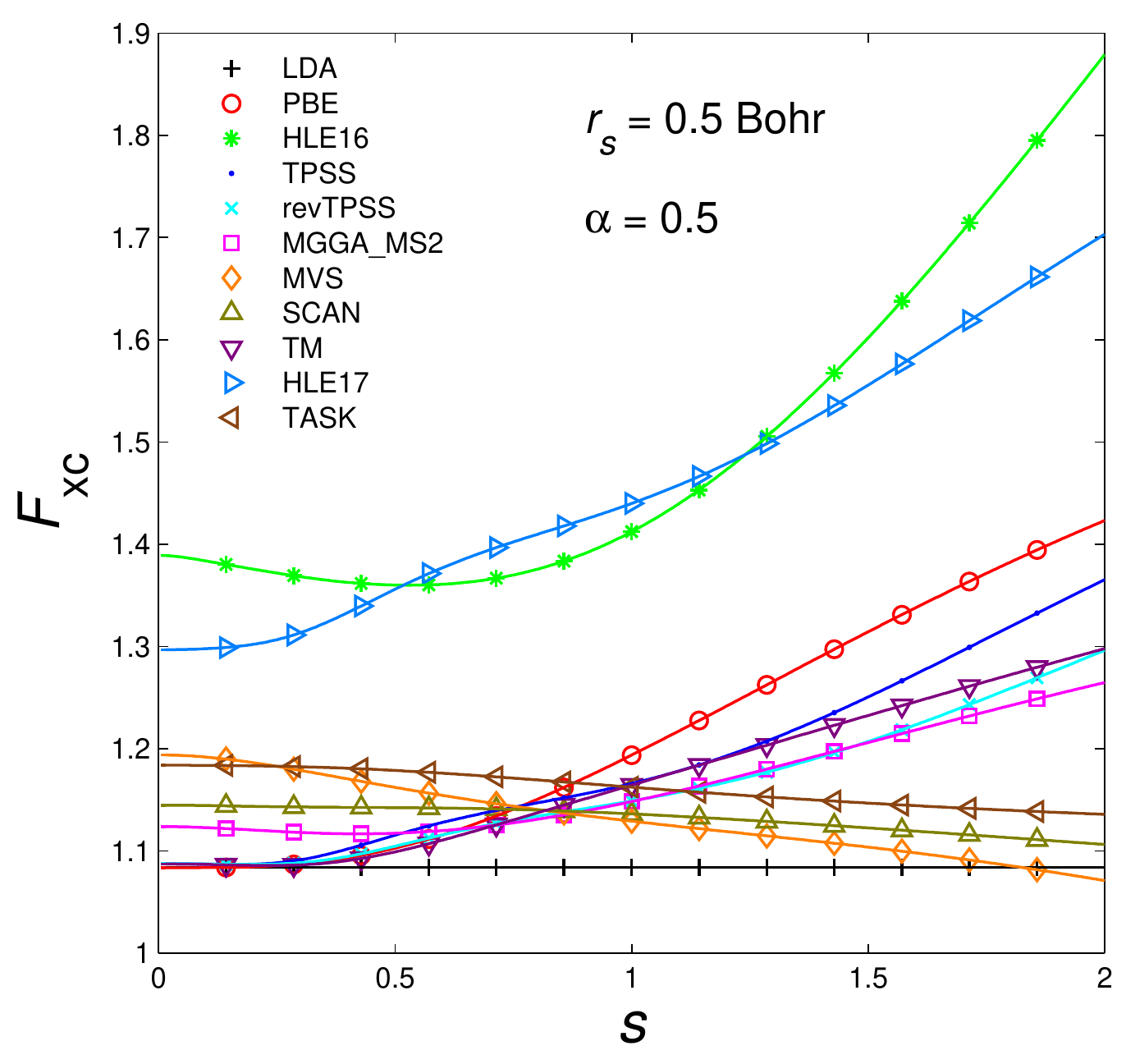}
\includegraphics[scale=0.42]{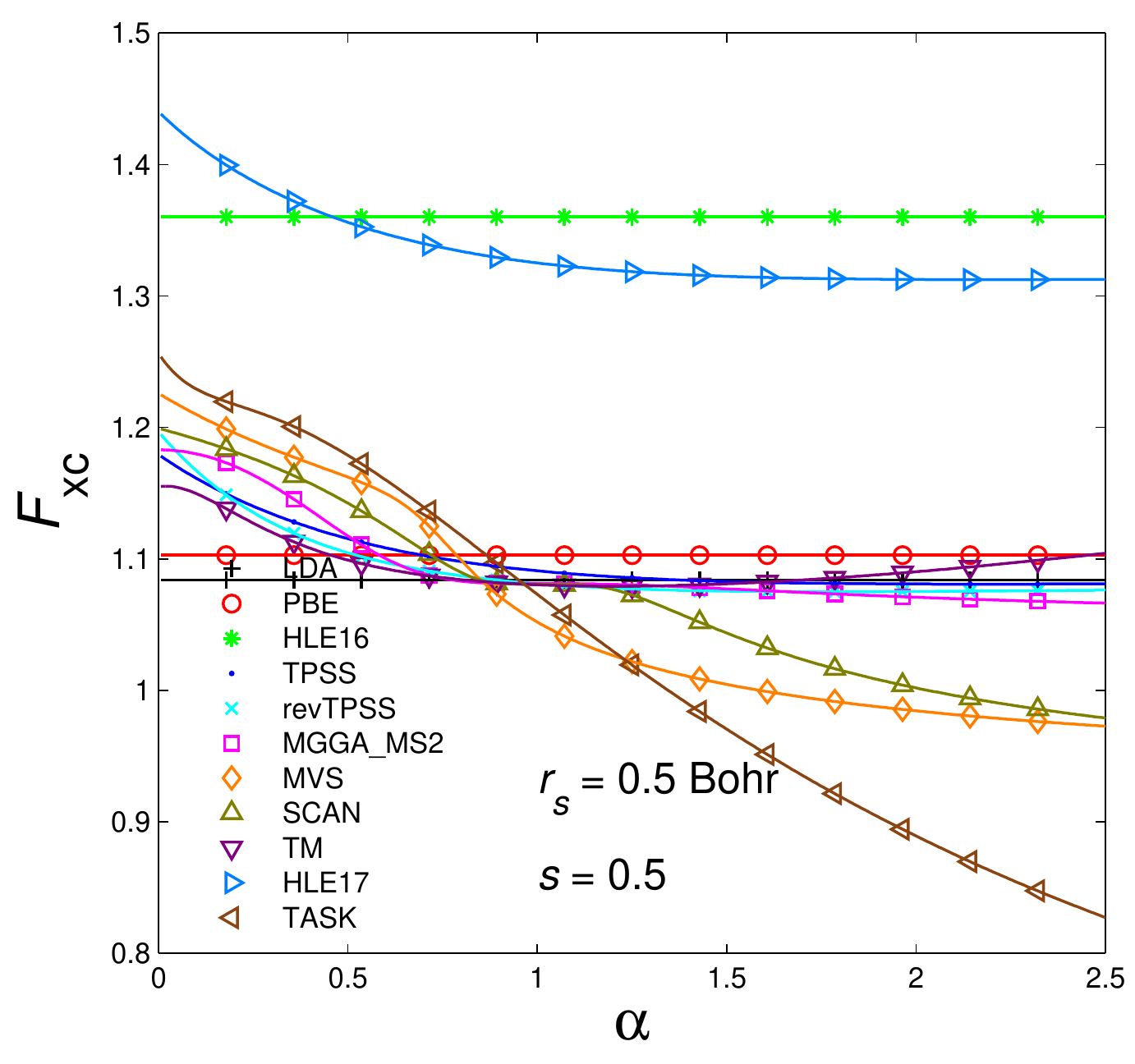}
\includegraphics[scale=0.42]{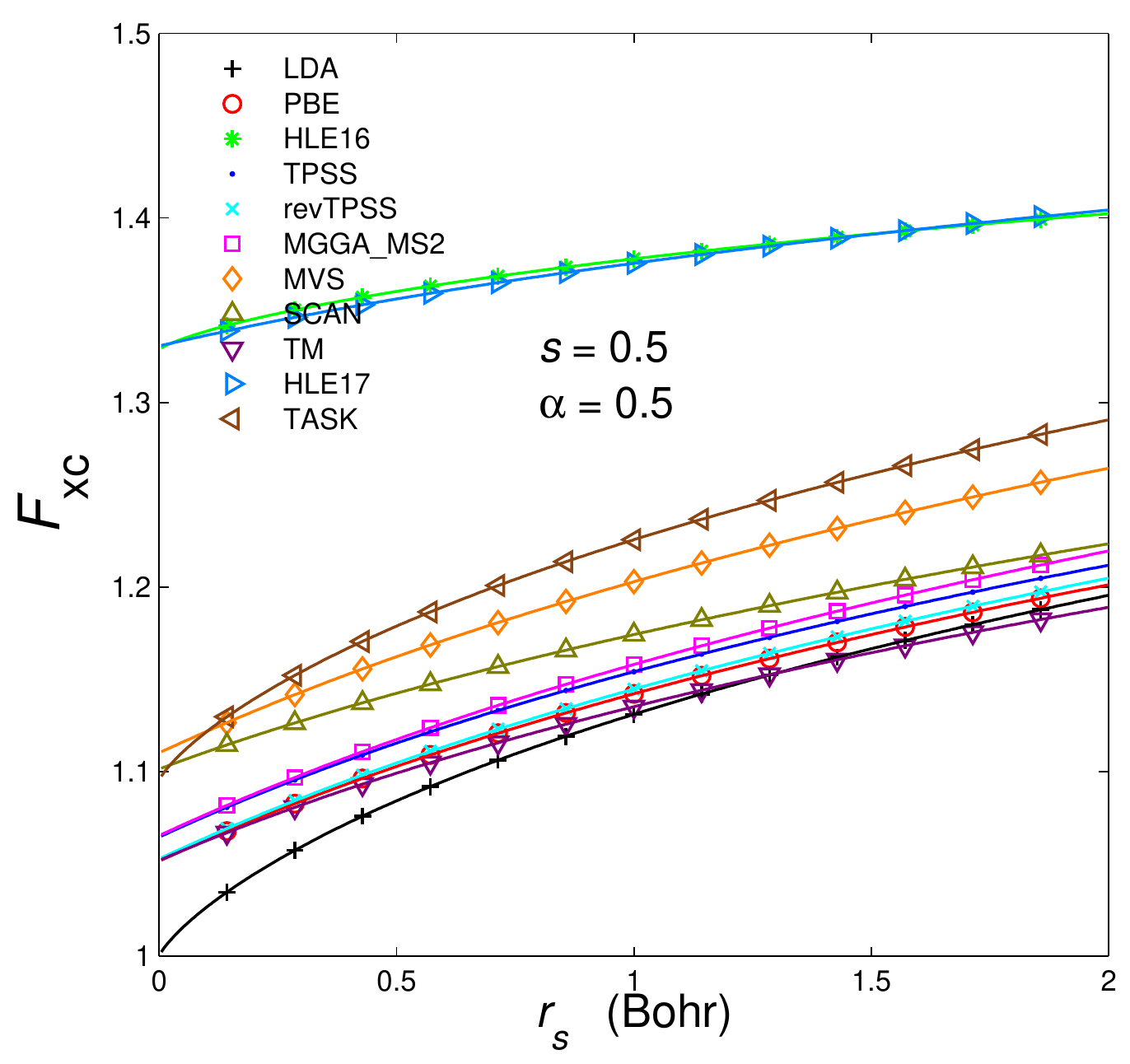}
\includegraphics[scale=0.42]{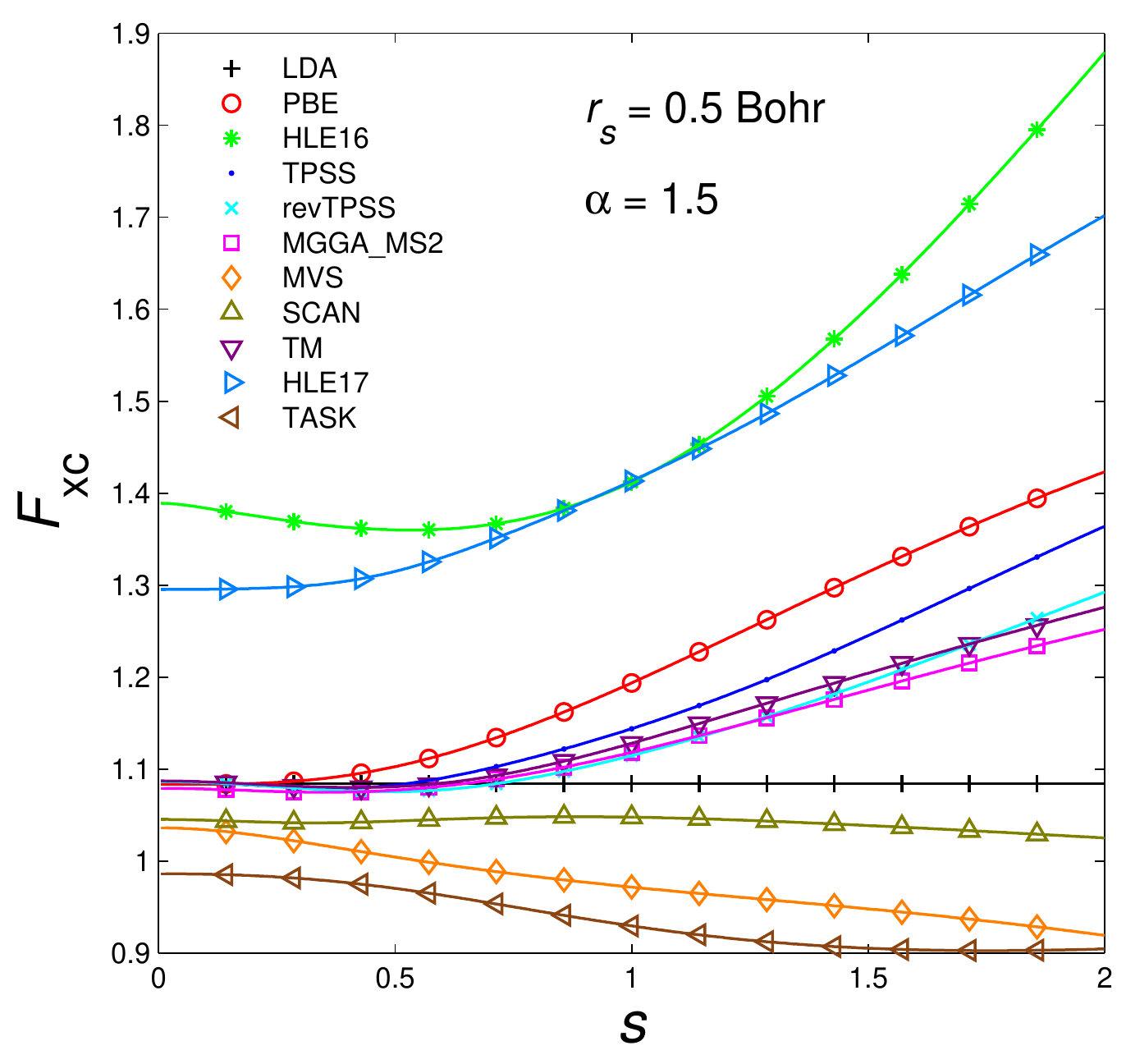}
\includegraphics[scale=0.42]{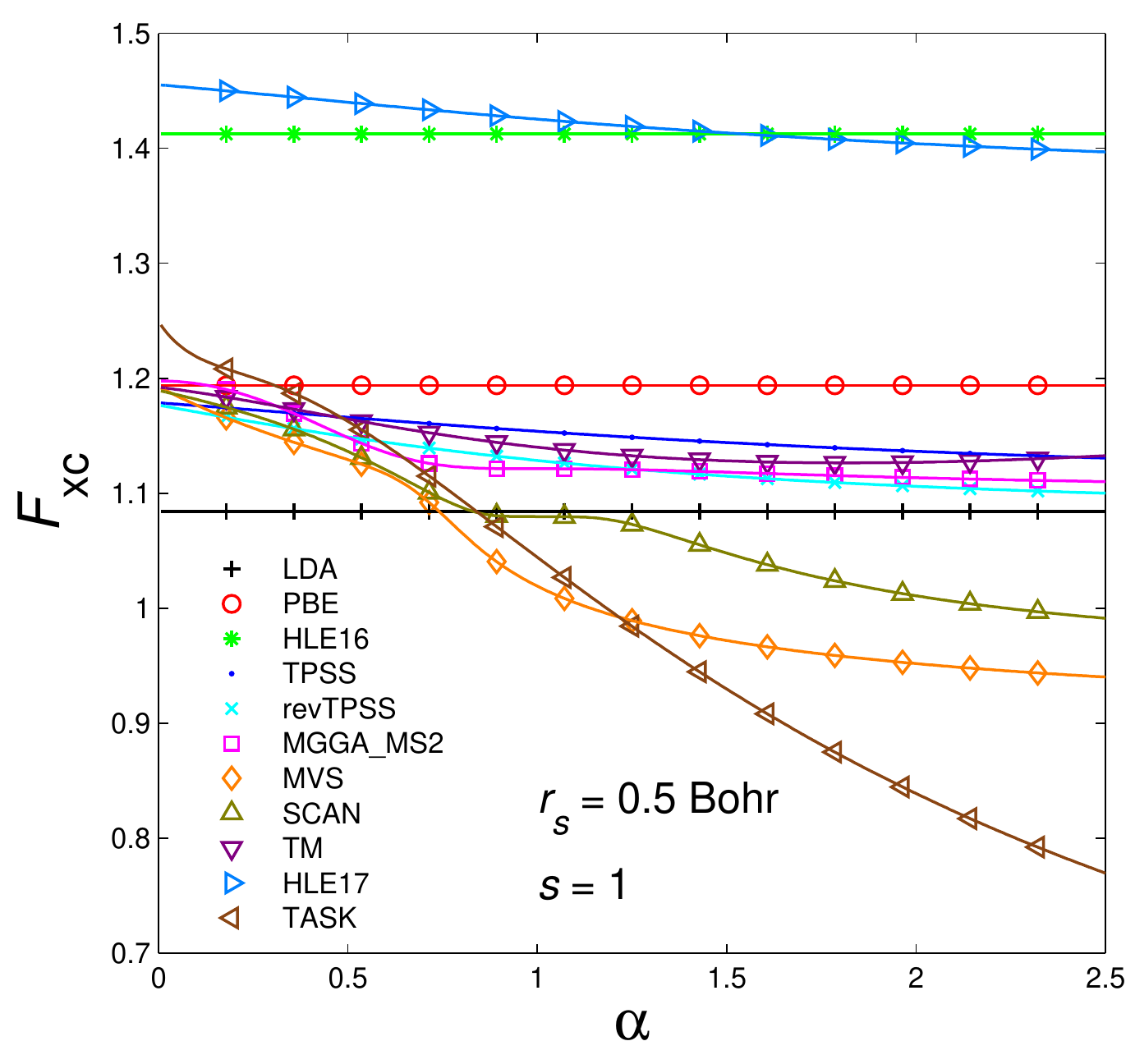}
\includegraphics[scale=0.42]{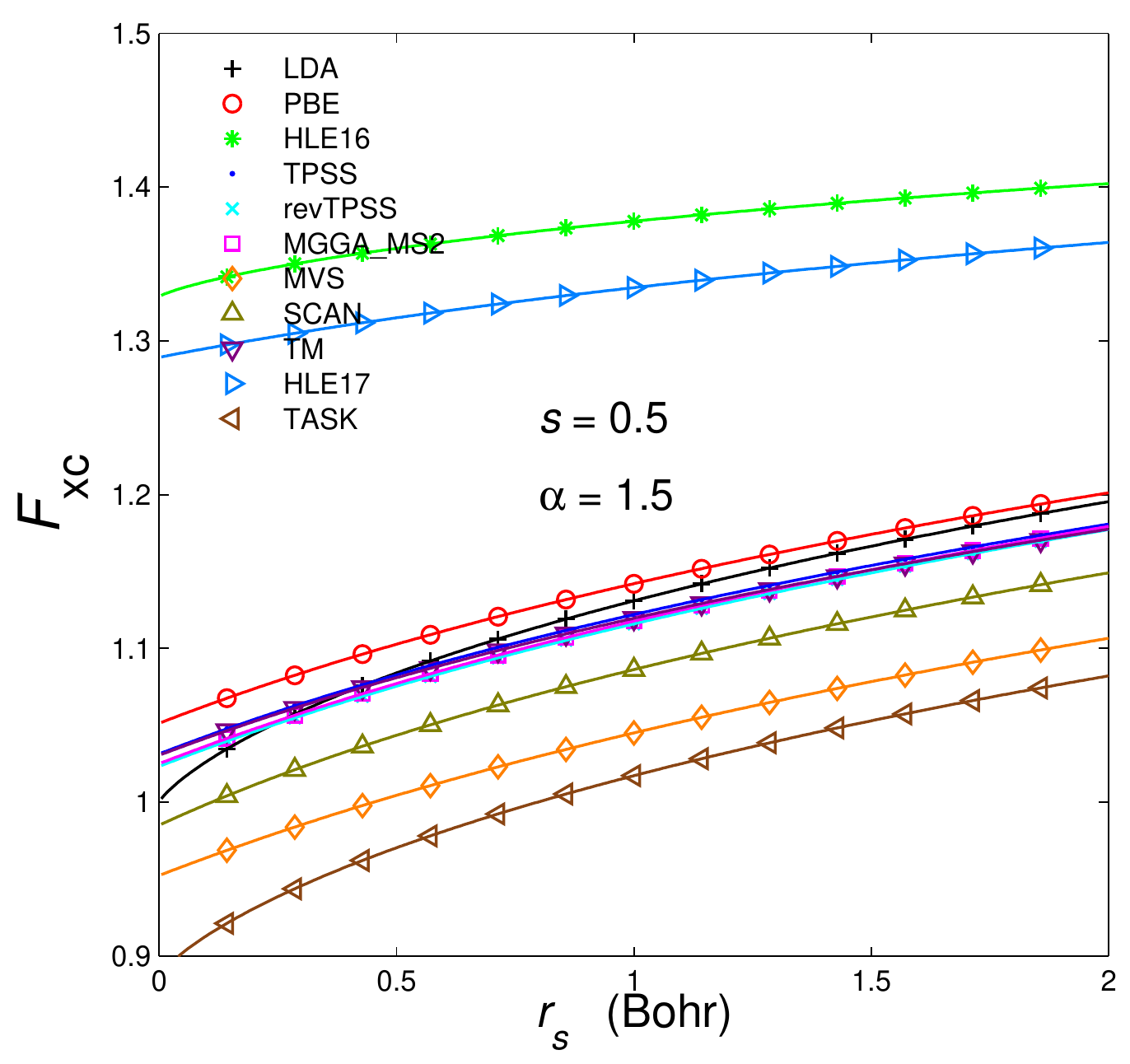}
\caption{\label{fig_Fxc}Enhancement factors $F_{\text{xc}}$ plotted as a
function of $s$ (left panels), $\alpha$ (middle panels), or $r_{s}$
(right panels). The value of the two other variables (that are kept fixed)
are indicated in the respective panels. Note the different scales on the
vertical axis.}
\end{figure*}

Figure~\ref{fig_Fxc} shows $F_{\text{xc}}$ plotted as a function of
$s$, $\alpha$, or $r_{s}$ for all functionals except SCAN-L and BR89.
These two functionals depend on $\nabla^{2}\rho$, which does not allow a
direct comparison with the other functionals.
We can see that the enhancement factors of the GGA HLE16 and MGGA HLE17 have
rather extreme shapes; both the value of $F_{\text{xc}}$ and its derivative
$\partial F_{\text{xc}}/\partial s$
are the largest. Such particular shapes of $F_{\text{xc}}$
lead to xc potentials with very large oscillations\cite{TranJPCA17}
and therefore erratic and somehow unpredictable behavior.
We also note that among the MGGAs,
TPSS, revTPSS, and TM have the weakest dependency on $\alpha$ and behave
nearly like GGAs. In contrast to them, $F_{\text{xc}}$ from the TASK
functional has by far the strongest
variation with respect to $\alpha$, so that starting at some value of
$\alpha$ it becomes the smallest enhancement factor among all those considered in
this work.
As discussed below, this particular behavior of the enhancement factor of
TASK, as well as the ones from SCAN and MVS which show a similar feature,
is related to the large magnetic moments obtained with these functionals.
Actually, for these three functionals, as well as HLE16 for small values of $s$,
Fig.~\ref{fig_Fxc} shows that $\partial F_{\text{xc}}/\partial s$ is also negative.
The $r_{s}$-dependency of the LDA and TASK enhancement factors are
also particular; depending on the value of $s$ and/or $\alpha$,
they increases faster than for all other functionals.

As a side note, we mention that a few additional calculations were done by
combining exchange of one functional (e.g., SCAN) with correlation of another
functional (e.g., TPSS). From the results (not shown), we concluded that
the choice of correlation has a rather minor effect on the magnetic moment.

\begin{figure}
\includegraphics[width=\columnwidth]{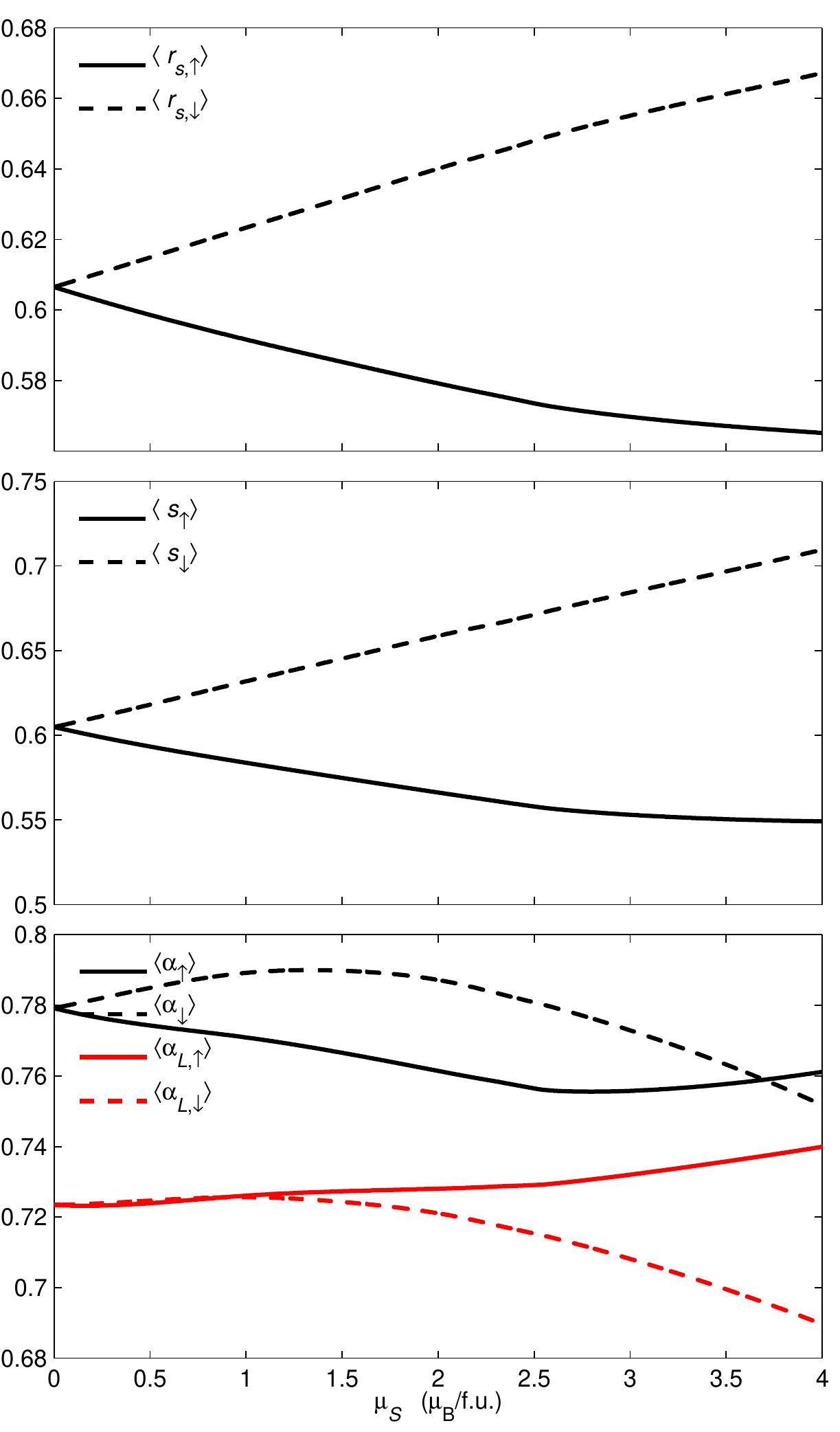}
\caption{\label{fig_bccfe_M_rs_s_alpha__Rmin000_Rmax125_1}Spatial average of
$r_{s,\sigma}$, $s_{\sigma}$,
$\alpha_{\sigma}$, and $\alpha_{L,\sigma}$ inside a sphere of radius 1.25~bohr
centered on the atom in Fe plotted as a function of $\mu_{S}$.
$\sigma=\uparrow$ corresponds to the majority spin.}
\end{figure}

MGGAs can be very different to each other in terms of enhancement factor and xc
magnetic energy density. Thus, the details of the mechanism leading to an
increase of the magnetic moment (with respect to PBE) may also differ from one
functional to the other. For instance, the two MGGAs HLE17 and TASK lead
(albeit not always with HLE17) to large magnetic moments, despite they have
extremely different analytical forms. As discussed above and in
Ref.~\onlinecite{MejiaRodriguezPRB19}, the analytical form of a functional
for densities close to the transition-metal atom determines the magnetic moment.
This is of course expected since this is where the $3d$ electrons are located,
as shown in Fig.~\ref{fig_rho3d} for FM Fe.
Figure~\ref{fig_bccfe_M_rs_s_alpha__Rmin000_Rmax125_1} shows the averages
of $r_{s,\sigma}$, $s_{\sigma}$, $\alpha_{\sigma}$, and $\alpha_{L,\sigma}$ in
a sphere of radius 1.25~bohr surrounding an atom in FM Fe plotted as function
of the magnetic moment $\mu_{S}$. We can see that for the majority spin
($\sigma=\uparrow$) $\langle r_{s,\uparrow}\rangle$, $\langle s_{\uparrow}\rangle$,
$\langle\alpha_{\uparrow}\rangle$ decrease when $\mu_{S}$ increases.
We just note that starting at $\mu_{S}\sim3$~$\mu_{\text{B}}$ $\langle\alpha_{\uparrow}\rangle$
increases, which is however not relevant since 3~$\mu_{\text{B}}$ is larger
than what all functionals give for Fe.
For the minority spin $\sigma=\downarrow$, the trends are more or less
the opposite.

Relating Fig.~\ref{fig_bccfe_M_rs_s_alpha__Rmin000_Rmax125_1} with 
the results for the magnetic moment, $\langle\alpha_{\sigma}\rangle$ is
particularly relevant for TASK, SCAN, and MVS as explained in the following.
When $\mu_{S}$ increases, $\langle\alpha_{\uparrow}\rangle$ and
$\langle\alpha_{\downarrow}\rangle$ decrease and increase, respectively.
Since the derivatives $\partial F_{\text{xc}}/\partial\alpha$ of these
three MGGAs are negative (see Fig.~\ref{fig_Fxc}),
an increase of $\mu_{S}$ leads to an increase and decrease of the
$\sigma=\uparrow$ and $\sigma=\downarrow$ exchange
enhancement factors, respectively. Thus, a negative
$\partial F_{\text{xc}}/\partial\alpha$ contributes to an increase of the
exchange splitting. This is not the case with TPSS, revTPSS, or TM which
have a very weak dependency on $\alpha$.
This is a very plausible explanation since TASK, SCAN, and MVS lead to
some of the largest moments, while TPSS, revTPSS, and TM to the smallest.
The same mechanism, but probably weaker, can be invoked with
$\langle s_{\sigma}\rangle$ since $\partial F_{\text{xc}}/\partial s$
is negative also only for TASK, SCAN, and MVS.

Aschebrock and K\"{u}mmel\cite{AschebrockPRR19}
showed that having a negative slope $\partial F_{\text{x}}/\partial\alpha$
leads to two desirable features: (a) the presence of a field-counteracting
term in the potential (as with exact exchange) and
(b) a derivative discontinuity that is larger and therefore leads to more accurate
band gaps. However, in the present context, magnetism, a
negative value of $\partial F_{\text{x}}/\partial\alpha$ does not seem
to be beneficial. Thus, MGGAs with such negative
$\partial F_{\text{x}}/\partial\alpha$ to a certain extent mimic exact exchange,
which also leads to too large magnetic moments in FM
metals.\cite{KotaniJPCM98,SchnellPRB03}
It was argued that a negative slope $\partial F_{\text{x}}/\partial\alpha$
increases the nonlocal character of the MGGA exchange.\cite{AschebrockPRR19}
A possible way to cure the over-magnetization
problem of TASK or SCAN would be to combine the exchange component with
a more compatible (and most likely more advanced) correlation component.
In the case of exact exchange, \textit{ab initio} correlations
can solve some of the problems of exact exchange.

Mej\'{\i}a-Rodr\'{\i}guez and Trickey\cite{MejiaRodriguezPRB19}
provided an explanation for the smaller moment obtained with SCAN-L
compared to SCAN. Here, a similar explanation is provided but with an
emphasis on the importance of $\partial F_{\text{xc}}/\partial\alpha$
as discussed above.
Compared to $\alpha_{\sigma}$, the deorbitalized $\alpha_{L,\sigma}$ is
smaller in magnitude for both spins, as shown in
Fig.~\ref{fig_bccfe_M_rs_s_alpha__Rmin000_Rmax125_1}.
Another important difference can be noted;
for the majority spin $\sigma=\uparrow$,
$\langle\alpha_{L,\uparrow}\rangle$ increases with $\mu_{S}$ instead
of decreasing, while for the minority spin, the increase for small
values of $\mu_{S}$ is strongly reduced.
Therefore, by substituting $\alpha_{\sigma}$ by $\alpha_{L,\sigma}$
in TASK, SCAN, or MVS, the effect on the exchange splitting due to
$\partial F_{\text{xc}}/\partial\alpha$ is strongly suppressed
(or maybe even reversed), which explains the reduction of magnetism.
Actually, a deorbitalization of TPSS leads to very small
change in the results (see Ref.~\onlinecite{MejiaRodriguezPRB19}),
which is due to the very weak dependency of TPSS on $\alpha$.

SCAN-L has been shown to be more appropriate than SCAN for magnetic and
non-magnetic itinerant metals.\cite{MejiaRodriguezPRB19}
However, the $\nabla^{2}\rho$-dependency of SCAN-L may also carry practical
disadvantages, since implementations of this family of functionals are less
common than for $t$-MGGAs. Furthermore, the third and fourth derivatives of the
density, that are required for the potential, may lead to numerical problems.
\cite{JemmerPRA95,NeumannCPL97,CancioIJQC12,MejiaRodriguezPRA17}
Therefore, it would be interesting to find a $t$-dependent alternative to
SCAN-L, that is, a slightly modified SCAN that leads to minimal changes for the
geometries and binding energies, but reduces the magnetic
moment. Mej\'{\i}a-Rodr\'{\i}guez and Trickey\cite{MejiaRodriguezPRB19}
reported such attempts, which were apparently unsuccessful. Our numerous own
attempts have all remained unsuccessful, as well.
The most simple ones consist of just changing the value of one of the
parameters in SCAN. Among them, $c_{1x}$ for instance, can be used to
vary the switching function and thus the magnetic moment
(see Ref.~\onlinecite{MejiaRodriguezPRB19}).
By increasing the value of $c_{1x}$ above $\sim2.5$
($c_{1x}=0.667$ in SCAN), the magnetic moments
of Fe, Co, and Ni get smaller and approach to some extent the SCAN-L values.
As discussed above, a negative slope $\partial F_{\text{xc}}/\partial\alpha$
favors a large moment. Since an increase of $c_{1x}$ makes
$\partial F_{\text{xc}}/\partial\alpha$ less negative for values of
$\alpha$ below 1, 
this should be (one of) the main reason(s) why the moments are smaller.
However, with such values for $c_{1x}$, the errors for the lattice constant
and cohesive energy (results now shown) are larger (by $\sim50$\%) than SCAN.
Another strategy that we have considered consists of slightly
modifying the expression of $\alpha$ in SCAN. Numerous expressions have been
tried, but none of them was useful to achieve our goal. A modification of
$\alpha$ can work for a system, but not for another.
Thus, the construction of a $t$-MGGA with similar performance as SCAN-L seems
far from trivial, as already reported by Mej\'{\i}a-Rodr\'{\i}guez and
Trickey.\cite{MejiaRodriguezPRB19}

\section{\label{summary}Summary}

The focus of this work has been on the description of magnetism in solids with
MGGA functionals. FM, NM, and AFM systems have been considered. The goal was to
provide an overview of the reliability of MGGAs and the possible improvement
with respect to standard GGA functionals like PBE. The most important
observations are the following. In the vast majority of cases, the tested MGGA
functionals lead to a magnetic moment that is larger than the value obtained
with LDA and PBE. This means that for the considered FM systems, which are itinerant
metals, the agreement with experiment can only be worse than with LDA and PBE,
which are known to already slightly (or even strongly when spin fluctuations
are important) overestimate the magnetic moment in FM solids. Consistent with this
trend, a certain number of NM metals can be wrongly described as FM with MGGAs,
and already with PBE in some cases.
In the case of the AFM oxides with localized $3d$ electrons, using a MGGA
leads to a more realistic value of the atomic moment, since PBE leads to too small
moments. Only in the case of Cr$_{2}$O$_{3}$ the LDA and PBE values seem to be within
the experimental range. Concerning the weakly correlated AFM CrSb and CrSb$_{2}$,
PBE is in agreement with experiment for the former, but strongly overestimates the
moment for the latter. Therefore, using a MGGA does not really seem to be beneficial
for such systems.

The MGGAs that we have considered can be split into two groups. Those which give
results that are qualitatively similar to PBE, namely BR89, TPSS, revTPSS, TM,
and SCAN-L. They lead to reasonable results for metals, but clearly 
underestimate the atomic magnetic moment in the AFM transition-metal oxides.
The other group consists of TASK, HLE17, SCAN, MVS, and MGGA$\_$MS2,
which lead to sizeably larger moments.
For the FM metals, TASK, SCAN, and MVS lead in many cases to the largest
magnetic moments, and therefore the largest disagreement with experiment.
They also lead to a non-zero magnetic moment for the NM metals that we have considered.
For the AFM systems, TASK, HLE17, as well as the mBJLDA potential and the GGA HLE16
lead to the largest values of the atomic moment. As just mentioned above,
this is beneficial for the transition-metal oxides, but not for the weakly correlated
CrSb and CrSb$_{2}$.
We have also shown that HLE16 and HLE17 lead to erratic and unpredictable
results. As a final short conclusion, no GGA and no MGGA leads to satisfying results,
even qualitatively, for itinerant metals and strongly correlated AFM oxides at the
same time. One goal that we have not been able to achieve is to propose a
\textit{non-deorbitalized} modification of SCAN that is good for the
magnetic moment of metals, while keeping the accuracy of SCAN for other
properties like the lattice constant.

In an attempt to provide an analysis of some of the observed trends, the
xc magnetic energy density was visualized and the analytical form of the
functionals compared. Some of the functionals, HLE16, HLE17, and TASK, have
very unusual shape for their enhancement factor.
However, it is not necessary to use a
functional with such an enhancement factor to get increased magnetism compared
to PBE; other functionals like SCAN or MVS also lead to larger magnetic moments.
Actually, we deduced
that a negative derivative $\partial F_{\text{xc}}/\partial\alpha$ of the 
enhancement factor should contribute in making the magnetic moment larger,
since the iso-orbital indicator $\alpha$ of the majority (minority)
spin decreases (increases) with the moment, leading to an enhanced exchange
splitting.

On a more technical side, we have also discussed the choice of the GGA potential
for generating the orbitals plugged into the MGGA functionals. We have shown
that when an appropriate GGA potential is chosen, the results are very close
to those obtained (from another code) self-consistently.
We also pointed out the importance of choosing an atomic
volume for the magnetic moment in AFM systems that is large enough,
for which we used the basin as defined in Bader's QTAIM.

\begin{acknowledgments}
P.B. acknowledges support from the Austrian Science Foundation (FWF) for
Project W1243 (Solids4Fun).

\end{acknowledgments}

\bibliography{references}

\begin{thebibliography}{147}%
\makeatletter
\providecommand \@ifxundefined [1]{%
 \@ifx{#1\undefined}
}%
\providecommand \@ifnum [1]{%
 \ifnum #1\expandafter \@firstoftwo
 \else \expandafter \@secondoftwo
 \fi
}%
\providecommand \@ifx [1]{%
 \ifx #1\expandafter \@firstoftwo
 \else \expandafter \@secondoftwo
 \fi
}%
\providecommand \natexlab [1]{#1}%
\providecommand \enquote  [1]{``#1''}%
\providecommand \bibnamefont  [1]{#1}%
\providecommand \bibfnamefont [1]{#1}%
\providecommand \citenamefont [1]{#1}%
\providecommand \href@noop [0]{\@secondoftwo}%
\providecommand \href [0]{\begingroup \@sanitize@url \@href}%
\providecommand \@href[1]{\@@startlink{#1}\@@href}%
\providecommand \@@href[1]{\endgroup#1\@@endlink}%
\providecommand \@sanitize@url [0]{\catcode `\\12\catcode `\$12\catcode
  `\&12\catcode `\#12\catcode `\^12\catcode `\_12\catcode `\%12\relax}%
\providecommand \@@startlink[1]{}%
\providecommand \@@endlink[0]{}%
\providecommand \url  [0]{\begingroup\@sanitize@url \@url }%
\providecommand \@url [1]{\endgroup\@href {#1}{\urlprefix }}%
\providecommand \urlprefix  [0]{URL }%
\providecommand \Eprint [0]{\href }%
\providecommand \doibase [0]{http://dx.doi.org/}%
\providecommand \selectlanguage [0]{\@gobble}%
\providecommand \bibinfo  [0]{\@secondoftwo}%
\providecommand \bibfield  [0]{\@secondoftwo}%
\providecommand \translation [1]{[#1]}%
\providecommand \BibitemOpen [0]{}%
\providecommand \bibitemStop [0]{}%
\providecommand \bibitemNoStop [0]{.\EOS\space}%
\providecommand \EOS [0]{\spacefactor3000\relax}%
\providecommand \BibitemShut  [1]{\csname bibitem#1\endcsname}%
\let\auto@bib@innerbib\@empty
\bibitem [{\citenamefont {Kohn}\ and\ \citenamefont {Sham}(1965)}]{KohnPR65}%
  \BibitemOpen
  \bibfield  {author} {\bibinfo {author} {\bibfnamefont {W.}~\bibnamefont
  {Kohn}}\ and\ \bibinfo {author} {\bibfnamefont {L.~J.}\ \bibnamefont
  {Sham}},\ }\href@noop {} {\bibfield  {journal} {\bibinfo  {journal} {Phys.
  Rev.}\ }\textbf {\bibinfo {volume} {140}},\ \bibinfo {pages} {A1133}
  (\bibinfo {year} {1965})}\BibitemShut {NoStop}%
\bibitem [{\citenamefont {Becke}(1988)}]{BeckePRA88}%
  \BibitemOpen
  \bibfield  {author} {\bibinfo {author} {\bibfnamefont {A.~D.}\ \bibnamefont
  {Becke}},\ }\href@noop {} {\bibfield  {journal} {\bibinfo  {journal} {Phys.
  Rev. A}\ }\textbf {\bibinfo {volume} {38}},\ \bibinfo {pages} {3098}
  (\bibinfo {year} {1988})}\BibitemShut {NoStop}%
\bibitem [{\citenamefont {Perdew}\ \emph {et~al.}(1992)\citenamefont {Perdew},
  \citenamefont {Chevary}, \citenamefont {Vosko}, \citenamefont {Jackson},
  \citenamefont {Pederson}, \citenamefont {Singh},\ and\ \citenamefont
  {Fiolhais}}]{PerdewPRB92b}%
  \BibitemOpen
  \bibfield  {author} {\bibinfo {author} {\bibfnamefont {J.~P.}\ \bibnamefont
  {Perdew}}, \bibinfo {author} {\bibfnamefont {J.~A.}\ \bibnamefont {Chevary}},
  \bibinfo {author} {\bibfnamefont {S.~H.}\ \bibnamefont {Vosko}}, \bibinfo
  {author} {\bibfnamefont {K.~A.}\ \bibnamefont {Jackson}}, \bibinfo {author}
  {\bibfnamefont {M.~R.}\ \bibnamefont {Pederson}}, \bibinfo {author}
  {\bibfnamefont {D.~J.}\ \bibnamefont {Singh}}, \ and\ \bibinfo {author}
  {\bibfnamefont {C.}~\bibnamefont {Fiolhais}},\ }\href@noop {} {\bibfield
  {journal} {\bibinfo  {journal} {Phys. Rev. B}\ }\textbf {\bibinfo {volume}
  {46}},\ \bibinfo {pages} {6671} (\bibinfo {year} {1992})},\ \bibinfo {note}
  {\textbf{48}, 4978(E) (1993)}\BibitemShut {NoStop}%
\bibitem [{\citenamefont {Hohenberg}\ and\ \citenamefont
  {Kohn}(1964)}]{HohenbergPR64}%
  \BibitemOpen
  \bibfield  {author} {\bibinfo {author} {\bibfnamefont {P.}~\bibnamefont
  {Hohenberg}}\ and\ \bibinfo {author} {\bibfnamefont {W.}~\bibnamefont
  {Kohn}},\ }\href@noop {} {\bibfield  {journal} {\bibinfo  {journal} {Phys.
  Rev.}\ }\textbf {\bibinfo {volume} {136}},\ \bibinfo {pages} {B864} (\bibinfo
  {year} {1964})}\BibitemShut {NoStop}%
\bibitem [{\citenamefont {Barbiellini}\ \emph {et~al.}(1990)\citenamefont
  {Barbiellini}, \citenamefont {Moroni},\ and\ \citenamefont
  {Jarlborg}}]{BarbielliniJPCM90}%
  \BibitemOpen
  \bibfield  {author} {\bibinfo {author} {\bibfnamefont {B.}~\bibnamefont
  {Barbiellini}}, \bibinfo {author} {\bibfnamefont {E.~G.}\ \bibnamefont
  {Moroni}}, \ and\ \bibinfo {author} {\bibfnamefont {T.}~\bibnamefont
  {Jarlborg}},\ }\href@noop {} {\bibfield  {journal} {\bibinfo  {journal} {J.
  Phys.: Condens. Matter}\ }\textbf {\bibinfo {volume} {2}},\ \bibinfo {pages}
  {7597} (\bibinfo {year} {1990})}\BibitemShut {NoStop}%
\bibitem [{\citenamefont {Singh}\ \emph {et~al.}(1991)\citenamefont {Singh},
  \citenamefont {Pickett},\ and\ \citenamefont {Krakauer}}]{SinghPRB91b}%
  \BibitemOpen
  \bibfield  {author} {\bibinfo {author} {\bibfnamefont {D.~J.}\ \bibnamefont
  {Singh}}, \bibinfo {author} {\bibfnamefont {W.~E.}\ \bibnamefont {Pickett}},
  \ and\ \bibinfo {author} {\bibfnamefont {H.}~\bibnamefont {Krakauer}},\
  }\href@noop {} {\bibfield  {journal} {\bibinfo  {journal} {Phys. Rev. B}\
  }\textbf {\bibinfo {volume} {43}},\ \bibinfo {pages} {11628} (\bibinfo {year}
  {1991})}\BibitemShut {NoStop}%
\bibitem [{\citenamefont {Sharma}\ \emph {et~al.}(2018)\citenamefont {Sharma},
  \citenamefont {Gross}, \citenamefont {Sanna},\ and\ \citenamefont
  {Dewhurst}}]{SharmaJCTC18}%
  \BibitemOpen
  \bibfield  {author} {\bibinfo {author} {\bibfnamefont {S.}~\bibnamefont
  {Sharma}}, \bibinfo {author} {\bibfnamefont {E.~K.~U.}\ \bibnamefont
  {Gross}}, \bibinfo {author} {\bibfnamefont {A.}~\bibnamefont {Sanna}}, \ and\
  \bibinfo {author} {\bibfnamefont {J.~K.}\ \bibnamefont {Dewhurst}},\
  }\href@noop {} {\bibfield  {journal} {\bibinfo  {journal} {J. Chem. Theory
  Comput.}\ }\textbf {\bibinfo {volume} {14}},\ \bibinfo {pages} {1247}
  (\bibinfo {year} {2018})}\BibitemShut {NoStop}%
\bibitem [{\citenamefont {Perdew}\ and\ \citenamefont
  {Zunger}(1981)}]{PerdewPRB81}%
  \BibitemOpen
  \bibfield  {author} {\bibinfo {author} {\bibfnamefont {J.~P.}\ \bibnamefont
  {Perdew}}\ and\ \bibinfo {author} {\bibfnamefont {A.}~\bibnamefont
  {Zunger}},\ }\href@noop {} {\bibfield  {journal} {\bibinfo  {journal} {Phys.
  Rev. B}\ }\textbf {\bibinfo {volume} {23}},\ \bibinfo {pages} {5048}
  (\bibinfo {year} {1981})}\BibitemShut {NoStop}%
\bibitem [{\citenamefont {Terakura}\ \emph {et~al.}(1984)\citenamefont
  {Terakura}, \citenamefont {Oguchi}, \citenamefont {Williams},\ and\
  \citenamefont {K\"{u}bler}}]{TerakuraPRB84}%
  \BibitemOpen
  \bibfield  {author} {\bibinfo {author} {\bibfnamefont {K.}~\bibnamefont
  {Terakura}}, \bibinfo {author} {\bibfnamefont {T.}~\bibnamefont {Oguchi}},
  \bibinfo {author} {\bibfnamefont {A.~R.}\ \bibnamefont {Williams}}, \ and\
  \bibinfo {author} {\bibfnamefont {J.}~\bibnamefont {K\"{u}bler}},\
  }\href@noop {} {\bibfield  {journal} {\bibinfo  {journal} {Phys. Rev. B}\
  }\textbf {\bibinfo {volume} {30}},\ \bibinfo {pages} {4734} (\bibinfo {year}
  {1984})}\BibitemShut {NoStop}%
\bibitem [{\citenamefont {Van~Voorhis}\ and\ \citenamefont
  {Scuseria}(1998)}]{VanVoorhisJCP98}%
  \BibitemOpen
  \bibfield  {author} {\bibinfo {author} {\bibfnamefont {T.}~\bibnamefont
  {Van~Voorhis}}\ and\ \bibinfo {author} {\bibfnamefont {G.~E.}\ \bibnamefont
  {Scuseria}},\ }\href@noop {} {\bibfield  {journal} {\bibinfo  {journal} {J.
  Chem. Phys.}\ }\textbf {\bibinfo {volume} {109}},\ \bibinfo {pages} {400}
  (\bibinfo {year} {1998})},\ \bibinfo {note} {\textbf{129}, 219901
  (2008)}\BibitemShut {NoStop}%
\bibitem [{\citenamefont {Staroverov}\ \emph {et~al.}(2003)\citenamefont
  {Staroverov}, \citenamefont {Scuseria}, \citenamefont {Tao},\ and\
  \citenamefont {Perdew}}]{StaroverovJCP03}%
  \BibitemOpen
  \bibfield  {author} {\bibinfo {author} {\bibfnamefont {V.~N.}\ \bibnamefont
  {Staroverov}}, \bibinfo {author} {\bibfnamefont {G.~E.}\ \bibnamefont
  {Scuseria}}, \bibinfo {author} {\bibfnamefont {J.}~\bibnamefont {Tao}}, \
  and\ \bibinfo {author} {\bibfnamefont {J.~P.}\ \bibnamefont {Perdew}},\
  }\href@noop {} {\bibfield  {journal} {\bibinfo  {journal} {J. Chem. Phys.}\
  }\textbf {\bibinfo {volume} {119}},\ \bibinfo {pages} {12129} (\bibinfo
  {year} {2003})},\ \bibinfo {note} {\textbf{121}, 11507 (2004)}\BibitemShut
  {NoStop}%
\bibitem [{\citenamefont {Staroverov}\ \emph {et~al.}(2004)\citenamefont
  {Staroverov}, \citenamefont {Scuseria}, \citenamefont {Tao},\ and\
  \citenamefont {Perdew}}]{StaroverovPRB04}%
  \BibitemOpen
  \bibfield  {author} {\bibinfo {author} {\bibfnamefont {V.~N.}\ \bibnamefont
  {Staroverov}}, \bibinfo {author} {\bibfnamefont {G.~E.}\ \bibnamefont
  {Scuseria}}, \bibinfo {author} {\bibfnamefont {J.}~\bibnamefont {Tao}}, \
  and\ \bibinfo {author} {\bibfnamefont {J.~P.}\ \bibnamefont {Perdew}},\
  }\href@noop {} {\bibfield  {journal} {\bibinfo  {journal} {Phys. Rev. B}\
  }\textbf {\bibinfo {volume} {69}},\ \bibinfo {pages} {075102} (\bibinfo
  {year} {2004})},\ \bibinfo {note} {\textbf{78}, 239907(E) (2008)}\BibitemShut
  {NoStop}%
\bibitem [{\citenamefont {Sun}\ \emph {et~al.}(2015{\natexlab{a}})\citenamefont
  {Sun}, \citenamefont {Ruzsinszky},\ and\ \citenamefont {Perdew}}]{SunPRL15}%
  \BibitemOpen
  \bibfield  {author} {\bibinfo {author} {\bibfnamefont {J.}~\bibnamefont
  {Sun}}, \bibinfo {author} {\bibfnamefont {A.}~\bibnamefont {Ruzsinszky}}, \
  and\ \bibinfo {author} {\bibfnamefont {J.~P.}\ \bibnamefont {Perdew}},\
  }\href@noop {} {\bibfield  {journal} {\bibinfo  {journal} {Phys. Rev. Lett.}\
  }\textbf {\bibinfo {volume} {115}},\ \bibinfo {pages} {036402} (\bibinfo
  {year} {2015}{\natexlab{a}})}\BibitemShut {NoStop}%
\bibitem [{\citenamefont {Tran}\ \emph {et~al.}(2016)\citenamefont {Tran},
  \citenamefont {Stelzl},\ and\ \citenamefont {Blaha}}]{TranJCP16}%
  \BibitemOpen
  \bibfield  {author} {\bibinfo {author} {\bibfnamefont {F.}~\bibnamefont
  {Tran}}, \bibinfo {author} {\bibfnamefont {J.}~\bibnamefont {Stelzl}}, \ and\
  \bibinfo {author} {\bibfnamefont {P.}~\bibnamefont {Blaha}},\ }\href@noop {}
  {\bibfield  {journal} {\bibinfo  {journal} {J. Chem. Phys.}\ }\textbf
  {\bibinfo {volume} {144}},\ \bibinfo {pages} {204120} (\bibinfo {year}
  {2016})}\BibitemShut {NoStop}%
\bibitem [{\citenamefont {Zhang}\ \emph {et~al.}(2018)\citenamefont {Zhang},
  \citenamefont {Kitchaev}, \citenamefont {Yang}, \citenamefont {Chen},
  \citenamefont {Dacek}, \citenamefont {Sarmiento-P\'{e}rez}, \citenamefont
  {Marques}, \citenamefont {Peng}, \citenamefont {Ceder}, \citenamefont
  {Perdew},\ and\ \citenamefont {Sun}}]{ZhangNPJCM18}%
  \BibitemOpen
  \bibfield  {author} {\bibinfo {author} {\bibfnamefont {Y.}~\bibnamefont
  {Zhang}}, \bibinfo {author} {\bibfnamefont {D.~A.}\ \bibnamefont {Kitchaev}},
  \bibinfo {author} {\bibfnamefont {J.}~\bibnamefont {Yang}}, \bibinfo {author}
  {\bibfnamefont {T.}~\bibnamefont {Chen}}, \bibinfo {author} {\bibfnamefont
  {S.~T.}\ \bibnamefont {Dacek}}, \bibinfo {author} {\bibfnamefont {R.~A.}\
  \bibnamefont {Sarmiento-P\'{e}rez}}, \bibinfo {author} {\bibfnamefont
  {M.~A.~L.}\ \bibnamefont {Marques}}, \bibinfo {author} {\bibfnamefont
  {H.}~\bibnamefont {Peng}}, \bibinfo {author} {\bibfnamefont {G.}~\bibnamefont
  {Ceder}}, \bibinfo {author} {\bibfnamefont {J.~P.}\ \bibnamefont {Perdew}}, \
  and\ \bibinfo {author} {\bibfnamefont {J.}~\bibnamefont {Sun}},\ }\href@noop
  {} {\bibfield  {journal} {\bibinfo  {journal} {npj Comput. Mater.}\ }\textbf
  {\bibinfo {volume} {4}},\ \bibinfo {pages} {9} (\bibinfo {year}
  {2018})}\BibitemShut {NoStop}%
\bibitem [{\citenamefont {Isaacs}\ and\ \citenamefont
  {Wolverton}(2018)}]{IsaacsPRM18}%
  \BibitemOpen
  \bibfield  {author} {\bibinfo {author} {\bibfnamefont {E.~B.}\ \bibnamefont
  {Isaacs}}\ and\ \bibinfo {author} {\bibfnamefont {C.}~\bibnamefont
  {Wolverton}},\ }\href@noop {} {\bibfield  {journal} {\bibinfo  {journal}
  {Phys. Rev. Materials}\ }\textbf {\bibinfo {volume} {2}},\ \bibinfo {pages}
  {063801} (\bibinfo {year} {2018})}\BibitemShut {NoStop}%
\bibitem [{\citenamefont {Peng}\ \emph {et~al.}(2016)\citenamefont {Peng},
  \citenamefont {Yang}, \citenamefont {Perdew},\ and\ \citenamefont
  {Sun}}]{PengPRX16}%
  \BibitemOpen
  \bibfield  {author} {\bibinfo {author} {\bibfnamefont {H.}~\bibnamefont
  {Peng}}, \bibinfo {author} {\bibfnamefont {Z.-H.}\ \bibnamefont {Yang}},
  \bibinfo {author} {\bibfnamefont {J.~P.}\ \bibnamefont {Perdew}}, \ and\
  \bibinfo {author} {\bibfnamefont {J.}~\bibnamefont {Sun}},\ }\href@noop {}
  {\bibfield  {journal} {\bibinfo  {journal} {Phys. Rev. X}\ }\textbf {\bibinfo
  {volume} {6}},\ \bibinfo {pages} {041005} (\bibinfo {year}
  {2016})}\BibitemShut {NoStop}%
\bibitem [{\citenamefont {Brandenburg}\ \emph {et~al.}(2016)\citenamefont
  {Brandenburg}, \citenamefont {Bates}, \citenamefont {Sun},\ and\
  \citenamefont {Perdew}}]{BrandenburgPRB16}%
  \BibitemOpen
  \bibfield  {author} {\bibinfo {author} {\bibfnamefont {J.~G.}\ \bibnamefont
  {Brandenburg}}, \bibinfo {author} {\bibfnamefont {J.~E.}\ \bibnamefont
  {Bates}}, \bibinfo {author} {\bibfnamefont {J.}~\bibnamefont {Sun}}, \ and\
  \bibinfo {author} {\bibfnamefont {J.~P.}\ \bibnamefont {Perdew}},\
  }\href@noop {} {\bibfield  {journal} {\bibinfo  {journal} {Phys. Rev. B}\
  }\textbf {\bibinfo {volume} {94}},\ \bibinfo {pages} {115144} (\bibinfo
  {year} {2016})}\BibitemShut {NoStop}%
\bibitem [{\citenamefont {Tran}\ \emph
  {et~al.}(2019{\natexlab{a}})\citenamefont {Tran}, \citenamefont {Kalantari},
  \citenamefont {Traor\'e}, \citenamefont {Rocquefelte},\ and\ \citenamefont
  {Blaha}}]{TranPRM19}%
  \BibitemOpen
  \bibfield  {author} {\bibinfo {author} {\bibfnamefont {F.}~\bibnamefont
  {Tran}}, \bibinfo {author} {\bibfnamefont {L.}~\bibnamefont {Kalantari}},
  \bibinfo {author} {\bibfnamefont {B.}~\bibnamefont {Traor\'e}}, \bibinfo
  {author} {\bibfnamefont {X.}~\bibnamefont {Rocquefelte}}, \ and\ \bibinfo
  {author} {\bibfnamefont {P.}~\bibnamefont {Blaha}},\ }\href@noop {}
  {\bibfield  {journal} {\bibinfo  {journal} {Phys. Rev. Materials}\ }\textbf
  {\bibinfo {volume} {3}},\ \bibinfo {pages} {063602} (\bibinfo {year}
  {2019}{\natexlab{a}})}\BibitemShut {NoStop}%
\bibitem [{\citenamefont {Jana}\ \emph {et~al.}(2018)\citenamefont {Jana},
  \citenamefont {Patra},\ and\ \citenamefont {Samal}}]{JanaJCP18a}%
  \BibitemOpen
  \bibfield  {author} {\bibinfo {author} {\bibfnamefont {S.}~\bibnamefont
  {Jana}}, \bibinfo {author} {\bibfnamefont {A.}~\bibnamefont {Patra}}, \ and\
  \bibinfo {author} {\bibfnamefont {P.}~\bibnamefont {Samal}},\ }\href@noop {}
  {\bibfield  {journal} {\bibinfo  {journal} {J. Chem. Phys.}\ }\textbf
  {\bibinfo {volume} {149}},\ \bibinfo {pages} {044120} (\bibinfo {year}
  {2018})}\BibitemShut {NoStop}%
\bibitem [{\citenamefont {Ekholm}\ \emph {et~al.}(2018)\citenamefont {Ekholm},
  \citenamefont {Gambino}, \citenamefont {J\"onsson}, \citenamefont
  {Tasn\'adi}, \citenamefont {Alling},\ and\ \citenamefont
  {Abrikosov}}]{EkholmPRB18}%
  \BibitemOpen
  \bibfield  {author} {\bibinfo {author} {\bibfnamefont {M.}~\bibnamefont
  {Ekholm}}, \bibinfo {author} {\bibfnamefont {D.}~\bibnamefont {Gambino}},
  \bibinfo {author} {\bibfnamefont {H.~J.~M.}\ \bibnamefont {J\"onsson}},
  \bibinfo {author} {\bibfnamefont {F.}~\bibnamefont {Tasn\'adi}}, \bibinfo
  {author} {\bibfnamefont {B.}~\bibnamefont {Alling}}, \ and\ \bibinfo {author}
  {\bibfnamefont {I.~A.}\ \bibnamefont {Abrikosov}},\ }\href@noop {} {\bibfield
   {journal} {\bibinfo  {journal} {Phys. Rev. B}\ }\textbf {\bibinfo {volume}
  {98}},\ \bibinfo {pages} {094413} (\bibinfo {year} {2018})}\BibitemShut
  {NoStop}%
\bibitem [{\citenamefont {Fu}\ and\ \citenamefont {Singh}(2018)}]{FuPRL18}%
  \BibitemOpen
  \bibfield  {author} {\bibinfo {author} {\bibfnamefont {Y.}~\bibnamefont
  {Fu}}\ and\ \bibinfo {author} {\bibfnamefont {D.~J.}\ \bibnamefont {Singh}},\
  }\href@noop {} {\bibfield  {journal} {\bibinfo  {journal} {Phys. Rev. Lett.}\
  }\textbf {\bibinfo {volume} {121}},\ \bibinfo {pages} {207201} (\bibinfo
  {year} {2018})}\BibitemShut {NoStop}%
\bibitem [{\citenamefont {Fu}\ and\ \citenamefont {Singh}(2019)}]{FuPRB19}%
  \BibitemOpen
  \bibfield  {author} {\bibinfo {author} {\bibfnamefont {Y.}~\bibnamefont
  {Fu}}\ and\ \bibinfo {author} {\bibfnamefont {D.~J.}\ \bibnamefont {Singh}},\
  }\href@noop {} {\bibfield  {journal} {\bibinfo  {journal} {Phys. Rev. B}\
  }\textbf {\bibinfo {volume} {100}},\ \bibinfo {pages} {045126} (\bibinfo
  {year} {2019})}\BibitemShut {NoStop}%
\bibitem [{\citenamefont {Mej\'{\i}a-Rodr\'{\i}guez}\ and\ \citenamefont
  {Trickey}(2019)}]{MejiaRodriguezPRB19}%
  \BibitemOpen
  \bibfield  {author} {\bibinfo {author} {\bibfnamefont {D.}~\bibnamefont
  {Mej\'{\i}a-Rodr\'{\i}guez}}\ and\ \bibinfo {author} {\bibfnamefont {S.~B.}\
  \bibnamefont {Trickey}},\ }\href@noop {} {\bibfield  {journal} {\bibinfo
  {journal} {Phys. Rev. B}\ }\textbf {\bibinfo {volume} {100}},\ \bibinfo
  {pages} {041113(R)} (\bibinfo {year} {2019})}\BibitemShut {NoStop}%
\bibitem [{\citenamefont {Romero}\ and\ \citenamefont
  {Verstraete}(2018)}]{RomeroEPJB18}%
  \BibitemOpen
  \bibfield  {author} {\bibinfo {author} {\bibfnamefont {A.~H.}\ \bibnamefont
  {Romero}}\ and\ \bibinfo {author} {\bibfnamefont {M.~J.}\ \bibnamefont
  {Verstraete}},\ }\href@noop {} {\bibfield  {journal} {\bibinfo  {journal}
  {Eur. Phys. J. B}\ }\textbf {\bibinfo {volume} {91}},\ \bibinfo {pages} {193}
  (\bibinfo {year} {2018})}\BibitemShut {NoStop}%
\bibitem [{\citenamefont {Buchelnikov}\ \emph {et~al.}(2019)\citenamefont
  {Buchelnikov}, \citenamefont {Sokolovskiy}, \citenamefont {Miroshkina},
  \citenamefont {Zagrebin}, \citenamefont {Nokelainen}, \citenamefont
  {Pulkkinen}, \citenamefont {Barbiellini},\ and\ \citenamefont
  {L\"ahderanta}}]{BuchelnikovPRB19}%
  \BibitemOpen
  \bibfield  {author} {\bibinfo {author} {\bibfnamefont {V.~D.}\ \bibnamefont
  {Buchelnikov}}, \bibinfo {author} {\bibfnamefont {V.~V.}\ \bibnamefont
  {Sokolovskiy}}, \bibinfo {author} {\bibfnamefont {O.~N.}\ \bibnamefont
  {Miroshkina}}, \bibinfo {author} {\bibfnamefont {M.~A.}\ \bibnamefont
  {Zagrebin}}, \bibinfo {author} {\bibfnamefont {J.}~\bibnamefont
  {Nokelainen}}, \bibinfo {author} {\bibfnamefont {A.}~\bibnamefont
  {Pulkkinen}}, \bibinfo {author} {\bibfnamefont {B.}~\bibnamefont
  {Barbiellini}}, \ and\ \bibinfo {author} {\bibfnamefont {E.}~\bibnamefont
  {L\"ahderanta}},\ }\href@noop {} {\bibfield  {journal} {\bibinfo  {journal}
  {Phys. Rev. B}\ }\textbf {\bibinfo {volume} {99}},\ \bibinfo {pages} {014426}
  (\bibinfo {year} {2019})}\BibitemShut {NoStop}%
\bibitem [{\citenamefont {Shepard}\ and\ \citenamefont
  {Smeu}(2019)}]{ShepardJCP19}%
  \BibitemOpen
  \bibfield  {author} {\bibinfo {author} {\bibfnamefont {S.}~\bibnamefont
  {Shepard}}\ and\ \bibinfo {author} {\bibfnamefont {M.}~\bibnamefont {Smeu}},\
  }\href@noop {} {\bibfield  {journal} {\bibinfo  {journal} {J. Chem. Phys.}\
  }\textbf {\bibinfo {volume} {150}},\ \bibinfo {pages} {154702} (\bibinfo
  {year} {2019})}\BibitemShut {NoStop}%
\bibitem [{\citenamefont {Tao}\ \emph {et~al.}(2003)\citenamefont {Tao},
  \citenamefont {Perdew}, \citenamefont {Staroverov},\ and\ \citenamefont
  {Scuseria}}]{TaoPRL03}%
  \BibitemOpen
  \bibfield  {author} {\bibinfo {author} {\bibfnamefont {J.}~\bibnamefont
  {Tao}}, \bibinfo {author} {\bibfnamefont {J.~P.}\ \bibnamefont {Perdew}},
  \bibinfo {author} {\bibfnamefont {V.~N.}\ \bibnamefont {Staroverov}}, \ and\
  \bibinfo {author} {\bibfnamefont {G.~E.}\ \bibnamefont {Scuseria}},\
  }\href@noop {} {\bibfield  {journal} {\bibinfo  {journal} {Phys. Rev. Lett.}\
  }\textbf {\bibinfo {volume} {91}},\ \bibinfo {pages} {146401} (\bibinfo
  {year} {2003})}\BibitemShut {NoStop}%
\bibitem [{\citenamefont {Perdew}\ \emph {et~al.}(2009)\citenamefont {Perdew},
  \citenamefont {Ruzsinszky}, \citenamefont {Csonka}, \citenamefont
  {Constantin},\ and\ \citenamefont {Sun}}]{PerdewPRL09}%
  \BibitemOpen
  \bibfield  {author} {\bibinfo {author} {\bibfnamefont {J.~P.}\ \bibnamefont
  {Perdew}}, \bibinfo {author} {\bibfnamefont {A.}~\bibnamefont {Ruzsinszky}},
  \bibinfo {author} {\bibfnamefont {G.~I.}\ \bibnamefont {Csonka}}, \bibinfo
  {author} {\bibfnamefont {L.~A.}\ \bibnamefont {Constantin}}, \ and\ \bibinfo
  {author} {\bibfnamefont {J.}~\bibnamefont {Sun}},\ }\href@noop {} {\bibfield
  {journal} {\bibinfo  {journal} {Phys. Rev. Lett.}\ }\textbf {\bibinfo
  {volume} {103}},\ \bibinfo {pages} {026403} (\bibinfo {year} {2009})},\
  \bibinfo {note} {\textbf{106}, 179902 (2011)}\BibitemShut {NoStop}%
\bibitem [{\citenamefont {Tao}\ and\ \citenamefont {Mo}(2016)}]{TaoPRL16}%
  \BibitemOpen
  \bibfield  {author} {\bibinfo {author} {\bibfnamefont {J.}~\bibnamefont
  {Tao}}\ and\ \bibinfo {author} {\bibfnamefont {Y.}~\bibnamefont {Mo}},\
  }\href@noop {} {\bibfield  {journal} {\bibinfo  {journal} {Phys. Rev. Lett.}\
  }\textbf {\bibinfo {volume} {117}},\ \bibinfo {pages} {073001} (\bibinfo
  {year} {2016})}\BibitemShut {NoStop}%
\bibitem [{\citenamefont {Sun}\ \emph {et~al.}(2011)\citenamefont {Sun},
  \citenamefont {Marsman}, \citenamefont {Csonka}, \citenamefont {Ruzsinszky},
  \citenamefont {Hao}, \citenamefont {Kim}, \citenamefont {Kresse},\ and\
  \citenamefont {Perdew}}]{SunPRB11b}%
  \BibitemOpen
  \bibfield  {author} {\bibinfo {author} {\bibfnamefont {J.}~\bibnamefont
  {Sun}}, \bibinfo {author} {\bibfnamefont {M.}~\bibnamefont {Marsman}},
  \bibinfo {author} {\bibfnamefont {G.~I.}\ \bibnamefont {Csonka}}, \bibinfo
  {author} {\bibfnamefont {A.}~\bibnamefont {Ruzsinszky}}, \bibinfo {author}
  {\bibfnamefont {P.}~\bibnamefont {Hao}}, \bibinfo {author} {\bibfnamefont
  {Y.-S.}\ \bibnamefont {Kim}}, \bibinfo {author} {\bibfnamefont
  {G.}~\bibnamefont {Kresse}}, \ and\ \bibinfo {author} {\bibfnamefont {J.~P.}\
  \bibnamefont {Perdew}},\ }\href@noop {} {\bibfield  {journal} {\bibinfo
  {journal} {Phys. Rev. B}\ }\textbf {\bibinfo {volume} {84}},\ \bibinfo
  {pages} {035117} (\bibinfo {year} {2011})}\BibitemShut {NoStop}%
\bibitem [{\citenamefont {Mejia-Rodriguez}\ and\ \citenamefont
  {Trickey}(2017)}]{MejiaRodriguezPRA17}%
  \BibitemOpen
  \bibfield  {author} {\bibinfo {author} {\bibfnamefont {D.}~\bibnamefont
  {Mejia-Rodriguez}}\ and\ \bibinfo {author} {\bibfnamefont {S.~B.}\
  \bibnamefont {Trickey}},\ }\href@noop {} {\bibfield  {journal} {\bibinfo
  {journal} {Phys. Rev. A}\ }\textbf {\bibinfo {volume} {96}},\ \bibinfo
  {pages} {052512} (\bibinfo {year} {2017})}\BibitemShut {NoStop}%
\bibitem [{\citenamefont {Mejia-Rodriguez}\ and\ \citenamefont
  {Trickey}(2018)}]{MejiaRodriguezPRB18}%
  \BibitemOpen
  \bibfield  {author} {\bibinfo {author} {\bibfnamefont {D.}~\bibnamefont
  {Mejia-Rodriguez}}\ and\ \bibinfo {author} {\bibfnamefont {S.~B.}\
  \bibnamefont {Trickey}},\ }\href@noop {} {\bibfield  {journal} {\bibinfo
  {journal} {Phys. Rev. B}\ }\textbf {\bibinfo {volume} {98}},\ \bibinfo
  {pages} {115161} (\bibinfo {year} {2018})}\BibitemShut {NoStop}%
\bibitem [{\citenamefont {Tran}\ \emph
  {et~al.}(2018{\natexlab{a}})\citenamefont {Tran}, \citenamefont {Kov\'{a}cs},
  \citenamefont {Kalantari}, \citenamefont {Madsen},\ and\ \citenamefont
  {Blaha}}]{TranJCP18}%
  \BibitemOpen
  \bibfield  {author} {\bibinfo {author} {\bibfnamefont {F.}~\bibnamefont
  {Tran}}, \bibinfo {author} {\bibfnamefont {P.}~\bibnamefont {Kov\'{a}cs}},
  \bibinfo {author} {\bibfnamefont {L.}~\bibnamefont {Kalantari}}, \bibinfo
  {author} {\bibfnamefont {G.~K.~H.}\ \bibnamefont {Madsen}}, \ and\ \bibinfo
  {author} {\bibfnamefont {P.}~\bibnamefont {Blaha}},\ }\href@noop {}
  {\bibfield  {journal} {\bibinfo  {journal} {J. Chem. Phys.}\ }\textbf
  {\bibinfo {volume} {149}},\ \bibinfo {pages} {144105} (\bibinfo {year}
  {2018}{\natexlab{a}})}\BibitemShut {NoStop}%
\bibitem [{\citenamefont {Sai~Gautam}\ and\ \citenamefont
  {Carter}(2018)}]{SaiGautamPRM18}%
  \BibitemOpen
  \bibfield  {author} {\bibinfo {author} {\bibfnamefont {G.}~\bibnamefont
  {Sai~Gautam}}\ and\ \bibinfo {author} {\bibfnamefont {E.~A.}\ \bibnamefont
  {Carter}},\ }\href@noop {} {\bibfield  {journal} {\bibinfo  {journal} {Phys.
  Rev. Materials}\ }\textbf {\bibinfo {volume} {2}},\ \bibinfo {pages} {095401}
  (\bibinfo {year} {2018})}\BibitemShut {NoStop}%
\bibitem [{\citenamefont {Long}\ \emph {et~al.}(2020)\citenamefont {Long},
  \citenamefont {Sai~Gautam},\ and\ \citenamefont {Carter}}]{LongPRM20}%
  \BibitemOpen
  \bibfield  {author} {\bibinfo {author} {\bibfnamefont {O.~Y.}\ \bibnamefont
  {Long}}, \bibinfo {author} {\bibfnamefont {G.}~\bibnamefont {Sai~Gautam}}, \
  and\ \bibinfo {author} {\bibfnamefont {E.~A.}\ \bibnamefont {Carter}},\
  }\href@noop {} {\bibfield  {journal} {\bibinfo  {journal} {Phys. Rev.
  Materials}\ }\textbf {\bibinfo {volume} {4}},\ \bibinfo {pages} {045401}
  (\bibinfo {year} {2020})}\BibitemShut {NoStop}%
\bibitem [{\citenamefont {Xiao}\ \emph {et~al.}(2014)\citenamefont {Xiao},
  \citenamefont {Sun}, \citenamefont {Ruzsinszky},\ and\ \citenamefont
  {Perdew}}]{XiaoPRB14}%
  \BibitemOpen
  \bibfield  {author} {\bibinfo {author} {\bibfnamefont {B.}~\bibnamefont
  {Xiao}}, \bibinfo {author} {\bibfnamefont {J.}~\bibnamefont {Sun}}, \bibinfo
  {author} {\bibfnamefont {A.}~\bibnamefont {Ruzsinszky}}, \ and\ \bibinfo
  {author} {\bibfnamefont {J.~P.}\ \bibnamefont {Perdew}},\ }\href@noop {}
  {\bibfield  {journal} {\bibinfo  {journal} {Phys. Rev. B}\ }\textbf {\bibinfo
  {volume} {90}},\ \bibinfo {pages} {085134} (\bibinfo {year}
  {2014})}\BibitemShut {NoStop}%
\bibitem [{\citenamefont {Kyl\"anp\"a\"a}\ \emph {et~al.}(2017)\citenamefont
  {Kyl\"anp\"a\"a}, \citenamefont {Balachandran}, \citenamefont {Ganesh},
  \citenamefont {Heinonen}, \citenamefont {Kent},\ and\ \citenamefont
  {Krogel}}]{KylanpaaPRM17}%
  \BibitemOpen
  \bibfield  {author} {\bibinfo {author} {\bibfnamefont {I.}~\bibnamefont
  {Kyl\"anp\"a\"a}}, \bibinfo {author} {\bibfnamefont {J.}~\bibnamefont
  {Balachandran}}, \bibinfo {author} {\bibfnamefont {P.}~\bibnamefont
  {Ganesh}}, \bibinfo {author} {\bibfnamefont {O.}~\bibnamefont {Heinonen}},
  \bibinfo {author} {\bibfnamefont {P.~R.~C.}\ \bibnamefont {Kent}}, \ and\
  \bibinfo {author} {\bibfnamefont {J.~T.}\ \bibnamefont {Krogel}},\
  }\href@noop {} {\bibfield  {journal} {\bibinfo  {journal} {Phys. Rev.
  Materials}\ }\textbf {\bibinfo {volume} {1}},\ \bibinfo {pages} {065408}
  (\bibinfo {year} {2017})}\BibitemShut {NoStop}%
\bibitem [{\citenamefont {Sun}\ \emph {et~al.}(2012)\citenamefont {Sun},
  \citenamefont {Xiao},\ and\ \citenamefont {Ruzsinszky}}]{SunJCP12}%
  \BibitemOpen
  \bibfield  {author} {\bibinfo {author} {\bibfnamefont {J.}~\bibnamefont
  {Sun}}, \bibinfo {author} {\bibfnamefont {B.}~\bibnamefont {Xiao}}, \ and\
  \bibinfo {author} {\bibfnamefont {A.}~\bibnamefont {Ruzsinszky}},\
  }\href@noop {} {\bibfield  {journal} {\bibinfo  {journal} {J. Chem. Phys.}\
  }\textbf {\bibinfo {volume} {137}},\ \bibinfo {pages} {051101} (\bibinfo
  {year} {2012})}\BibitemShut {NoStop}%
\bibitem [{\citenamefont {Sun}\ \emph {et~al.}(2013)\citenamefont {Sun},
  \citenamefont {Haunschild}, \citenamefont {Xiao}, \citenamefont {Bulik},
  \citenamefont {Scuseria},\ and\ \citenamefont {Perdew}}]{SunJCP13}%
  \BibitemOpen
  \bibfield  {author} {\bibinfo {author} {\bibfnamefont {J.}~\bibnamefont
  {Sun}}, \bibinfo {author} {\bibfnamefont {R.}~\bibnamefont {Haunschild}},
  \bibinfo {author} {\bibfnamefont {B.}~\bibnamefont {Xiao}}, \bibinfo {author}
  {\bibfnamefont {I.~W.}\ \bibnamefont {Bulik}}, \bibinfo {author}
  {\bibfnamefont {G.~E.}\ \bibnamefont {Scuseria}}, \ and\ \bibinfo {author}
  {\bibfnamefont {J.~P.}\ \bibnamefont {Perdew}},\ }\href@noop {} {\bibfield
  {journal} {\bibinfo  {journal} {J. Chem. Phys.}\ }\textbf {\bibinfo {volume}
  {138}},\ \bibinfo {pages} {044113} (\bibinfo {year} {2013})}\BibitemShut
  {NoStop}%
\bibitem [{\citenamefont {Lane}\ \emph {et~al.}(2018)\citenamefont {Lane},
  \citenamefont {Furness}, \citenamefont {Buda}, \citenamefont {Zhang},
  \citenamefont {Markiewicz}, \citenamefont {Barbiellini}, \citenamefont
  {Sun},\ and\ \citenamefont {Bansil}}]{LanePRB18}%
  \BibitemOpen
  \bibfield  {author} {\bibinfo {author} {\bibfnamefont {C.}~\bibnamefont
  {Lane}}, \bibinfo {author} {\bibfnamefont {J.~W.}\ \bibnamefont {Furness}},
  \bibinfo {author} {\bibfnamefont {I.~G.}\ \bibnamefont {Buda}}, \bibinfo
  {author} {\bibfnamefont {Y.}~\bibnamefont {Zhang}}, \bibinfo {author}
  {\bibfnamefont {R.~S.}\ \bibnamefont {Markiewicz}}, \bibinfo {author}
  {\bibfnamefont {B.}~\bibnamefont {Barbiellini}}, \bibinfo {author}
  {\bibfnamefont {J.}~\bibnamefont {Sun}}, \ and\ \bibinfo {author}
  {\bibfnamefont {A.}~\bibnamefont {Bansil}},\ }\href@noop {} {\bibfield
  {journal} {\bibinfo  {journal} {Phys. Rev. B}\ }\textbf {\bibinfo {volume}
  {98}},\ \bibinfo {pages} {125140} (\bibinfo {year} {2018})}\BibitemShut
  {NoStop}%
\bibitem [{\citenamefont {Zhang}\ \emph {et~al.}(2019)\citenamefont {Zhang},
  \citenamefont {Furness}, \citenamefont {Zhang}, \citenamefont {Wang},
  \citenamefont {Zunger},\ and\ \citenamefont {Sun}}]{Zhang19}%
  \BibitemOpen
  \bibfield  {author} {\bibinfo {author} {\bibfnamefont {Y.}~\bibnamefont
  {Zhang}}, \bibinfo {author} {\bibfnamefont {J.}~\bibnamefont {Furness}},
  \bibinfo {author} {\bibfnamefont {R.}~\bibnamefont {Zhang}}, \bibinfo
  {author} {\bibfnamefont {Z.}~\bibnamefont {Wang}}, \bibinfo {author}
  {\bibfnamefont {A.}~\bibnamefont {Zunger}}, \ and\ \bibinfo {author}
  {\bibfnamefont {J.}~\bibnamefont {Sun}},\ }\href@noop {} {\bibfield
  {journal} {\bibinfo  {journal} {arXiv e-prints}\ ,\ \bibinfo {pages}
  {arXiv:1906.06467}} (\bibinfo {year} {2019})}\BibitemShut {NoStop}%
\bibitem [{\citenamefont {Pulkkinen}\ \emph {et~al.}(2020)\citenamefont
  {Pulkkinen}, \citenamefont {Barbiellini}, \citenamefont {Nokelainen},
  \citenamefont {Sokolovskiy}, \citenamefont {Baigutlin}, \citenamefont
  {Miroshkina}, \citenamefont {Zagrebin}, \citenamefont {Buchelnikov},
  \citenamefont {Lane}, \citenamefont {Markiewicz}, \citenamefont {Bansil},
  \citenamefont {Sun}, \citenamefont {Pussi},\ and\ \citenamefont
  {L\"ahderanta}}]{PulkkinenPRB20}%
  \BibitemOpen
  \bibfield  {author} {\bibinfo {author} {\bibfnamefont {A.}~\bibnamefont
  {Pulkkinen}}, \bibinfo {author} {\bibfnamefont {B.}~\bibnamefont
  {Barbiellini}}, \bibinfo {author} {\bibfnamefont {J.}~\bibnamefont
  {Nokelainen}}, \bibinfo {author} {\bibfnamefont {V.}~\bibnamefont
  {Sokolovskiy}}, \bibinfo {author} {\bibfnamefont {D.}~\bibnamefont
  {Baigutlin}}, \bibinfo {author} {\bibfnamefont {O.}~\bibnamefont
  {Miroshkina}}, \bibinfo {author} {\bibfnamefont {M.}~\bibnamefont
  {Zagrebin}}, \bibinfo {author} {\bibfnamefont {V.}~\bibnamefont
  {Buchelnikov}}, \bibinfo {author} {\bibfnamefont {C.}~\bibnamefont {Lane}},
  \bibinfo {author} {\bibfnamefont {R.~S.}\ \bibnamefont {Markiewicz}},
  \bibinfo {author} {\bibfnamefont {A.}~\bibnamefont {Bansil}}, \bibinfo
  {author} {\bibfnamefont {J.}~\bibnamefont {Sun}}, \bibinfo {author}
  {\bibfnamefont {K.}~\bibnamefont {Pussi}}, \ and\ \bibinfo {author}
  {\bibfnamefont {E.}~\bibnamefont {L\"ahderanta}},\ }\href@noop {} {\bibfield
  {journal} {\bibinfo  {journal} {Phys. Rev. B}\ }\textbf {\bibinfo {volume}
  {101}},\ \bibinfo {pages} {075115} (\bibinfo {year} {2020})}\BibitemShut
  {NoStop}%
\bibitem [{\citenamefont {Tran}\ \emph {et~al.}(2012)\citenamefont {Tran},
  \citenamefont {Koller},\ and\ \citenamefont {Blaha}}]{TranPRB12}%
  \BibitemOpen
  \bibfield  {author} {\bibinfo {author} {\bibfnamefont {F.}~\bibnamefont
  {Tran}}, \bibinfo {author} {\bibfnamefont {D.}~\bibnamefont {Koller}}, \ and\
  \bibinfo {author} {\bibfnamefont {P.}~\bibnamefont {Blaha}},\ }\href@noop {}
  {\bibfield  {journal} {\bibinfo  {journal} {Phys. Rev. B}\ }\textbf {\bibinfo
  {volume} {86}},\ \bibinfo {pages} {134406} (\bibinfo {year}
  {2012})}\BibitemShut {NoStop}%
\bibitem [{\citenamefont {Della~Sala}\ \emph {et~al.}(2016)\citenamefont
  {Della~Sala}, \citenamefont {Fabiano},\ and\ \citenamefont
  {Constantin}}]{DellaSalaIJQC16}%
  \BibitemOpen
  \bibfield  {author} {\bibinfo {author} {\bibfnamefont {F.}~\bibnamefont
  {Della~Sala}}, \bibinfo {author} {\bibfnamefont {E.}~\bibnamefont {Fabiano}},
  \ and\ \bibinfo {author} {\bibfnamefont {L.~A.}\ \bibnamefont {Constantin}},\
  }\href@noop {} {\bibfield  {journal} {\bibinfo  {journal} {Int. J. Quantum
  Chem.}\ }\textbf {\bibinfo {volume} {116}},\ \bibinfo {pages} {1641}
  (\bibinfo {year} {2016})}\BibitemShut {NoStop}%
\bibitem [{\citenamefont {Becke}\ and\ \citenamefont
  {Roussel}(1989)}]{BeckePRA89}%
  \BibitemOpen
  \bibfield  {author} {\bibinfo {author} {\bibfnamefont {A.~D.}\ \bibnamefont
  {Becke}}\ and\ \bibinfo {author} {\bibfnamefont {M.~R.}\ \bibnamefont
  {Roussel}},\ }\href@noop {} {\bibfield  {journal} {\bibinfo  {journal} {Phys.
  Rev. A}\ }\textbf {\bibinfo {volume} {39}},\ \bibinfo {pages} {3761}
  (\bibinfo {year} {1989})}\BibitemShut {NoStop}%
\bibitem [{\citenamefont {Sun}\ \emph {et~al.}(2015{\natexlab{b}})\citenamefont
  {Sun}, \citenamefont {Perdew},\ and\ \citenamefont {Ruzsinszky}}]{SunPNAS15}%
  \BibitemOpen
  \bibfield  {author} {\bibinfo {author} {\bibfnamefont {J.}~\bibnamefont
  {Sun}}, \bibinfo {author} {\bibfnamefont {J.~P.}\ \bibnamefont {Perdew}}, \
  and\ \bibinfo {author} {\bibfnamefont {A.}~\bibnamefont {Ruzsinszky}},\
  }\href@noop {} {\bibfield  {journal} {\bibinfo  {journal} {Proc. Natl. Acad.
  Sci. U.S.A.}\ }\textbf {\bibinfo {volume} {112}},\ \bibinfo {pages} {685}
  (\bibinfo {year} {2015}{\natexlab{b}})}\BibitemShut {NoStop}%
\bibitem [{\citenamefont {Verma}\ and\ \citenamefont
  {Truhlar}(2017{\natexlab{a}})}]{VermaJPCC17}%
  \BibitemOpen
  \bibfield  {author} {\bibinfo {author} {\bibfnamefont {P.}~\bibnamefont
  {Verma}}\ and\ \bibinfo {author} {\bibfnamefont {D.~G.}\ \bibnamefont
  {Truhlar}},\ }\href@noop {} {\bibfield  {journal} {\bibinfo  {journal} {J.
  Phys. Chem. C}\ }\textbf {\bibinfo {volume} {121}},\ \bibinfo {pages} {7144}
  (\bibinfo {year} {2017}{\natexlab{a}})}\BibitemShut {NoStop}%
\bibitem [{\citenamefont {Aschebrock}\ and\ \citenamefont
  {K\"ummel}(2019)}]{AschebrockPRR19}%
  \BibitemOpen
  \bibfield  {author} {\bibinfo {author} {\bibfnamefont {T.}~\bibnamefont
  {Aschebrock}}\ and\ \bibinfo {author} {\bibfnamefont {S.}~\bibnamefont
  {K\"ummel}},\ }\href@noop {} {\bibfield  {journal} {\bibinfo  {journal}
  {Phys. Rev. Research}\ }\textbf {\bibinfo {volume} {1}},\ \bibinfo {pages}
  {033082} (\bibinfo {year} {2019})}\BibitemShut {NoStop}%
\bibitem [{\citenamefont {Perdew}\ and\ \citenamefont
  {Wang}(1992)}]{PerdewPRB92a}%
  \BibitemOpen
  \bibfield  {author} {\bibinfo {author} {\bibfnamefont {J.~P.}\ \bibnamefont
  {Perdew}}\ and\ \bibinfo {author} {\bibfnamefont {Y.}~\bibnamefont {Wang}},\
  }\href@noop {} {\bibfield  {journal} {\bibinfo  {journal} {Phys. Rev. B}\
  }\textbf {\bibinfo {volume} {45}},\ \bibinfo {pages} {13244} (\bibinfo {year}
  {1992})},\ \bibinfo {note} {\textbf{98}, 079904(E) (2018)}\BibitemShut
  {NoStop}%
\bibitem [{\citenamefont {Perdew}\ and\ \citenamefont
  {Constantin}(2007)}]{PerdewPRB07}%
  \BibitemOpen
  \bibfield  {author} {\bibinfo {author} {\bibfnamefont {J.~P.}\ \bibnamefont
  {Perdew}}\ and\ \bibinfo {author} {\bibfnamefont {L.~A.}\ \bibnamefont
  {Constantin}},\ }\href@noop {} {\bibfield  {journal} {\bibinfo  {journal}
  {Phys. Rev. B}\ }\textbf {\bibinfo {volume} {75}},\ \bibinfo {pages} {155109}
  (\bibinfo {year} {2007})}\BibitemShut {NoStop}%
\bibitem [{\citenamefont {Bienvenu}\ and\ \citenamefont
  {Knizia}(2018)}]{BienvenuJCTC18}%
  \BibitemOpen
  \bibfield  {author} {\bibinfo {author} {\bibfnamefont {A.~V.}\ \bibnamefont
  {Bienvenu}}\ and\ \bibinfo {author} {\bibfnamefont {G.}~\bibnamefont
  {Knizia}},\ }\href@noop {} {\bibfield  {journal} {\bibinfo  {journal} {J.
  Chem. Theory Comput.}\ }\textbf {\bibinfo {volume} {14}},\ \bibinfo {pages}
  {1297} (\bibinfo {year} {2018})}\BibitemShut {NoStop}%
\bibitem [{\citenamefont {Tran}\ and\ \citenamefont {Blaha}(2009)}]{TranPRL09}%
  \BibitemOpen
  \bibfield  {author} {\bibinfo {author} {\bibfnamefont {F.}~\bibnamefont
  {Tran}}\ and\ \bibinfo {author} {\bibfnamefont {P.}~\bibnamefont {Blaha}},\
  }\href@noop {} {\bibfield  {journal} {\bibinfo  {journal} {Phys. Rev. Lett.}\
  }\textbf {\bibinfo {volume} {102}},\ \bibinfo {pages} {226401} (\bibinfo
  {year} {2009})}\BibitemShut {NoStop}%
\bibitem [{\citenamefont {Perdew}\ \emph {et~al.}(1996)\citenamefont {Perdew},
  \citenamefont {Burke},\ and\ \citenamefont {Ernzerhof}}]{PerdewPRL96}%
  \BibitemOpen
  \bibfield  {author} {\bibinfo {author} {\bibfnamefont {J.~P.}\ \bibnamefont
  {Perdew}}, \bibinfo {author} {\bibfnamefont {K.}~\bibnamefont {Burke}}, \
  and\ \bibinfo {author} {\bibfnamefont {M.}~\bibnamefont {Ernzerhof}},\
  }\href@noop {} {\bibfield  {journal} {\bibinfo  {journal} {Phys. Rev. Lett.}\
  }\textbf {\bibinfo {volume} {77}},\ \bibinfo {pages} {3865} (\bibinfo {year}
  {1996})},\ \bibinfo {note} {\textbf{78}, 1396(E) (1997)}\BibitemShut
  {NoStop}%
\bibitem [{\citenamefont {Verma}\ and\ \citenamefont
  {Truhlar}(2017{\natexlab{b}})}]{VermaJPCL17}%
  \BibitemOpen
  \bibfield  {author} {\bibinfo {author} {\bibfnamefont {P.}~\bibnamefont
  {Verma}}\ and\ \bibinfo {author} {\bibfnamefont {D.~G.}\ \bibnamefont
  {Truhlar}},\ }\href@noop {} {\bibfield  {journal} {\bibinfo  {journal} {J.
  Phys. Chem. Lett.}\ }\textbf {\bibinfo {volume} {8}},\ \bibinfo {pages} {380}
  (\bibinfo {year} {2017}{\natexlab{b}})}\BibitemShut {NoStop}%
\bibitem [{\citenamefont {Boese}\ and\ \citenamefont
  {Handy}(2001)}]{BoeseJCP01}%
  \BibitemOpen
  \bibfield  {author} {\bibinfo {author} {\bibfnamefont {A.~D.}\ \bibnamefont
  {Boese}}\ and\ \bibinfo {author} {\bibfnamefont {N.~C.}\ \bibnamefont
  {Handy}},\ }\href@noop {} {\bibfield  {journal} {\bibinfo  {journal} {J.
  Chem. Phys.}\ }\textbf {\bibinfo {volume} {114}},\ \bibinfo {pages} {5497}
  (\bibinfo {year} {2001})}\BibitemShut {NoStop}%
\bibitem [{\citenamefont {Tran}\ and\ \citenamefont
  {Blaha}(2017)}]{TranJPCA17}%
  \BibitemOpen
  \bibfield  {author} {\bibinfo {author} {\bibfnamefont {F.}~\bibnamefont
  {Tran}}\ and\ \bibinfo {author} {\bibfnamefont {P.}~\bibnamefont {Blaha}},\
  }\href@noop {} {\bibfield  {journal} {\bibinfo  {journal} {J. Phys. Chem. A}\
  }\textbf {\bibinfo {volume} {121}},\ \bibinfo {pages} {3318} (\bibinfo {year}
  {2017})}\BibitemShut {NoStop}%
\bibitem [{\citenamefont {Borlido}\ \emph {et~al.}(2019)\citenamefont
  {Borlido}, \citenamefont {Aull}, \citenamefont {Huran}, \citenamefont {Tran},
  \citenamefont {Marques},\ and\ \citenamefont {Botti}}]{BorlidoJCTC19}%
  \BibitemOpen
  \bibfield  {author} {\bibinfo {author} {\bibfnamefont {P.}~\bibnamefont
  {Borlido}}, \bibinfo {author} {\bibfnamefont {T.}~\bibnamefont {Aull}},
  \bibinfo {author} {\bibfnamefont {A.~W.}\ \bibnamefont {Huran}}, \bibinfo
  {author} {\bibfnamefont {F.}~\bibnamefont {Tran}}, \bibinfo {author}
  {\bibfnamefont {M.~A.~L.}\ \bibnamefont {Marques}}, \ and\ \bibinfo {author}
  {\bibfnamefont {S.}~\bibnamefont {Botti}},\ }\href@noop {} {\bibfield
  {journal} {\bibinfo  {journal} {J. Chem. Theory Comput.}\ }\textbf {\bibinfo
  {volume} {15}},\ \bibinfo {pages} {5069} (\bibinfo {year}
  {2019})}\BibitemShut {NoStop}%
\bibitem [{\citenamefont {Tran}\ \emph
  {et~al.}(2019{\natexlab{b}})\citenamefont {Tran}, \citenamefont {Doumont},
  \citenamefont {Kalantari}, \citenamefont {Huran}, \citenamefont {Marques},\
  and\ \citenamefont {Blaha}}]{TranJAP19}%
  \BibitemOpen
  \bibfield  {author} {\bibinfo {author} {\bibfnamefont {F.}~\bibnamefont
  {Tran}}, \bibinfo {author} {\bibfnamefont {J.}~\bibnamefont {Doumont}},
  \bibinfo {author} {\bibfnamefont {L.}~\bibnamefont {Kalantari}}, \bibinfo
  {author} {\bibfnamefont {A.~W.}\ \bibnamefont {Huran}}, \bibinfo {author}
  {\bibfnamefont {M.~A.~L.}\ \bibnamefont {Marques}}, \ and\ \bibinfo {author}
  {\bibfnamefont {P.}~\bibnamefont {Blaha}},\ }\href@noop {} {\bibfield
  {journal} {\bibinfo  {journal} {J. Appl. Phys.}\ }\textbf {\bibinfo {volume}
  {126}},\ \bibinfo {pages} {110902} (\bibinfo {year}
  {2019}{\natexlab{b}})}\BibitemShut {NoStop}%
\bibitem [{\citenamefont {Blaha}\ \emph {et~al.}(2018)\citenamefont {Blaha},
  \citenamefont {Schwarz}, \citenamefont {Madsen}, \citenamefont {Kvasnicka},
  \citenamefont {Luitz}, \citenamefont {Laskowski}, \citenamefont {Tran},\ and\
  \citenamefont {Marks}}]{WIEN2k}%
  \BibitemOpen
  \bibfield  {author} {\bibinfo {author} {\bibfnamefont {P.}~\bibnamefont
  {Blaha}}, \bibinfo {author} {\bibfnamefont {K.}~\bibnamefont {Schwarz}},
  \bibinfo {author} {\bibfnamefont {G.~K.~H.}\ \bibnamefont {Madsen}}, \bibinfo
  {author} {\bibfnamefont {D.}~\bibnamefont {Kvasnicka}}, \bibinfo {author}
  {\bibfnamefont {J.}~\bibnamefont {Luitz}}, \bibinfo {author} {\bibfnamefont
  {R.}~\bibnamefont {Laskowski}}, \bibinfo {author} {\bibfnamefont
  {F.}~\bibnamefont {Tran}}, \ and\ \bibinfo {author} {\bibfnamefont {L.~D.}\
  \bibnamefont {Marks}},\ }\href@noop {} {\emph {\bibinfo {title} {WIEN2k: An
  Augmented Plane Wave plus Local Orbitals Program for Calculating Crystal
  Properties}}}\ (\bibinfo  {publisher} {Vienna University of Technology},\
  \bibinfo {address} {Austria},\ \bibinfo {year} {2018})\BibitemShut {NoStop}%
\bibitem [{\citenamefont {Blaha}\ \emph {et~al.}(2020)\citenamefont {Blaha},
  \citenamefont {Schwarz}, \citenamefont {Tran}, \citenamefont {Laskowski},
  \citenamefont {Madsen},\ and\ \citenamefont {Marks}}]{BlahaJCP20}%
  \BibitemOpen
  \bibfield  {author} {\bibinfo {author} {\bibfnamefont {P.}~\bibnamefont
  {Blaha}}, \bibinfo {author} {\bibfnamefont {K.}~\bibnamefont {Schwarz}},
  \bibinfo {author} {\bibfnamefont {F.}~\bibnamefont {Tran}}, \bibinfo {author}
  {\bibfnamefont {R.}~\bibnamefont {Laskowski}}, \bibinfo {author}
  {\bibfnamefont {G.~K.~H.}\ \bibnamefont {Madsen}}, \ and\ \bibinfo {author}
  {\bibfnamefont {L.~D.}\ \bibnamefont {Marks}},\ }\href@noop {} {\bibfield
  {journal} {\bibinfo  {journal} {J. Chem. Phys.}\ }\textbf {\bibinfo {volume}
  {152}},\ \bibinfo {pages} {074101} (\bibinfo {year} {2020})}\BibitemShut
  {NoStop}%
\bibitem [{\citenamefont {Andersen}(1975)}]{AndersenPRB75}%
  \BibitemOpen
  \bibfield  {author} {\bibinfo {author} {\bibfnamefont {O.~K.}\ \bibnamefont
  {Andersen}},\ }\href@noop {} {\bibfield  {journal} {\bibinfo  {journal}
  {Phys. Rev. B}\ }\textbf {\bibinfo {volume} {12}},\ \bibinfo {pages} {3060}
  (\bibinfo {year} {1975})}\BibitemShut {NoStop}%
\bibitem [{\citenamefont {Singh}\ and\ \citenamefont
  {Nordstr{\"{o}}m}(2006)}]{Singh}%
  \BibitemOpen
  \bibfield  {author} {\bibinfo {author} {\bibfnamefont {D.~J.}\ \bibnamefont
  {Singh}}\ and\ \bibinfo {author} {\bibfnamefont {L.}~\bibnamefont
  {Nordstr{\"{o}}m}},\ }\href@noop {} {\emph {\bibinfo {title} {Planewaves,
  Pseudopotentials, and the LAPW Method, 2nd ed.}}}\ (\bibinfo  {publisher}
  {Springer},\ \bibinfo {address} {New York},\ \bibinfo {year}
  {2006})\BibitemShut {NoStop}%
\bibitem [{\citenamefont {Karsai}\ \emph {et~al.}(2017)\citenamefont {Karsai},
  \citenamefont {Tran},\ and\ \citenamefont {Blaha}}]{KarsaiCPC17}%
  \BibitemOpen
  \bibfield  {author} {\bibinfo {author} {\bibfnamefont {F.}~\bibnamefont
  {Karsai}}, \bibinfo {author} {\bibfnamefont {F.}~\bibnamefont {Tran}}, \ and\
  \bibinfo {author} {\bibfnamefont {P.}~\bibnamefont {Blaha}},\ }\href@noop {}
  {\bibfield  {journal} {\bibinfo  {journal} {Comput. Phys. Commun.}\ }\textbf
  {\bibinfo {volume} {220}},\ \bibinfo {pages} {230} (\bibinfo {year}
  {2017})}\BibitemShut {NoStop}%
\bibitem [{\citenamefont {Marques}\ \emph {et~al.}(2012)\citenamefont
  {Marques}, \citenamefont {Oliveira},\ and\ \citenamefont
  {Burnus}}]{MarquesCPC12}%
  \BibitemOpen
  \bibfield  {author} {\bibinfo {author} {\bibfnamefont {M.~A.~L.}\
  \bibnamefont {Marques}}, \bibinfo {author} {\bibfnamefont {M.~J.~T.}\
  \bibnamefont {Oliveira}}, \ and\ \bibinfo {author} {\bibfnamefont
  {T.}~\bibnamefont {Burnus}},\ }\href@noop {} {\bibfield  {journal} {\bibinfo
  {journal} {Comput. Phys. Commun.}\ }\textbf {\bibinfo {volume} {183}},\
  \bibinfo {pages} {2272} (\bibinfo {year} {2012})}\BibitemShut {NoStop}%
\bibitem [{\citenamefont {Lehtola}\ \emph {et~al.}(2018)\citenamefont
  {Lehtola}, \citenamefont {Steigemann}, \citenamefont {Oliveira},\ and\
  \citenamefont {Marques}}]{LehtolaSX18}%
  \BibitemOpen
  \bibfield  {author} {\bibinfo {author} {\bibfnamefont {S.}~\bibnamefont
  {Lehtola}}, \bibinfo {author} {\bibfnamefont {C.}~\bibnamefont {Steigemann}},
  \bibinfo {author} {\bibfnamefont {M.~J.~T.}\ \bibnamefont {Oliveira}}, \ and\
  \bibinfo {author} {\bibfnamefont {M.~A.~L.}\ \bibnamefont {Marques}},\
  }\href@noop {} {\bibfield  {journal} {\bibinfo  {journal} {SoftwareX}\
  }\textbf {\bibinfo {volume} {7}},\ \bibinfo {pages} {1} (\bibinfo {year}
  {2018})}\BibitemShut {NoStop}%
\bibitem [{\citenamefont {Schwarz}\ and\ \citenamefont
  {Mohn}(1984)}]{SchwarzJPF84}%
  \BibitemOpen
  \bibfield  {author} {\bibinfo {author} {\bibfnamefont {K.}~\bibnamefont
  {Schwarz}}\ and\ \bibinfo {author} {\bibfnamefont {P.}~\bibnamefont {Mohn}},\
  }\href@noop {} {\bibfield  {journal} {\bibinfo  {journal} {J. Phys. F: Met.
  Phys.}\ }\textbf {\bibinfo {volume} {14}},\ \bibinfo {pages} {L129} (\bibinfo
  {year} {1984})}\BibitemShut {NoStop}%
\bibitem [{\citenamefont {Tran}\ \emph
  {et~al.}(2019{\natexlab{c}})\citenamefont {Tran}, \citenamefont {Doumont},
  \citenamefont {Blaha}, \citenamefont {Marques}, \citenamefont {Botti},\ and\
  \citenamefont {Bart\'{o}k}}]{TranJCP19}%
  \BibitemOpen
  \bibfield  {author} {\bibinfo {author} {\bibfnamefont {F.}~\bibnamefont
  {Tran}}, \bibinfo {author} {\bibfnamefont {J.}~\bibnamefont {Doumont}},
  \bibinfo {author} {\bibfnamefont {P.}~\bibnamefont {Blaha}}, \bibinfo
  {author} {\bibfnamefont {M.~A.~L.}\ \bibnamefont {Marques}}, \bibinfo
  {author} {\bibfnamefont {S.}~\bibnamefont {Botti}}, \ and\ \bibinfo {author}
  {\bibfnamefont {A.~P.}\ \bibnamefont {Bart\'{o}k}},\ }\href@noop {}
  {\bibfield  {journal} {\bibinfo  {journal} {J. Chem. Phys.}\ }\textbf
  {\bibinfo {volume} {151}},\ \bibinfo {pages} {161102} (\bibinfo {year}
  {2019}{\natexlab{c}})}\BibitemShut {NoStop}%
\bibitem [{\citenamefont {Hammer}\ \emph {et~al.}(1999)\citenamefont {Hammer},
  \citenamefont {Hansen},\ and\ \citenamefont {N{\o}rskov}}]{HammerPRB99}%
  \BibitemOpen
  \bibfield  {author} {\bibinfo {author} {\bibfnamefont {B.}~\bibnamefont
  {Hammer}}, \bibinfo {author} {\bibfnamefont {L.~B.}\ \bibnamefont {Hansen}},
  \ and\ \bibinfo {author} {\bibfnamefont {J.~K.}\ \bibnamefont {N{\o}rskov}},\
  }\href@noop {} {\bibfield  {journal} {\bibinfo  {journal} {Phys. Rev. B}\
  }\textbf {\bibinfo {volume} {59}},\ \bibinfo {pages} {7413} (\bibinfo {year}
  {1999})}\BibitemShut {NoStop}%
\bibitem [{\citenamefont {Engel}\ and\ \citenamefont
  {Vosko}(1993)}]{EngelPRB93}%
  \BibitemOpen
  \bibfield  {author} {\bibinfo {author} {\bibfnamefont {E.}~\bibnamefont
  {Engel}}\ and\ \bibinfo {author} {\bibfnamefont {S.~H.}\ \bibnamefont
  {Vosko}},\ }\href@noop {} {\bibfield  {journal} {\bibinfo  {journal} {Phys.
  Rev. B}\ }\textbf {\bibinfo {volume} {47}},\ \bibinfo {pages} {13164}
  (\bibinfo {year} {1993})}\BibitemShut {NoStop}%
\bibitem [{\citenamefont {Kresse}\ and\ \citenamefont
  {Furthm\"uller}(1996)}]{KressePRB96}%
  \BibitemOpen
  \bibfield  {author} {\bibinfo {author} {\bibfnamefont {G.}~\bibnamefont
  {Kresse}}\ and\ \bibinfo {author} {\bibfnamefont {J.}~\bibnamefont
  {Furthm\"uller}},\ }\href@noop {} {\bibfield  {journal} {\bibinfo  {journal}
  {Phys. Rev. B}\ }\textbf {\bibinfo {volume} {54}},\ \bibinfo {pages} {11169}
  (\bibinfo {year} {1996})}\BibitemShut {NoStop}%
\bibitem [{\citenamefont {Enkovaara}\ \emph {et~al.}(2010)\citenamefont
  {Enkovaara}, \citenamefont {Rostgaard}, \citenamefont {Mortensen},
  \citenamefont {Chen}, \citenamefont {Du{\l}ak}, \citenamefont {Ferrighi},
  \citenamefont {Gavnholt}, \citenamefont {Glinsvad}, \citenamefont {Haikola},
  \citenamefont {Hansen}, \citenamefont {Kristoffersen}, \citenamefont
  {Kuisma}, \citenamefont {Larsen}, \citenamefont {Lehtovaara}, \citenamefont
  {Ljungberg}, \citenamefont {Lopez-Acevedo}, \citenamefont {Moses},
  \citenamefont {Ojanen}, \citenamefont {Olsen}, \citenamefont {Petzold},
  \citenamefont {Romero}, \citenamefont {Stausholm-M{\o}ller}, \citenamefont
  {Strange}, \citenamefont {Tritsaris}, \citenamefont {Vanin}, \citenamefont
  {Walter}, \citenamefont {Hammer}, \citenamefont {H\"{a}kkinen}, \citenamefont
  {Madsen}, \citenamefont {Nieminen}, \citenamefont {N{\o}rskov}, \citenamefont
  {Puska}, \citenamefont {Rantala}, \citenamefont {Schi{\o}tz}, \citenamefont
  {Thygesen},\ and\ \citenamefont {Jacobsen}}]{EnkovaaraJPCM10}%
  \BibitemOpen
  \bibfield  {author} {\bibinfo {author} {\bibfnamefont {J.}~\bibnamefont
  {Enkovaara}}, \bibinfo {author} {\bibfnamefont {C.}~\bibnamefont
  {Rostgaard}}, \bibinfo {author} {\bibfnamefont {J.~J.}\ \bibnamefont
  {Mortensen}}, \bibinfo {author} {\bibfnamefont {J.}~\bibnamefont {Chen}},
  \bibinfo {author} {\bibfnamefont {M.}~\bibnamefont {Du{\l}ak}}, \bibinfo
  {author} {\bibfnamefont {L.}~\bibnamefont {Ferrighi}}, \bibinfo {author}
  {\bibfnamefont {J.}~\bibnamefont {Gavnholt}}, \bibinfo {author}
  {\bibfnamefont {C.}~\bibnamefont {Glinsvad}}, \bibinfo {author}
  {\bibfnamefont {V.}~\bibnamefont {Haikola}}, \bibinfo {author} {\bibfnamefont
  {H.~A.}\ \bibnamefont {Hansen}}, \bibinfo {author} {\bibfnamefont {H.~H.}\
  \bibnamefont {Kristoffersen}}, \bibinfo {author} {\bibfnamefont
  {M.}~\bibnamefont {Kuisma}}, \bibinfo {author} {\bibfnamefont {A.~H.}\
  \bibnamefont {Larsen}}, \bibinfo {author} {\bibfnamefont {L.}~\bibnamefont
  {Lehtovaara}}, \bibinfo {author} {\bibfnamefont {M.}~\bibnamefont
  {Ljungberg}}, \bibinfo {author} {\bibfnamefont {O.}~\bibnamefont
  {Lopez-Acevedo}}, \bibinfo {author} {\bibfnamefont {P.~G.}\ \bibnamefont
  {Moses}}, \bibinfo {author} {\bibfnamefont {J.}~\bibnamefont {Ojanen}},
  \bibinfo {author} {\bibfnamefont {T.}~\bibnamefont {Olsen}}, \bibinfo
  {author} {\bibfnamefont {V.}~\bibnamefont {Petzold}}, \bibinfo {author}
  {\bibfnamefont {N.~A.}\ \bibnamefont {Romero}}, \bibinfo {author}
  {\bibfnamefont {J.}~\bibnamefont {Stausholm-M{\o}ller}}, \bibinfo {author}
  {\bibfnamefont {M.}~\bibnamefont {Strange}}, \bibinfo {author} {\bibfnamefont
  {G.~A.}\ \bibnamefont {Tritsaris}}, \bibinfo {author} {\bibfnamefont
  {M.}~\bibnamefont {Vanin}}, \bibinfo {author} {\bibfnamefont
  {M.}~\bibnamefont {Walter}}, \bibinfo {author} {\bibfnamefont
  {B.}~\bibnamefont {Hammer}}, \bibinfo {author} {\bibfnamefont
  {H.}~\bibnamefont {H\"{a}kkinen}}, \bibinfo {author} {\bibfnamefont
  {G.~K.~H.}\ \bibnamefont {Madsen}}, \bibinfo {author} {\bibfnamefont {R.~M.}\
  \bibnamefont {Nieminen}}, \bibinfo {author} {\bibfnamefont {J.~K.}\
  \bibnamefont {N{\o}rskov}}, \bibinfo {author} {\bibfnamefont
  {M.}~\bibnamefont {Puska}}, \bibinfo {author} {\bibfnamefont {T.~T.}\
  \bibnamefont {Rantala}}, \bibinfo {author} {\bibfnamefont {J.}~\bibnamefont
  {Schi{\o}tz}}, \bibinfo {author} {\bibfnamefont {K.~S.}\ \bibnamefont
  {Thygesen}}, \ and\ \bibinfo {author} {\bibfnamefont {K.~W.}\ \bibnamefont
  {Jacobsen}},\ }\href@noop {} {\bibfield  {journal} {\bibinfo  {journal} {J.
  Phys.: Condens. Matter}\ }\textbf {\bibinfo {volume} {22}},\ \bibinfo {pages}
  {253202} (\bibinfo {year} {2010})}\BibitemShut {NoStop}%
\bibitem [{\citenamefont {Ferrighi}\ \emph {et~al.}(2011)\citenamefont
  {Ferrighi}, \citenamefont {Madsen},\ and\ \citenamefont
  {Hammer}}]{FerrighiJCP11}%
  \BibitemOpen
  \bibfield  {author} {\bibinfo {author} {\bibfnamefont {L.}~\bibnamefont
  {Ferrighi}}, \bibinfo {author} {\bibfnamefont {G.~K.~H.}\ \bibnamefont
  {Madsen}}, \ and\ \bibinfo {author} {\bibfnamefont {B.}~\bibnamefont
  {Hammer}},\ }\href@noop {} {\bibfield  {journal} {\bibinfo  {journal} {J.
  Chem. Phys.}\ }\textbf {\bibinfo {volume} {135}},\ \bibinfo {pages} {084704}
  (\bibinfo {year} {2011})}\BibitemShut {NoStop}%
\bibitem [{\citenamefont {Bl\"{o}chl}(1994)}]{BlochlPRB94b}%
  \BibitemOpen
  \bibfield  {author} {\bibinfo {author} {\bibfnamefont {P.~E.}\ \bibnamefont
  {Bl\"{o}chl}},\ }\href@noop {} {\bibfield  {journal} {\bibinfo  {journal}
  {Phys. Rev. B}\ }\textbf {\bibinfo {volume} {50}},\ \bibinfo {pages} {17953}
  (\bibinfo {year} {1994})}\BibitemShut {NoStop}%
\bibitem [{\citenamefont {Bader}(1990)}]{Bader90}%
  \BibitemOpen
  \bibfield  {author} {\bibinfo {author} {\bibfnamefont {R.~F.~W.}\
  \bibnamefont {Bader}},\ }\href@noop {} {\emph {\bibinfo {title} {Atoms in
  Molecules: A Quantum Theory}}}\ (\bibinfo  {publisher} {Oxford University
  Press},\ \bibinfo {address} {Oxford},\ \bibinfo {year} {1990})\BibitemShut
  {NoStop}%
\bibitem [{\citenamefont {Bader}(1991)}]{BaderCR91}%
  \BibitemOpen
  \bibfield  {author} {\bibinfo {author} {\bibfnamefont {R.~F.~W.}\
  \bibnamefont {Bader}},\ }\href@noop {} {\bibfield  {journal} {\bibinfo
  {journal} {Chem. Rev.}\ }\textbf {\bibinfo {volume} {91}},\ \bibinfo {pages}
  {893} (\bibinfo {year} {1991})}\BibitemShut {NoStop}%
\bibitem [{\citenamefont {{Otero-de-la-Roza}}\ \emph
  {et~al.}(2009)\citenamefont {{Otero-de-la-Roza}}, \citenamefont {Blanco},
  \citenamefont {Mart\'{i}n~Pend\'{a}s},\ and\ \citenamefont
  {Lua{\~{n}}a}}]{OterodelaRozaCPC09}%
  \BibitemOpen
  \bibfield  {author} {\bibinfo {author} {\bibfnamefont {A.}~\bibnamefont
  {{Otero-de-la-Roza}}}, \bibinfo {author} {\bibfnamefont {M.~A.}\ \bibnamefont
  {Blanco}}, \bibinfo {author} {\bibfnamefont {A.}~\bibnamefont
  {Mart\'{i}n~Pend\'{a}s}}, \ and\ \bibinfo {author} {\bibfnamefont
  {V.}~\bibnamefont {Lua{\~{n}}a}},\ }\href@noop {} {\bibfield  {journal}
  {\bibinfo  {journal} {Comput. Phys. Commun.}\ }\textbf {\bibinfo {volume}
  {180}},\ \bibinfo {pages} {157} (\bibinfo {year} {2009})}\BibitemShut
  {NoStop}%
\bibitem [{\citenamefont {{Otero-de-la-Roza}}\ \emph
  {et~al.}(2014)\citenamefont {{Otero-de-la-Roza}}, \citenamefont {Johnson},\
  and\ \citenamefont {Lua{\~{n}}a}}]{OterodelaRozaCPC14}%
  \BibitemOpen
  \bibfield  {author} {\bibinfo {author} {\bibfnamefont {A.}~\bibnamefont
  {{Otero-de-la-Roza}}}, \bibinfo {author} {\bibfnamefont {E.~R.}\ \bibnamefont
  {Johnson}}, \ and\ \bibinfo {author} {\bibfnamefont {V.}~\bibnamefont
  {Lua{\~{n}}a}},\ }\href@noop {} {\bibfield  {journal} {\bibinfo  {journal}
  {Comput. Phys. Commun.}\ }\textbf {\bibinfo {volume} {185}},\ \bibinfo
  {pages} {1007} (\bibinfo {year} {2014})}\BibitemShut {NoStop}%
\bibitem [{\citenamefont {Bergerhoff}\ \emph {et~al.}(1983)\citenamefont
  {Bergerhoff}, \citenamefont {Hundt}, \citenamefont {Sievers},\ and\
  \citenamefont {Brown}}]{BergerhoffJCICS83}%
  \BibitemOpen
  \bibfield  {author} {\bibinfo {author} {\bibfnamefont {G.}~\bibnamefont
  {Bergerhoff}}, \bibinfo {author} {\bibfnamefont {R.}~\bibnamefont {Hundt}},
  \bibinfo {author} {\bibfnamefont {R.}~\bibnamefont {Sievers}}, \ and\
  \bibinfo {author} {\bibfnamefont {I.~D.}\ \bibnamefont {Brown}},\ }\href@noop
  {} {\bibfield  {journal} {\bibinfo  {journal} {J. Chem. Inf. Comput. Sci.}\
  }\textbf {\bibinfo {volume} {23}},\ \bibinfo {pages} {66} (\bibinfo {year}
  {1983})}\BibitemShut {NoStop}%
\bibitem [{\citenamefont {Belsky}\ \emph {et~al.}(2002)\citenamefont {Belsky},
  \citenamefont {Hellenbrandt}, \citenamefont {Karen},\ and\ \citenamefont
  {Luksch}}]{BelskyAC02}%
  \BibitemOpen
  \bibfield  {author} {\bibinfo {author} {\bibfnamefont {A.}~\bibnamefont
  {Belsky}}, \bibinfo {author} {\bibfnamefont {M.}~\bibnamefont
  {Hellenbrandt}}, \bibinfo {author} {\bibfnamefont {V.~L.}\ \bibnamefont
  {Karen}}, \ and\ \bibinfo {author} {\bibfnamefont {P.}~\bibnamefont
  {Luksch}},\ }\href@noop {} {\bibfield  {journal} {\bibinfo  {journal} {Acta
  Cryst.}\ }\textbf {\bibinfo {volume} {B58}},\ \bibinfo {pages} {364}
  (\bibinfo {year} {2002})}\BibitemShut {NoStop}%
\bibitem [{\citenamefont {Chen}\ \emph {et~al.}(1995)\citenamefont {Chen},
  \citenamefont {Idzerda}, \citenamefont {Lin}, \citenamefont {Smith},
  \citenamefont {Meigs}, \citenamefont {Chaban}, \citenamefont {Ho},
  \citenamefont {Pellegrin},\ and\ \citenamefont {Sette}}]{ChenPRL95}%
  \BibitemOpen
  \bibfield  {author} {\bibinfo {author} {\bibfnamefont {C.~T.}\ \bibnamefont
  {Chen}}, \bibinfo {author} {\bibfnamefont {Y.~U.}\ \bibnamefont {Idzerda}},
  \bibinfo {author} {\bibfnamefont {H.-J.}\ \bibnamefont {Lin}}, \bibinfo
  {author} {\bibfnamefont {N.~V.}\ \bibnamefont {Smith}}, \bibinfo {author}
  {\bibfnamefont {G.}~\bibnamefont {Meigs}}, \bibinfo {author} {\bibfnamefont
  {E.}~\bibnamefont {Chaban}}, \bibinfo {author} {\bibfnamefont {G.~H.}\
  \bibnamefont {Ho}}, \bibinfo {author} {\bibfnamefont {E.}~\bibnamefont
  {Pellegrin}}, \ and\ \bibinfo {author} {\bibfnamefont {F.}~\bibnamefont
  {Sette}},\ }\href@noop {} {\bibfield  {journal} {\bibinfo  {journal} {Phys.
  Rev. Lett.}\ }\textbf {\bibinfo {volume} {75}},\ \bibinfo {pages} {152}
  (\bibinfo {year} {1995})}\BibitemShut {NoStop}%
\bibitem [{\citenamefont {Scherz}(2003)}]{Scherz03}%
  \BibitemOpen
  \bibfield  {author} {\bibinfo {author} {\bibfnamefont {A.}~\bibnamefont
  {Scherz}},\ }\href@noop {} {}\bibinfo {howpublished} {Ph.D. thesis, Free
  University of Berlin} (\bibinfo {year} {2003})\BibitemShut {NoStop}%
\bibitem [{\citenamefont {Reck}\ and\ \citenamefont {Fry}(1969)}]{ReckPR69}%
  \BibitemOpen
  \bibfield  {author} {\bibinfo {author} {\bibfnamefont {R.~A.}\ \bibnamefont
  {Reck}}\ and\ \bibinfo {author} {\bibfnamefont {D.~L.}\ \bibnamefont {Fry}},\
  }\href@noop {} {\bibfield  {journal} {\bibinfo  {journal} {Phys. Rev.}\
  }\textbf {\bibinfo {volume} {184}},\ \bibinfo {pages} {492} (\bibinfo {year}
  {1969})}\BibitemShut {NoStop}%
\bibitem [{\citenamefont {Moon}(1964)}]{MoonPR64}%
  \BibitemOpen
  \bibfield  {author} {\bibinfo {author} {\bibfnamefont {R.~M.}\ \bibnamefont
  {Moon}},\ }\href@noop {} {\bibfield  {journal} {\bibinfo  {journal} {Phys.
  Rev.}\ }\textbf {\bibinfo {volume} {136}},\ \bibinfo {pages} {A195} (\bibinfo
  {year} {1964})}\BibitemShut {NoStop}%
\bibitem [{\citenamefont {Mook}\ and\ \citenamefont {Shull}(1966)}]{MookJAP66}%
  \BibitemOpen
  \bibfield  {author} {\bibinfo {author} {\bibfnamefont {H.~A.}\ \bibnamefont
  {Mook}}\ and\ \bibinfo {author} {\bibfnamefont {C.~G.}\ \bibnamefont
  {Shull}},\ }\href@noop {} {\bibfield  {journal} {\bibinfo  {journal} {J.
  Appl. Phys.}\ }\textbf {\bibinfo {volume} {37}},\ \bibinfo {pages} {1034}
  (\bibinfo {year} {1966})}\BibitemShut {NoStop}%
\bibitem [{\citenamefont {Di~Fabrizio}\ \emph {et~al.}(1989)\citenamefont
  {Di~Fabrizio}, \citenamefont {Mazzone}, \citenamefont {Petrillo},\ and\
  \citenamefont {Sacchetti}}]{DiFabrizioPRB89}%
  \BibitemOpen
  \bibfield  {author} {\bibinfo {author} {\bibfnamefont {E.}~\bibnamefont
  {Di~Fabrizio}}, \bibinfo {author} {\bibfnamefont {G.}~\bibnamefont
  {Mazzone}}, \bibinfo {author} {\bibfnamefont {C.}~\bibnamefont {Petrillo}}, \
  and\ \bibinfo {author} {\bibfnamefont {F.}~\bibnamefont {Sacchetti}},\
  }\href@noop {} {\bibfield  {journal} {\bibinfo  {journal} {Phys. Rev. B}\
  }\textbf {\bibinfo {volume} {40}},\ \bibinfo {pages} {9502} (\bibinfo {year}
  {1989})}\BibitemShut {NoStop}%
\bibitem [{\citenamefont {Buschow}\ and\ \citenamefont {van
  Stapele}(1970)}]{BuschowJAP70}%
  \BibitemOpen
  \bibfield  {author} {\bibinfo {author} {\bibfnamefont {K.~H.~J.}\
  \bibnamefont {Buschow}}\ and\ \bibinfo {author} {\bibfnamefont {R.~P.}\
  \bibnamefont {van Stapele}},\ }\href@noop {} {\bibfield  {journal} {\bibinfo
  {journal} {J. Appl. Phys.}\ }\textbf {\bibinfo {volume} {41}},\ \bibinfo
  {pages} {4066} (\bibinfo {year} {1970})}\BibitemShut {NoStop}%
\bibitem [{\citenamefont {Uhlarz}\ \emph {et~al.}(2004)\citenamefont {Uhlarz},
  \citenamefont {Pfleiderer},\ and\ \citenamefont {Hayden}}]{UhlarzPRL04}%
  \BibitemOpen
  \bibfield  {author} {\bibinfo {author} {\bibfnamefont {M.}~\bibnamefont
  {Uhlarz}}, \bibinfo {author} {\bibfnamefont {C.}~\bibnamefont {Pfleiderer}},
  \ and\ \bibinfo {author} {\bibfnamefont {S.~M.}\ \bibnamefont {Hayden}},\
  }\href@noop {} {\bibfield  {journal} {\bibinfo  {journal} {Phys. Rev. Lett.}\
  }\textbf {\bibinfo {volume} {93}},\ \bibinfo {pages} {256404} (\bibinfo
  {year} {2004})}\BibitemShut {NoStop}%
\bibitem [{\citenamefont {Yelland}\ \emph {et~al.}(2005)\citenamefont
  {Yelland}, \citenamefont {Yates}, \citenamefont {Taylor}, \citenamefont
  {Griffiths}, \citenamefont {Hayden},\ and\ \citenamefont
  {Carrington}}]{YellandPRB05}%
  \BibitemOpen
  \bibfield  {author} {\bibinfo {author} {\bibfnamefont {E.~A.}\ \bibnamefont
  {Yelland}}, \bibinfo {author} {\bibfnamefont {S.~J.~C.}\ \bibnamefont
  {Yates}}, \bibinfo {author} {\bibfnamefont {O.}~\bibnamefont {Taylor}},
  \bibinfo {author} {\bibfnamefont {A.}~\bibnamefont {Griffiths}}, \bibinfo
  {author} {\bibfnamefont {S.~M.}\ \bibnamefont {Hayden}}, \ and\ \bibinfo
  {author} {\bibfnamefont {A.}~\bibnamefont {Carrington}},\ }\href@noop {}
  {\bibfield  {journal} {\bibinfo  {journal} {Phys. Rev. B}\ }\textbf {\bibinfo
  {volume} {72}},\ \bibinfo {pages} {184436} (\bibinfo {year}
  {2005})}\BibitemShut {NoStop}%
\bibitem [{\citenamefont {de~Boer}\ \emph {et~al.}(1969)\citenamefont
  {de~Boer}, \citenamefont {Schinkel}, \citenamefont {Biesterbos},\ and\
  \citenamefont {Proost}}]{DeBoerJAP69}%
  \BibitemOpen
  \bibfield  {author} {\bibinfo {author} {\bibfnamefont {F.~R.}\ \bibnamefont
  {de~Boer}}, \bibinfo {author} {\bibfnamefont {C.~J.}\ \bibnamefont
  {Schinkel}}, \bibinfo {author} {\bibfnamefont {J.}~\bibnamefont
  {Biesterbos}}, \ and\ \bibinfo {author} {\bibfnamefont {S.}~\bibnamefont
  {Proost}},\ }\href@noop {} {\bibfield  {journal} {\bibinfo  {journal} {J.
  Appl. Phys.}\ }\textbf {\bibinfo {volume} {40}},\ \bibinfo {pages} {1049}
  (\bibinfo {year} {1969})}\BibitemShut {NoStop}%
\bibitem [{\citenamefont {Mazin}\ and\ \citenamefont
  {Singh}(2004)}]{MazinPRB04}%
  \BibitemOpen
  \bibfield  {author} {\bibinfo {author} {\bibfnamefont {I.~I.}\ \bibnamefont
  {Mazin}}\ and\ \bibinfo {author} {\bibfnamefont {D.~J.}\ \bibnamefont
  {Singh}},\ }\href@noop {} {\bibfield  {journal} {\bibinfo  {journal} {Phys.
  Rev. B}\ }\textbf {\bibinfo {volume} {69}},\ \bibinfo {pages} {020402(R)}
  (\bibinfo {year} {2004})}\BibitemShut {NoStop}%
\bibitem [{\citenamefont {Aguayo}\ \emph {et~al.}(2004)\citenamefont {Aguayo},
  \citenamefont {Mazin},\ and\ \citenamefont {Singh}}]{AguayoPRL04}%
  \BibitemOpen
  \bibfield  {author} {\bibinfo {author} {\bibfnamefont {A.}~\bibnamefont
  {Aguayo}}, \bibinfo {author} {\bibfnamefont {I.~I.}\ \bibnamefont {Mazin}}, \
  and\ \bibinfo {author} {\bibfnamefont {D.~J.}\ \bibnamefont {Singh}},\
  }\href@noop {} {\bibfield  {journal} {\bibinfo  {journal} {Phys. Rev. Lett.}\
  }\textbf {\bibinfo {volume} {92}},\ \bibinfo {pages} {147201} (\bibinfo
  {year} {2004})}\BibitemShut {NoStop}%
\bibitem [{\citenamefont {Ortenzi}\ \emph {et~al.}(2012)\citenamefont
  {Ortenzi}, \citenamefont {Mazin}, \citenamefont {Blaha},\ and\ \citenamefont
  {Boeri}}]{OrtenziPRB12}%
  \BibitemOpen
  \bibfield  {author} {\bibinfo {author} {\bibfnamefont {L.}~\bibnamefont
  {Ortenzi}}, \bibinfo {author} {\bibfnamefont {I.~I.}\ \bibnamefont {Mazin}},
  \bibinfo {author} {\bibfnamefont {P.}~\bibnamefont {Blaha}}, \ and\ \bibinfo
  {author} {\bibfnamefont {L.}~\bibnamefont {Boeri}},\ }\href@noop {}
  {\bibfield  {journal} {\bibinfo  {journal} {Phys. Rev. B}\ }\textbf {\bibinfo
  {volume} {86}},\ \bibinfo {pages} {064437} (\bibinfo {year}
  {2012})}\BibitemShut {NoStop}%
\bibitem [{\citenamefont {Kov\'{a}cs}\ \emph {et~al.}(2019)\citenamefont
  {Kov\'{a}cs}, \citenamefont {Tran}, \citenamefont {Blaha},\ and\
  \citenamefont {Madsen}}]{KovacsJCP19}%
  \BibitemOpen
  \bibfield  {author} {\bibinfo {author} {\bibfnamefont {P.}~\bibnamefont
  {Kov\'{a}cs}}, \bibinfo {author} {\bibfnamefont {F.}~\bibnamefont {Tran}},
  \bibinfo {author} {\bibfnamefont {P.}~\bibnamefont {Blaha}}, \ and\ \bibinfo
  {author} {\bibfnamefont {G.~K.~H.}\ \bibnamefont {Madsen}},\ }\href@noop {}
  {\bibfield  {journal} {\bibinfo  {journal} {J. Chem. Phys.}\ }\textbf
  {\bibinfo {volume} {150}},\ \bibinfo {pages} {164119} (\bibinfo {year}
  {2019})}\BibitemShut {NoStop}%
\bibitem [{\citenamefont {Yang}\ \emph {et~al.}(2019)\citenamefont {Yang},
  \citenamefont {Kitchaev},\ and\ \citenamefont {Ceder}}]{YangPRB19}%
  \BibitemOpen
  \bibfield  {author} {\bibinfo {author} {\bibfnamefont {J.~H.}\ \bibnamefont
  {Yang}}, \bibinfo {author} {\bibfnamefont {D.~A.}\ \bibnamefont {Kitchaev}},
  \ and\ \bibinfo {author} {\bibfnamefont {G.}~\bibnamefont {Ceder}},\
  }\href@noop {} {\bibfield  {journal} {\bibinfo  {journal} {Phys. Rev. B}\
  }\textbf {\bibinfo {volume} {100}},\ \bibinfo {pages} {035132} (\bibinfo
  {year} {2019})}\BibitemShut {NoStop}%
\bibitem [{\citenamefont {Paier}\ \emph {et~al.}(2006)\citenamefont {Paier},
  \citenamefont {Marsman}, \citenamefont {Hummer}, \citenamefont {Kresse},
  \citenamefont {Gerber},\ and\ \citenamefont {\'{A}ngy\'{a}n}}]{PaierJCP06}%
  \BibitemOpen
  \bibfield  {author} {\bibinfo {author} {\bibfnamefont {J.}~\bibnamefont
  {Paier}}, \bibinfo {author} {\bibfnamefont {M.}~\bibnamefont {Marsman}},
  \bibinfo {author} {\bibfnamefont {K.}~\bibnamefont {Hummer}}, \bibinfo
  {author} {\bibfnamefont {G.}~\bibnamefont {Kresse}}, \bibinfo {author}
  {\bibfnamefont {I.~C.}\ \bibnamefont {Gerber}}, \ and\ \bibinfo {author}
  {\bibfnamefont {J.~G.}\ \bibnamefont {\'{A}ngy\'{a}n}},\ }\href@noop {}
  {\bibfield  {journal} {\bibinfo  {journal} {J. Chem. Phys.}\ }\textbf
  {\bibinfo {volume} {124}},\ \bibinfo {pages} {154709} (\bibinfo {year}
  {2006})},\ \bibinfo {note} {\textbf{125}, 249901 (2006)}\BibitemShut
  {NoStop}%
\bibitem [{\citenamefont {Jang}\ and\ \citenamefont {Yu}(2011)}]{JangJM11}%
  \BibitemOpen
  \bibfield  {author} {\bibinfo {author} {\bibfnamefont {Y.-R.}\ \bibnamefont
  {Jang}}\ and\ \bibinfo {author} {\bibfnamefont {B.~D.}\ \bibnamefont {Yu}},\
  }\href@noop {} {\bibfield  {journal} {\bibinfo  {journal} {J. Magn.}\
  }\textbf {\bibinfo {volume} {16}},\ \bibinfo {pages} {201} (\bibinfo {year}
  {2011})}\BibitemShut {NoStop}%
\bibitem [{\citenamefont {Jang}\ and\ \citenamefont {Yu}(2012)}]{JangJPSJ12}%
  \BibitemOpen
  \bibfield  {author} {\bibinfo {author} {\bibfnamefont {Y.-R.}\ \bibnamefont
  {Jang}}\ and\ \bibinfo {author} {\bibfnamefont {B.~D.}\ \bibnamefont {Yu}},\
  }\href@noop {} {\bibfield  {journal} {\bibinfo  {journal} {J. Phys. Soc.
  Jpn.}\ }\textbf {\bibinfo {volume} {81}},\ \bibinfo {pages} {114715}
  (\bibinfo {year} {2012})}\BibitemShut {NoStop}%
\bibitem [{\citenamefont {Janthon}\ \emph {et~al.}(2014)\citenamefont
  {Janthon}, \citenamefont {Luo}, \citenamefont {Kozlov}, \citenamefont
  {Vi\~{n}es}, \citenamefont {Limtrakul}, \citenamefont {Truhlar},\ and\
  \citenamefont {Illas}}]{JanthonJCTC14}%
  \BibitemOpen
  \bibfield  {author} {\bibinfo {author} {\bibfnamefont {P.}~\bibnamefont
  {Janthon}}, \bibinfo {author} {\bibfnamefont {S.}~\bibnamefont {Luo}},
  \bibinfo {author} {\bibfnamefont {S.~M.}\ \bibnamefont {Kozlov}}, \bibinfo
  {author} {\bibfnamefont {F.}~\bibnamefont {Vi\~{n}es}}, \bibinfo {author}
  {\bibfnamefont {J.}~\bibnamefont {Limtrakul}}, \bibinfo {author}
  {\bibfnamefont {D.~G.}\ \bibnamefont {Truhlar}}, \ and\ \bibinfo {author}
  {\bibfnamefont {F.}~\bibnamefont {Illas}},\ }\href@noop {} {\bibfield
  {journal} {\bibinfo  {journal} {J. Chem. Theory Comput.}\ }\textbf {\bibinfo
  {volume} {10}},\ \bibinfo {pages} {3832} (\bibinfo {year}
  {2014})}\BibitemShut {NoStop}%
\bibitem [{\citenamefont {Gao}\ \emph {et~al.}(2016)\citenamefont {Gao},
  \citenamefont {Abtew}, \citenamefont {Cai}, \citenamefont {Sun},
  \citenamefont {Zhang},\ and\ \citenamefont {Zhang}}]{GaoSSC16}%
  \BibitemOpen
  \bibfield  {author} {\bibinfo {author} {\bibfnamefont {W.}~\bibnamefont
  {Gao}}, \bibinfo {author} {\bibfnamefont {T.~A.}\ \bibnamefont {Abtew}},
  \bibinfo {author} {\bibfnamefont {T.}~\bibnamefont {Cai}}, \bibinfo {author}
  {\bibfnamefont {Y.-Y.}\ \bibnamefont {Sun}}, \bibinfo {author} {\bibfnamefont
  {S.}~\bibnamefont {Zhang}}, \ and\ \bibinfo {author} {\bibfnamefont
  {P.}~\bibnamefont {Zhang}},\ }\href@noop {} {\bibfield  {journal} {\bibinfo
  {journal} {Solid State Commun.}\ }\textbf {\bibinfo {volume} {234-235}},\
  \bibinfo {pages} {10} (\bibinfo {year} {2016})}\BibitemShut {NoStop}%
\bibitem [{\citenamefont {Tran}\ \emph
  {et~al.}(2018{\natexlab{b}})\citenamefont {Tran}, \citenamefont {Ehsan},\
  and\ \citenamefont {Blaha}}]{TranPRM18}%
  \BibitemOpen
  \bibfield  {author} {\bibinfo {author} {\bibfnamefont {F.}~\bibnamefont
  {Tran}}, \bibinfo {author} {\bibfnamefont {S.}~\bibnamefont {Ehsan}}, \ and\
  \bibinfo {author} {\bibfnamefont {P.}~\bibnamefont {Blaha}},\ }\href@noop {}
  {\bibfield  {journal} {\bibinfo  {journal} {Phys. Rev. Materials}\ }\textbf
  {\bibinfo {volume} {2}},\ \bibinfo {pages} {023802} (\bibinfo {year}
  {2018}{\natexlab{b}})}\BibitemShut {NoStop}%
\bibitem [{\citenamefont {Jana}\ \emph {et~al.}(2020)\citenamefont {Jana},
  \citenamefont {Patra}, \citenamefont {Constantin},\ and\ \citenamefont
  {Samal}}]{JanaJCP20}%
  \BibitemOpen
  \bibfield  {author} {\bibinfo {author} {\bibfnamefont {S.}~\bibnamefont
  {Jana}}, \bibinfo {author} {\bibfnamefont {A.}~\bibnamefont {Patra}},
  \bibinfo {author} {\bibfnamefont {L.~A.}\ \bibnamefont {Constantin}}, \ and\
  \bibinfo {author} {\bibfnamefont {P.}~\bibnamefont {Samal}},\ }\href@noop {}
  {\bibfield  {journal} {\bibinfo  {journal} {J. Chem. Phys.}\ }\textbf
  {\bibinfo {volume} {152}},\ \bibinfo {pages} {044111} (\bibinfo {year}
  {2020})}\BibitemShut {NoStop}%
\bibitem [{\citenamefont {Heyd}\ \emph {et~al.}(2003)\citenamefont {Heyd},
  \citenamefont {Scuseria},\ and\ \citenamefont {Ernzerhof}}]{HeydJCP03}%
  \BibitemOpen
  \bibfield  {author} {\bibinfo {author} {\bibfnamefont {J.}~\bibnamefont
  {Heyd}}, \bibinfo {author} {\bibfnamefont {G.~E.}\ \bibnamefont {Scuseria}},
  \ and\ \bibinfo {author} {\bibfnamefont {M.}~\bibnamefont {Ernzerhof}},\
  }\href@noop {} {\bibfield  {journal} {\bibinfo  {journal} {J. Chem. Phys.}\
  }\textbf {\bibinfo {volume} {118}},\ \bibinfo {pages} {8207} (\bibinfo {year}
  {2003})},\ \bibinfo {note} {\textbf{124}, 219906 (2006)}\BibitemShut
  {NoStop}%
\bibitem [{\citenamefont {Krukau}\ \emph {et~al.}(2006)\citenamefont {Krukau},
  \citenamefont {Vydrov}, \citenamefont {Izmaylov},\ and\ \citenamefont
  {Scuseria}}]{KrukauJCP06}%
  \BibitemOpen
  \bibfield  {author} {\bibinfo {author} {\bibfnamefont {A.~V.}\ \bibnamefont
  {Krukau}}, \bibinfo {author} {\bibfnamefont {O.~A.}\ \bibnamefont {Vydrov}},
  \bibinfo {author} {\bibfnamefont {A.~F.}\ \bibnamefont {Izmaylov}}, \ and\
  \bibinfo {author} {\bibfnamefont {G.~E.}\ \bibnamefont {Scuseria}},\
  }\href@noop {} {\bibfield  {journal} {\bibinfo  {journal} {J. Chem. Phys.}\
  }\textbf {\bibinfo {volume} {125}},\ \bibinfo {pages} {224106} (\bibinfo
  {year} {2006})}\BibitemShut {NoStop}%
\bibitem [{\citenamefont {Kuisma}\ \emph {et~al.}(2010)\citenamefont {Kuisma},
  \citenamefont {Ojanen}, \citenamefont {Enkovaara},\ and\ \citenamefont
  {Rantala}}]{KuismaPRB10}%
  \BibitemOpen
  \bibfield  {author} {\bibinfo {author} {\bibfnamefont {M.}~\bibnamefont
  {Kuisma}}, \bibinfo {author} {\bibfnamefont {J.}~\bibnamefont {Ojanen}},
  \bibinfo {author} {\bibfnamefont {J.}~\bibnamefont {Enkovaara}}, \ and\
  \bibinfo {author} {\bibfnamefont {T.~T.}\ \bibnamefont {Rantala}},\
  }\href@noop {} {\bibfield  {journal} {\bibinfo  {journal} {Phys. Rev. B}\
  }\textbf {\bibinfo {volume} {82}},\ \bibinfo {pages} {115106} (\bibinfo
  {year} {2010})}\BibitemShut {NoStop}%
\bibitem [{\citenamefont {Karolewski}\ \emph {et~al.}(2009)\citenamefont
  {Karolewski}, \citenamefont {Armiento},\ and\ \citenamefont
  {K\"{u}mmel}}]{KarolewskiJCTC09}%
  \BibitemOpen
  \bibfield  {author} {\bibinfo {author} {\bibfnamefont {A.}~\bibnamefont
  {Karolewski}}, \bibinfo {author} {\bibfnamefont {R.}~\bibnamefont
  {Armiento}}, \ and\ \bibinfo {author} {\bibfnamefont {S.}~\bibnamefont
  {K\"{u}mmel}},\ }\href@noop {} {\bibfield  {journal} {\bibinfo  {journal} {J.
  Chem. Theory Comput.}\ }\textbf {\bibinfo {volume} {5}},\ \bibinfo {pages}
  {712} (\bibinfo {year} {2009})}\BibitemShut {NoStop}%
\bibitem [{\citenamefont {Gaiduk}\ and\ \citenamefont
  {Staroverov}(2009)}]{GaidukJCP09}%
  \BibitemOpen
  \bibfield  {author} {\bibinfo {author} {\bibfnamefont {A.~P.}\ \bibnamefont
  {Gaiduk}}\ and\ \bibinfo {author} {\bibfnamefont {V.~N.}\ \bibnamefont
  {Staroverov}},\ }\href@noop {} {\bibfield  {journal} {\bibinfo  {journal} {J.
  Chem. Phys.}\ }\textbf {\bibinfo {volume} {131}},\ \bibinfo {pages} {044107}
  (\bibinfo {year} {2009})}\BibitemShut {NoStop}%
\bibitem [{\citenamefont {Svane}\ and\ \citenamefont
  {Gunnarsson}(1990)}]{SvanePRL90}%
  \BibitemOpen
  \bibfield  {author} {\bibinfo {author} {\bibfnamefont {A.}~\bibnamefont
  {Svane}}\ and\ \bibinfo {author} {\bibfnamefont {O.}~\bibnamefont
  {Gunnarsson}},\ }\href@noop {} {\bibfield  {journal} {\bibinfo  {journal}
  {Phys. Rev. Lett.}\ }\textbf {\bibinfo {volume} {65}},\ \bibinfo {pages}
  {1148} (\bibinfo {year} {1990})}\BibitemShut {NoStop}%
\bibitem [{\citenamefont {Tran}\ \emph {et~al.}(2006)\citenamefont {Tran},
  \citenamefont {Blaha}, \citenamefont {Schwarz},\ and\ \citenamefont
  {Nov\'{a}k}}]{TranPRB06}%
  \BibitemOpen
  \bibfield  {author} {\bibinfo {author} {\bibfnamefont {F.}~\bibnamefont
  {Tran}}, \bibinfo {author} {\bibfnamefont {P.}~\bibnamefont {Blaha}},
  \bibinfo {author} {\bibfnamefont {K.}~\bibnamefont {Schwarz}}, \ and\
  \bibinfo {author} {\bibfnamefont {P.}~\bibnamefont {Nov\'{a}k}},\ }\href@noop
  {} {\bibfield  {journal} {\bibinfo  {journal} {Phys. Rev. B}\ }\textbf
  {\bibinfo {volume} {74}},\ \bibinfo {pages} {155108} (\bibinfo {year}
  {2006})}\BibitemShut {NoStop}%
\bibitem [{\citenamefont {Radwanski}\ and\ \citenamefont
  {Ropka}(2008)}]{RadwanskiPB08}%
  \BibitemOpen
  \bibfield  {author} {\bibinfo {author} {\bibfnamefont {R.~J.}\ \bibnamefont
  {Radwanski}}\ and\ \bibinfo {author} {\bibfnamefont {Z.}~\bibnamefont
  {Ropka}},\ }\href@noop {} {\bibfield  {journal} {\bibinfo  {journal} {Physica
  B}\ }\textbf {\bibinfo {volume} {403}},\ \bibinfo {pages} {1453} (\bibinfo
  {year} {2008})}\BibitemShut {NoStop}%
\bibitem [{\citenamefont {Schr\"{o}n}\ and\ \citenamefont
  {Bechstedt}(2013)}]{SchronJPCM13}%
  \BibitemOpen
  \bibfield  {author} {\bibinfo {author} {\bibfnamefont {A.}~\bibnamefont
  {Schr\"{o}n}}\ and\ \bibinfo {author} {\bibfnamefont {F.}~\bibnamefont
  {Bechstedt}},\ }\href@noop {} {\bibfield  {journal} {\bibinfo  {journal} {J.
  Phys.: Condens. Matter}\ }\textbf {\bibinfo {volume} {25}},\ \bibinfo {pages}
  {486002} (\bibinfo {year} {2013})}\BibitemShut {NoStop}%
\bibitem [{\citenamefont {Solovyev}\ \emph {et~al.}(1998)\citenamefont
  {Solovyev}, \citenamefont {Liechtenstein},\ and\ \citenamefont
  {Terakura}}]{SolovyevPRL98}%
  \BibitemOpen
  \bibfield  {author} {\bibinfo {author} {\bibfnamefont {I.~V.}\ \bibnamefont
  {Solovyev}}, \bibinfo {author} {\bibfnamefont {A.~I.}\ \bibnamefont
  {Liechtenstein}}, \ and\ \bibinfo {author} {\bibfnamefont {K.}~\bibnamefont
  {Terakura}},\ }\href@noop {} {\bibfield  {journal} {\bibinfo  {journal}
  {Phys. Rev. Lett.}\ }\textbf {\bibinfo {volume} {80}},\ \bibinfo {pages}
  {5758} (\bibinfo {year} {1998})}\BibitemShut {NoStop}%
\bibitem [{\citenamefont {Shishidou}\ and\ \citenamefont
  {Jo}(1998)}]{ShishidouJPSJ98}%
  \BibitemOpen
  \bibfield  {author} {\bibinfo {author} {\bibfnamefont {T.}~\bibnamefont
  {Shishidou}}\ and\ \bibinfo {author} {\bibfnamefont {T.}~\bibnamefont {Jo}},\
  }\href@noop {} {\bibfield  {journal} {\bibinfo  {journal} {J. Phys. Soc.
  Jpn.}\ }\textbf {\bibinfo {volume} {67}},\ \bibinfo {pages} {2637} (\bibinfo
  {year} {1998})}\BibitemShut {NoStop}%
\bibitem [{\citenamefont {Neubeck}\ \emph {et~al.}(2001)\citenamefont
  {Neubeck}, \citenamefont {Vettier}, \citenamefont {de~Bergevin},
  \citenamefont {Yakhou}, \citenamefont {Mannix}, \citenamefont {Ranno},\ and\
  \citenamefont {Chatterji}}]{NeubeckJPCS01}%
  \BibitemOpen
  \bibfield  {author} {\bibinfo {author} {\bibfnamefont {W.}~\bibnamefont
  {Neubeck}}, \bibinfo {author} {\bibfnamefont {C.}~\bibnamefont {Vettier}},
  \bibinfo {author} {\bibfnamefont {F.}~\bibnamefont {de~Bergevin}}, \bibinfo
  {author} {\bibfnamefont {F.}~\bibnamefont {Yakhou}}, \bibinfo {author}
  {\bibfnamefont {D.}~\bibnamefont {Mannix}}, \bibinfo {author} {\bibfnamefont
  {L.}~\bibnamefont {Ranno}}, \ and\ \bibinfo {author} {\bibfnamefont
  {T.}~\bibnamefont {Chatterji}},\ }\href@noop {} {\bibfield  {journal}
  {\bibinfo  {journal} {J. Phys. Chem. Solids}\ }\textbf {\bibinfo {volume}
  {62}},\ \bibinfo {pages} {2173} (\bibinfo {year} {2001})}\BibitemShut
  {NoStop}%
\bibitem [{\citenamefont {Jauch}\ and\ \citenamefont
  {Reehuis}(2002)}]{JauchPRB02}%
  \BibitemOpen
  \bibfield  {author} {\bibinfo {author} {\bibfnamefont {W.}~\bibnamefont
  {Jauch}}\ and\ \bibinfo {author} {\bibfnamefont {M.}~\bibnamefont
  {Reehuis}},\ }\href@noop {} {\bibfield  {journal} {\bibinfo  {journal} {Phys.
  Rev. B}\ }\textbf {\bibinfo {volume} {65}},\ \bibinfo {pages} {125111}
  (\bibinfo {year} {2002})}\BibitemShut {NoStop}%
\bibitem [{\citenamefont {Ghiringhelli}\ \emph {et~al.}(2002)\citenamefont
  {Ghiringhelli}, \citenamefont {Tjeng}, \citenamefont {Tanaka}, \citenamefont
  {Tjernberg}, \citenamefont {Mizokawa}, \citenamefont {de~Boer},\ and\
  \citenamefont {Brookes}}]{GhiringhelliPRB02}%
  \BibitemOpen
  \bibfield  {author} {\bibinfo {author} {\bibfnamefont {G.}~\bibnamefont
  {Ghiringhelli}}, \bibinfo {author} {\bibfnamefont {L.~H.}\ \bibnamefont
  {Tjeng}}, \bibinfo {author} {\bibfnamefont {A.}~\bibnamefont {Tanaka}},
  \bibinfo {author} {\bibfnamefont {O.}~\bibnamefont {Tjernberg}}, \bibinfo
  {author} {\bibfnamefont {T.}~\bibnamefont {Mizokawa}}, \bibinfo {author}
  {\bibfnamefont {J.~L.}\ \bibnamefont {de~Boer}}, \ and\ \bibinfo {author}
  {\bibfnamefont {N.~B.}\ \bibnamefont {Brookes}},\ }\href@noop {} {\bibfield
  {journal} {\bibinfo  {journal} {Phys. Rev. B}\ }\textbf {\bibinfo {volume}
  {66}},\ \bibinfo {pages} {075101} (\bibinfo {year} {2002})}\BibitemShut
  {NoStop}%
\bibitem [{\citenamefont {Radwanski}\ and\ \citenamefont
  {Ropka}(2004)}]{RadwanskiPB04}%
  \BibitemOpen
  \bibfield  {author} {\bibinfo {author} {\bibfnamefont {R.~J.}\ \bibnamefont
  {Radwanski}}\ and\ \bibinfo {author} {\bibfnamefont {Z.}~\bibnamefont
  {Ropka}},\ }\href@noop {} {\bibfield  {journal} {\bibinfo  {journal} {Physica
  B}\ }\textbf {\bibinfo {volume} {345}},\ \bibinfo {pages} {107} (\bibinfo
  {year} {2004})}\BibitemShut {NoStop}%
\bibitem [{\citenamefont {Boussendel}\ \emph {et~al.}(2010)\citenamefont
  {Boussendel}, \citenamefont {Baadji}, \citenamefont {Haroun}, \citenamefont
  {Dreyss\'e},\ and\ \citenamefont {Alouani}}]{BoussendelPRB10}%
  \BibitemOpen
  \bibfield  {author} {\bibinfo {author} {\bibfnamefont {A.}~\bibnamefont
  {Boussendel}}, \bibinfo {author} {\bibfnamefont {N.}~\bibnamefont {Baadji}},
  \bibinfo {author} {\bibfnamefont {A.}~\bibnamefont {Haroun}}, \bibinfo
  {author} {\bibfnamefont {H.}~\bibnamefont {Dreyss\'e}}, \ and\ \bibinfo
  {author} {\bibfnamefont {M.}~\bibnamefont {Alouani}},\ }\href@noop {}
  {\bibfield  {journal} {\bibinfo  {journal} {Phys. Rev. B}\ }\textbf {\bibinfo
  {volume} {81}},\ \bibinfo {pages} {184432} (\bibinfo {year}
  {2010})}\BibitemShut {NoStop}%
\bibitem [{\citenamefont {Fernandez}\ \emph {et~al.}(1998)\citenamefont
  {Fernandez}, \citenamefont {Vettier}, \citenamefont {de~Bergevin},
  \citenamefont {Giles},\ and\ \citenamefont {Neubeck}}]{FernandezPRB98}%
  \BibitemOpen
  \bibfield  {author} {\bibinfo {author} {\bibfnamefont {V.}~\bibnamefont
  {Fernandez}}, \bibinfo {author} {\bibfnamefont {C.}~\bibnamefont {Vettier}},
  \bibinfo {author} {\bibfnamefont {F.}~\bibnamefont {de~Bergevin}}, \bibinfo
  {author} {\bibfnamefont {C.}~\bibnamefont {Giles}}, \ and\ \bibinfo {author}
  {\bibfnamefont {W.}~\bibnamefont {Neubeck}},\ }\href@noop {} {\bibfield
  {journal} {\bibinfo  {journal} {Phys. Rev. B}\ }\textbf {\bibinfo {volume}
  {57}},\ \bibinfo {pages} {7870} (\bibinfo {year} {1998})}\BibitemShut
  {NoStop}%
\bibitem [{\citenamefont {Cheetham}\ and\ \citenamefont
  {Hope}(1983)}]{CheethamPRB83}%
  \BibitemOpen
  \bibfield  {author} {\bibinfo {author} {\bibfnamefont {A.~K.}\ \bibnamefont
  {Cheetham}}\ and\ \bibinfo {author} {\bibfnamefont {D.~A.~O.}\ \bibnamefont
  {Hope}},\ }\href@noop {} {\bibfield  {journal} {\bibinfo  {journal} {Phys.
  Rev. B}\ }\textbf {\bibinfo {volume} {27}},\ \bibinfo {pages} {6964}
  (\bibinfo {year} {1983})}\BibitemShut {NoStop}%
\bibitem [{\citenamefont {Roth}(1958)}]{RothPR58}%
  \BibitemOpen
  \bibfield  {author} {\bibinfo {author} {\bibfnamefont {W.~L.}\ \bibnamefont
  {Roth}},\ }\href@noop {} {\bibfield  {journal} {\bibinfo  {journal} {Phys.
  Rev.}\ }\textbf {\bibinfo {volume} {110}},\ \bibinfo {pages} {1333} (\bibinfo
  {year} {1958})}\BibitemShut {NoStop}%
\bibitem [{\citenamefont {Battle}\ and\ \citenamefont
  {Cheetham}(1979)}]{BattleJPC79}%
  \BibitemOpen
  \bibfield  {author} {\bibinfo {author} {\bibfnamefont {P.~D.}\ \bibnamefont
  {Battle}}\ and\ \bibinfo {author} {\bibfnamefont {A.~K.}\ \bibnamefont
  {Cheetham}},\ }\href@noop {} {\bibfield  {journal} {\bibinfo  {journal} {J.
  Phys. C: Solid State Phys.}\ }\textbf {\bibinfo {volume} {12}},\ \bibinfo
  {pages} {337} (\bibinfo {year} {1979})}\BibitemShut {NoStop}%
\bibitem [{\citenamefont {Fjellv\r{a}g}\ \emph {et~al.}(1996)\citenamefont
  {Fjellv\r{a}g}, \citenamefont {Gr{\o}nvold}, \citenamefont {St{\o}len},\ and\
  \citenamefont {Hauback}}]{FjellvagJSSC96}%
  \BibitemOpen
  \bibfield  {author} {\bibinfo {author} {\bibfnamefont {H.}~\bibnamefont
  {Fjellv\r{a}g}}, \bibinfo {author} {\bibfnamefont {F.}~\bibnamefont
  {Gr{\o}nvold}}, \bibinfo {author} {\bibfnamefont {S.}~\bibnamefont
  {St{\o}len}}, \ and\ \bibinfo {author} {\bibfnamefont {B.}~\bibnamefont
  {Hauback}},\ }\href@noop {} {\bibfield  {journal} {\bibinfo  {journal} {J.
  Solid State Chem.}\ }\textbf {\bibinfo {volume} {124}},\ \bibinfo {pages}
  {52} (\bibinfo {year} {1996})}\BibitemShut {NoStop}%
\bibitem [{\citenamefont {Khan}\ and\ \citenamefont
  {Erickson}(1970)}]{KhanPRB70}%
  \BibitemOpen
  \bibfield  {author} {\bibinfo {author} {\bibfnamefont {D.~C.}\ \bibnamefont
  {Khan}}\ and\ \bibinfo {author} {\bibfnamefont {R.~A.}\ \bibnamefont
  {Erickson}},\ }\href@noop {} {\bibfield  {journal} {\bibinfo  {journal}
  {Phys. Rev. B}\ }\textbf {\bibinfo {volume} {1}},\ \bibinfo {pages} {2243}
  (\bibinfo {year} {1970})}\BibitemShut {NoStop}%
\bibitem [{\citenamefont {Herrmann-Ronzaud}\ \emph {et~al.}(1978)\citenamefont
  {Herrmann-Ronzaud}, \citenamefont {Burlet},\ and\ \citenamefont
  {Rossat-Mignod}}]{HerrmannRonzaudJPC78}%
  \BibitemOpen
  \bibfield  {author} {\bibinfo {author} {\bibfnamefont {D.}~\bibnamefont
  {Herrmann-Ronzaud}}, \bibinfo {author} {\bibfnamefont {P.}~\bibnamefont
  {Burlet}}, \ and\ \bibinfo {author} {\bibfnamefont {J.}~\bibnamefont
  {Rossat-Mignod}},\ }\href@noop {} {\bibfield  {journal} {\bibinfo  {journal}
  {J. Phys. C: Solid State Phys.}\ }\textbf {\bibinfo {volume} {11}},\ \bibinfo
  {pages} {2123} (\bibinfo {year} {1978})}\BibitemShut {NoStop}%
\bibitem [{\citenamefont {Jauch}\ \emph {et~al.}(2001)\citenamefont {Jauch},
  \citenamefont {Reehuis}, \citenamefont {Bleif}, \citenamefont {Kubanek},\
  and\ \citenamefont {Pattison}}]{JauchPRB01}%
  \BibitemOpen
  \bibfield  {author} {\bibinfo {author} {\bibfnamefont {W.}~\bibnamefont
  {Jauch}}, \bibinfo {author} {\bibfnamefont {M.}~\bibnamefont {Reehuis}},
  \bibinfo {author} {\bibfnamefont {H.~J.}\ \bibnamefont {Bleif}}, \bibinfo
  {author} {\bibfnamefont {F.}~\bibnamefont {Kubanek}}, \ and\ \bibinfo
  {author} {\bibfnamefont {P.}~\bibnamefont {Pattison}},\ }\href@noop {}
  {\bibfield  {journal} {\bibinfo  {journal} {Phys. Rev. B}\ }\textbf {\bibinfo
  {volume} {64}},\ \bibinfo {pages} {052102} (\bibinfo {year}
  {2001})}\BibitemShut {NoStop}%
\bibitem [{\citenamefont {Neubeck}\ \emph {et~al.}(1999)\citenamefont
  {Neubeck}, \citenamefont {Vettier}, \citenamefont {Fernandez}, \citenamefont
  {de~Bergevin},\ and\ \citenamefont {Giles}}]{NeubeckJAP99}%
  \BibitemOpen
  \bibfield  {author} {\bibinfo {author} {\bibfnamefont {W.}~\bibnamefont
  {Neubeck}}, \bibinfo {author} {\bibfnamefont {C.}~\bibnamefont {Vettier}},
  \bibinfo {author} {\bibfnamefont {V.}~\bibnamefont {Fernandez}}, \bibinfo
  {author} {\bibfnamefont {F.}~\bibnamefont {de~Bergevin}}, \ and\ \bibinfo
  {author} {\bibfnamefont {C.}~\bibnamefont {Giles}},\ }\href@noop {}
  {\bibfield  {journal} {\bibinfo  {journal} {J. Appl. Phys.}\ }\textbf
  {\bibinfo {volume} {85}},\ \bibinfo {pages} {4847} (\bibinfo {year}
  {1999})}\BibitemShut {NoStop}%
\bibitem [{\citenamefont {Forsyth}\ \emph {et~al.}(1988)\citenamefont
  {Forsyth}, \citenamefont {Brown},\ and\ \citenamefont
  {Wanklyn}}]{ForsythJPC88}%
  \BibitemOpen
  \bibfield  {author} {\bibinfo {author} {\bibfnamefont {J.~B.}\ \bibnamefont
  {Forsyth}}, \bibinfo {author} {\bibfnamefont {P.~J.}\ \bibnamefont {Brown}},
  \ and\ \bibinfo {author} {\bibfnamefont {B.~M.}\ \bibnamefont {Wanklyn}},\
  }\href@noop {} {\bibfield  {journal} {\bibinfo  {journal} {J. Phys. C: Solid
  State Phys.}\ }\textbf {\bibinfo {volume} {21}},\ \bibinfo {pages} {2917}
  (\bibinfo {year} {1988})}\BibitemShut {NoStop}%
\bibitem [{\citenamefont {Golosova}\ \emph {et~al.}(2017)\citenamefont
  {Golosova}, \citenamefont {Kozlenko}, \citenamefont {Kichanov}, \citenamefont
  {Lukin}, \citenamefont {Liermann}, \citenamefont {Glazyrin},\ and\
  \citenamefont {Savenko}}]{GolosovaJAC17}%
  \BibitemOpen
  \bibfield  {author} {\bibinfo {author} {\bibfnamefont {N.~O.}\ \bibnamefont
  {Golosova}}, \bibinfo {author} {\bibfnamefont {D.~P.}\ \bibnamefont
  {Kozlenko}}, \bibinfo {author} {\bibfnamefont {S.~E.}\ \bibnamefont
  {Kichanov}}, \bibinfo {author} {\bibfnamefont {E.~V.}\ \bibnamefont {Lukin}},
  \bibinfo {author} {\bibfnamefont {H.-P.}\ \bibnamefont {Liermann}}, \bibinfo
  {author} {\bibfnamefont {K.~V.}\ \bibnamefont {Glazyrin}}, \ and\ \bibinfo
  {author} {\bibfnamefont {B.~N.}\ \bibnamefont {Savenko}},\ }\href@noop {}
  {\bibfield  {journal} {\bibinfo  {journal} {J. Alloys Compd.}\ }\textbf
  {\bibinfo {volume} {722}},\ \bibinfo {pages} {593} (\bibinfo {year}
  {2017})}\BibitemShut {NoStop}%
\bibitem [{\citenamefont {Brown}\ \emph {et~al.}(2002)\citenamefont {Brown},
  \citenamefont {Forsyth}, \citenamefont {Leli\`{e}vre-Berna},\ and\
  \citenamefont {Tasset}}]{BrownJPCM02}%
  \BibitemOpen
  \bibfield  {author} {\bibinfo {author} {\bibfnamefont {P.~J.}\ \bibnamefont
  {Brown}}, \bibinfo {author} {\bibfnamefont {J.~B.}\ \bibnamefont {Forsyth}},
  \bibinfo {author} {\bibfnamefont {E.}~\bibnamefont {Leli\`{e}vre-Berna}}, \
  and\ \bibinfo {author} {\bibfnamefont {F.}~\bibnamefont {Tasset}},\
  }\href@noop {} {\bibfield  {journal} {\bibinfo  {journal} {J. Phys.: Condens.
  Matter}\ }\textbf {\bibinfo {volume} {14}},\ \bibinfo {pages} {1957}
  (\bibinfo {year} {2002})}\BibitemShut {NoStop}%
\bibitem [{\citenamefont {Corliss}\ \emph {et~al.}(1965)\citenamefont
  {Corliss}, \citenamefont {Hastings}, \citenamefont {Nathans},\ and\
  \citenamefont {Shirane}}]{CorlissJAP65}%
  \BibitemOpen
  \bibfield  {author} {\bibinfo {author} {\bibfnamefont {L.~M.}\ \bibnamefont
  {Corliss}}, \bibinfo {author} {\bibfnamefont {J.~M.}\ \bibnamefont
  {Hastings}}, \bibinfo {author} {\bibfnamefont {R.}~\bibnamefont {Nathans}}, \
  and\ \bibinfo {author} {\bibfnamefont {G.}~\bibnamefont {Shirane}},\
  }\href@noop {} {\bibfield  {journal} {\bibinfo  {journal} {J. Appl. Phys.}\
  }\textbf {\bibinfo {volume} {36}},\ \bibinfo {pages} {1099} (\bibinfo {year}
  {1965})}\BibitemShut {NoStop}%
\bibitem [{\citenamefont {Baron}\ \emph {et~al.}(2005)\citenamefont {Baron},
  \citenamefont {Gutzmer}, \citenamefont {Rundl\"{o}f},\ and\ \citenamefont
  {Tellgren}}]{BaronSSS05}%
  \BibitemOpen
  \bibfield  {author} {\bibinfo {author} {\bibfnamefont {V.}~\bibnamefont
  {Baron}}, \bibinfo {author} {\bibfnamefont {J.}~\bibnamefont {Gutzmer}},
  \bibinfo {author} {\bibfnamefont {H.}~\bibnamefont {Rundl\"{o}f}}, \ and\
  \bibinfo {author} {\bibfnamefont {R.}~\bibnamefont {Tellgren}},\ }\href@noop
  {} {\bibfield  {journal} {\bibinfo  {journal} {Solid State Sci.}\ }\textbf
  {\bibinfo {volume} {7}},\ \bibinfo {pages} {753} (\bibinfo {year}
  {2005})}\BibitemShut {NoStop}%
\bibitem [{\citenamefont {Hill}\ \emph {et~al.}(2008)\citenamefont {Hill},
  \citenamefont {Jiao}, \citenamefont {Bruce}, \citenamefont {Harrison},
  \citenamefont {Kockelmann},\ and\ \citenamefont {Ritter}}]{HillCM08}%
  \BibitemOpen
  \bibfield  {author} {\bibinfo {author} {\bibfnamefont {A.~H.}\ \bibnamefont
  {Hill}}, \bibinfo {author} {\bibfnamefont {F.}~\bibnamefont {Jiao}}, \bibinfo
  {author} {\bibfnamefont {P.~G.}\ \bibnamefont {Bruce}}, \bibinfo {author}
  {\bibfnamefont {A.}~\bibnamefont {Harrison}}, \bibinfo {author}
  {\bibfnamefont {W.}~\bibnamefont {Kockelmann}}, \ and\ \bibinfo {author}
  {\bibfnamefont {C.}~\bibnamefont {Ritter}},\ }\href@noop {} {\bibfield
  {journal} {\bibinfo  {journal} {Chem. Mater.}\ }\textbf {\bibinfo {volume}
  {20}},\ \bibinfo {pages} {4891} (\bibinfo {year} {2008})}\BibitemShut
  {NoStop}%
\bibitem [{\citenamefont {Takei}\ \emph {et~al.}(1963)\citenamefont {Takei},
  \citenamefont {Cox},\ and\ \citenamefont {Shirane}}]{TakeiPR63}%
  \BibitemOpen
  \bibfield  {author} {\bibinfo {author} {\bibfnamefont {W.~J.}\ \bibnamefont
  {Takei}}, \bibinfo {author} {\bibfnamefont {D.~E.}\ \bibnamefont {Cox}}, \
  and\ \bibinfo {author} {\bibfnamefont {G.}~\bibnamefont {Shirane}},\
  }\href@noop {} {\bibfield  {journal} {\bibinfo  {journal} {Phys. Rev.}\
  }\textbf {\bibinfo {volume} {129}},\ \bibinfo {pages} {2008} (\bibinfo {year}
  {1963})}\BibitemShut {NoStop}%
\bibitem [{\citenamefont {Holseth}\ \emph {et~al.}(1970)\citenamefont
  {Holseth}, \citenamefont {Kjekshus},\ and\ \citenamefont
  {Andresen}}]{HolsethACS70}%
  \BibitemOpen
  \bibfield  {author} {\bibinfo {author} {\bibfnamefont {H.}~\bibnamefont
  {Holseth}}, \bibinfo {author} {\bibfnamefont {A.}~\bibnamefont {Kjekshus}}, \
  and\ \bibinfo {author} {\bibfnamefont {A.~F.}\ \bibnamefont {Andresen}},\
  }\href@noop {} {\bibfield  {journal} {\bibinfo  {journal} {Acta Chem.
  Scand.}\ }\textbf {\bibinfo {volume} {24}},\ \bibinfo {pages} {3309}
  (\bibinfo {year} {1970})}\BibitemShut {NoStop}%
\bibitem [{\citenamefont {Kuhn}\ \emph {et~al.}(2013)\citenamefont {Kuhn},
  \citenamefont {Mankovsky}, \citenamefont {Ebert}, \citenamefont {Regus},\
  and\ \citenamefont {Bensch}}]{KuhnPRB13}%
  \BibitemOpen
  \bibfield  {author} {\bibinfo {author} {\bibfnamefont {G.}~\bibnamefont
  {Kuhn}}, \bibinfo {author} {\bibfnamefont {S.}~\bibnamefont {Mankovsky}},
  \bibinfo {author} {\bibfnamefont {H.}~\bibnamefont {Ebert}}, \bibinfo
  {author} {\bibfnamefont {M.}~\bibnamefont {Regus}}, \ and\ \bibinfo {author}
  {\bibfnamefont {W.}~\bibnamefont {Bensch}},\ }\href@noop {} {\bibfield
  {journal} {\bibinfo  {journal} {Phys. Rev. B}\ }\textbf {\bibinfo {volume}
  {87}},\ \bibinfo {pages} {085113} (\bibinfo {year} {2013})}\BibitemShut
  {NoStop}%
\bibitem [{\citenamefont {Czy\ifmmode~\dot{z}\else \.{z}\fi{}yk}\ and\
  \citenamefont {Sawatzky}(1994)}]{CzyzykPRB94}%
  \BibitemOpen
  \bibfield  {author} {\bibinfo {author} {\bibfnamefont {M.~T.}\ \bibnamefont
  {Czy\ifmmode~\dot{z}\else \.{z}\fi{}yk}}\ and\ \bibinfo {author}
  {\bibfnamefont {G.~A.}\ \bibnamefont {Sawatzky}},\ }\href@noop {} {\bibfield
  {journal} {\bibinfo  {journal} {Phys. Rev. B}\ }\textbf {\bibinfo {volume}
  {49}},\ \bibinfo {pages} {14211} (\bibinfo {year} {1994})}\BibitemShut
  {NoStop}%
\bibitem [{\citenamefont {Petukhov}\ \emph {et~al.}(2003)\citenamefont
  {Petukhov}, \citenamefont {Mazin}, \citenamefont {Chioncel},\ and\
  \citenamefont {Lichtenstein}}]{PetukhovPRB03}%
  \BibitemOpen
  \bibfield  {author} {\bibinfo {author} {\bibfnamefont {A.~G.}\ \bibnamefont
  {Petukhov}}, \bibinfo {author} {\bibfnamefont {I.~I.}\ \bibnamefont {Mazin}},
  \bibinfo {author} {\bibfnamefont {L.}~\bibnamefont {Chioncel}}, \ and\
  \bibinfo {author} {\bibfnamefont {A.~I.}\ \bibnamefont {Lichtenstein}},\
  }\href@noop {} {\bibfield  {journal} {\bibinfo  {journal} {Phys. Rev. B}\
  }\textbf {\bibinfo {volume} {67}},\ \bibinfo {pages} {153106} (\bibinfo
  {year} {2003})}\BibitemShut {NoStop}%
\bibitem [{\citenamefont {Mohn}\ \emph {et~al.}(2001)\citenamefont {Mohn},
  \citenamefont {Persson}, \citenamefont {Blaha}, \citenamefont {Schwarz},
  \citenamefont {Nov\'ak},\ and\ \citenamefont {Eschrig}}]{MohnPRL01}%
  \BibitemOpen
  \bibfield  {author} {\bibinfo {author} {\bibfnamefont {P.}~\bibnamefont
  {Mohn}}, \bibinfo {author} {\bibfnamefont {C.}~\bibnamefont {Persson}},
  \bibinfo {author} {\bibfnamefont {P.}~\bibnamefont {Blaha}}, \bibinfo
  {author} {\bibfnamefont {K.}~\bibnamefont {Schwarz}}, \bibinfo {author}
  {\bibfnamefont {P.}~\bibnamefont {Nov\'ak}}, \ and\ \bibinfo {author}
  {\bibfnamefont {H.}~\bibnamefont {Eschrig}},\ }\href@noop {} {\bibfield
  {journal} {\bibinfo  {journal} {Phys. Rev. Lett.}\ }\textbf {\bibinfo
  {volume} {87}},\ \bibinfo {pages} {196401} (\bibinfo {year}
  {2001})}\BibitemShut {NoStop}%
\bibitem [{\citenamefont {Thomas}(1927)}]{ThomasPCPS27}%
  \BibitemOpen
  \bibfield  {author} {\bibinfo {author} {\bibfnamefont {L.~H.}\ \bibnamefont
  {Thomas}},\ }\href@noop {} {\bibfield  {journal} {\bibinfo  {journal} {Proc.
  Cambridge Philos. Soc.}\ }\textbf {\bibinfo {volume} {23}},\ \bibinfo {pages}
  {542} (\bibinfo {year} {1927})}\BibitemShut {NoStop}%
\bibitem [{\citenamefont {Fermi}(1927)}]{FermiRANL27}%
  \BibitemOpen
  \bibfield  {author} {\bibinfo {author} {\bibfnamefont {E.}~\bibnamefont
  {Fermi}},\ }\href@noop {} {\bibfield  {journal} {\bibinfo  {journal} {Rend.
  Accad. Naz. Lincei}\ }\textbf {\bibinfo {volume} {6}},\ \bibinfo {pages}
  {602} (\bibinfo {year} {1927})}\BibitemShut {NoStop}%
\bibitem [{\citenamefont {von Weizs\"{a}cker}(1935)}]{vonWeizsackerZP35}%
  \BibitemOpen
  \bibfield  {author} {\bibinfo {author} {\bibfnamefont {C.~F.}\ \bibnamefont
  {von Weizs\"{a}cker}},\ }\href@noop {} {\bibfield  {journal} {\bibinfo
  {journal} {Z. Phys.}\ }\textbf {\bibinfo {volume} {96}},\ \bibinfo {pages}
  {431} (\bibinfo {year} {1935})}\BibitemShut {NoStop}%
\bibitem [{\citenamefont {Kotani}(1998)}]{KotaniJPCM98}%
  \BibitemOpen
  \bibfield  {author} {\bibinfo {author} {\bibfnamefont {T.}~\bibnamefont
  {Kotani}},\ }\href@noop {} {\bibfield  {journal} {\bibinfo  {journal} {J.
  Phys.: Condens. Matter}\ }\textbf {\bibinfo {volume} {10}},\ \bibinfo {pages}
  {9241} (\bibinfo {year} {1998})}\BibitemShut {NoStop}%
\bibitem [{\citenamefont {Schnell}\ \emph {et~al.}(2003)\citenamefont
  {Schnell}, \citenamefont {Czycholl},\ and\ \citenamefont
  {Albers}}]{SchnellPRB03}%
  \BibitemOpen
  \bibfield  {author} {\bibinfo {author} {\bibfnamefont {I.}~\bibnamefont
  {Schnell}}, \bibinfo {author} {\bibfnamefont {G.}~\bibnamefont {Czycholl}}, \
  and\ \bibinfo {author} {\bibfnamefont {R.~C.}\ \bibnamefont {Albers}},\
  }\href@noop {} {\bibfield  {journal} {\bibinfo  {journal} {Phys. Rev. B}\
  }\textbf {\bibinfo {volume} {68}},\ \bibinfo {pages} {245102} (\bibinfo
  {year} {2003})}\BibitemShut {NoStop}%
\bibitem [{\citenamefont {\"{J}emmer}\ and\ \citenamefont
  {Knowles}(1995)}]{JemmerPRA95}%
  \BibitemOpen
  \bibfield  {author} {\bibinfo {author} {\bibfnamefont {P.}~\bibnamefont
  {\"{J}emmer}}\ and\ \bibinfo {author} {\bibfnamefont {P.~J.}\ \bibnamefont
  {Knowles}},\ }\href@noop {} {\bibfield  {journal} {\bibinfo  {journal} {Phys.
  Rev. A}\ }\textbf {\bibinfo {volume} {51}},\ \bibinfo {pages} {3571}
  (\bibinfo {year} {1995})}\BibitemShut {NoStop}%
\bibitem [{\citenamefont {Neumann}\ and\ \citenamefont
  {Handy}(1997)}]{NeumannCPL97}%
  \BibitemOpen
  \bibfield  {author} {\bibinfo {author} {\bibfnamefont {R.}~\bibnamefont
  {Neumann}}\ and\ \bibinfo {author} {\bibfnamefont {N.~C.}\ \bibnamefont
  {Handy}},\ }\href@noop {} {\bibfield  {journal} {\bibinfo  {journal} {Chem.
  Phys. Lett.}\ }\textbf {\bibinfo {volume} {266}},\ \bibinfo {pages} {16}
  (\bibinfo {year} {1997})}\BibitemShut {NoStop}%
\bibitem [{\citenamefont {Cancio}\ \emph {et~al.}(2012)\citenamefont {Cancio},
  \citenamefont {Wagner},\ and\ \citenamefont {Wood}}]{CancioIJQC12}%
  \BibitemOpen
  \bibfield  {author} {\bibinfo {author} {\bibfnamefont {A.~C.}\ \bibnamefont
  {Cancio}}, \bibinfo {author} {\bibfnamefont {C.~E.}\ \bibnamefont {Wagner}},
  \ and\ \bibinfo {author} {\bibfnamefont {S.~A.}\ \bibnamefont {Wood}},\
  }\href@noop {} {\bibfield  {journal} {\bibinfo  {journal} {Int. J. Quantum
  Chem.}\ }\textbf {\bibinfo {volume} {112}},\ \bibinfo {pages} {3796}
  (\bibinfo {year} {2012})}\BibitemShut {NoStop}%
\end{thebibliography}%

\end{document}